\documentclass[twoside,a4paper]{book} %blank pages may be inserted

\usepackage{amsmath,amssymb,bm,tabularx}
\usepackage{float}
\usepackage[dvips]{graphicx,color}
\usepackage{fancybox} %For table of contents
\usepackage{fancyhdr} %For header and footer
\usepackage[square,comma,numbers,sort&compress]{natbib}
\title{\LARGE\textsf{Ph.D. Thesis}\\
\vskip 0.7cm
\Large\bf{Anisotropic Ground States of the Quantum Hall System \\ with Currents}
\vskip 3.5cm
}

\author{\bf{Kazumi Tsuda} \\ {\it Department of Physics, Hokkaido
University, Sapporo 060-0810, Japan}}

\date{March, 2008}

\begin{document}

%=== Front matter ===%
\frontmatter %page numbers are i, ii, iii...
\maketitle
%\begin{titlepage}
%\begin{center}
%\LARGE{\textsf{Ph.D. Thesis}}\\
%\vskip 6zh
%\LARGE{\bf{Anisotropic Ground States  of  the Quantum Hall System \\ with Currents}}
%\vskip 15zh
%\Large\bf{Kazumi Tsuda} \\ 
%{\n Department of Physics, Hokkaido University, Sapporo 060-0810,
% Japan} \\
%\vskip 1zh
%March, 2008
%\end{center}
%\end{titlepage}

%=== Acknowledgment ===%
\chapter{Acknowledgment}
\noindent
First, I thank my supervisor, 
Professor Kenzo Ishikawa for his
continuous support in the Ph.D program. 
When I faced formidable obstacles in study, 
discussions with him were always very helpful. 
He showed me the different ways to 
approach a research problem and always
insisted on working hard. 
It has really been a wonderful experience working with him.

\vskip 0.7cm %insert vertical space

\noindent 
I thank my collaborator, Dr.~Nobuki Maeda 
for his continuous support throughout the Ph.D program. 
He was always there to listen and give advice. 
Interactions with him were always so friendly that 
it gave me ample opportunity to ask questions and express my ideas. 

\vskip 0.7cm 

\noindent
I am very grateful to all the other members of 
the theoretical particle physics group in Hokkaido university for 
their support and encouragement.  

\vskip 0.7cm 

\noindent
As for ex-members of our group, 
I thank Dr.~T. Aoyama and Dr.~J. Goryo 
for wonderful time we spent at a restaurant discussing physics 
and other interesting topics. 

\vskip 0.7cm 

\noindent
Last but not the least, I thank my family: my parents, brother, sister, 
aunts and grandparents for unconditional support and
encouragement to pursue my interests.

%=== Abstract ===%
\chapter{Abstract}
Anisotropic states at half-filled third and higher Landau levels are 
investigated in the system with a finite electric current. 
We study the response of the striped Hall state and the anisotropic 
charge density wave (ACDW) state against the injected current 
using the effective action.
Current distributions and a current dependence of the 
total energy are determined for both states. 
With no injected current, 
the energy of the ACDW state is lower 
than that of the striped Hall state.
We find that the energy of the ACDW state increases faster than 
that of the striped Hall state as the injected current increases. 
Hence, the striped Hall state becomes the lower energy state 
when the current exceeds the critical value. 
The critical value is estimated at about $0.04$ - $0.05$ nA, 
which is much smaller than the current used in the experiments.

%=== Main matter ===%
\mainmatter %page numbers are 1,2,3...

\tableofcontents

\chapter{Introduction}
\label{chapter1}
%%%%%%%%%%%%%%%%%%%%%%%%%%%%%%%%%%%%%%%%%%%%%%%%%%%%%%%%%%%
%%%%%%%%%%%%%%%%% Chapter1 Introduction %%%%%%%%%%%%%%%%%%%
%%%%%%%%%%%%%%%%%%%%%%%%%%%%%%%%%%%%%%%%%%%%%%%%%%%%%%%%%%%
Since the discovery of the integer quantum Hall effect by von
Klitzing et al.\ \cite{Klitzing} in 1980, 
a two-dimensional (2D) 
electron system in a
perpendicular magnetic field has attracted much attention 
%both experimentally and theoretically, 
for more than a quarter of a century. 
In 1980, von Klizing et al.\ have measured the Hall conductivity 
of the 2D electron system in strong perpendicular
magnetic fields at low temperatures, and have found that the Hall conductivity 
exhibits plateaus just at integer multiples of $e^2/h$ 
as the magnitude of the magnetic field is varied, 
where $e(>0)$ is the
electron charge and $h$ is the Plank constant. 
This is the {\it integer quantum Hall effect} (IQHE), 
and after its discovery, a 2D electron system subjected to a
perpendicular magnetic field is called a {\it quantum Hall system}. 
The integer quantization of the Hall conductivity is closely related 
to a Landau level quantization of an energy spectrum of 
free electrons as follows. 
In classical mechanics, a free electron performs a cyclotron motion 
in a uniform magnetic field $B$. Since this cyclotron motion corresponds to 
a harmonic oscillation with the cyclotron frequency $\omega_c=eB/m_e$ 
($m_e$: electron mass) in quantum mechanics, 
the energy spectrum is quantized to $E_l=\hbar\omega_c(l+1/2)$ 
($l=0,1,2,\dots$). This quantized energy level is called 
{\it Landau level} (LL)
named after Landau who solved this problem for the first time. 
The number of states in each LL is proportional to $B$, 
and when the magnetic field is strong and the lower few LLs are fully
occupied by electrons, the Hall conductivity is quantized. 
Thus, the IQHE is closely related to the LL quantization. 
In order to explain 
the emergency of the plateau structure, a localization effect of
electrons trapped by impurities should be taken into account 
and this provides another interesting topic. 
For more details about the IQHE, 
see the reviews by Ando et al.\ \cite{Ando} and St\"{o}rmer \cite{Stormer}. 

Only two years later after the discovery of the IQHE, another surprising
phenomenon, a {\it fractional quantum Hall effect} (FQHE), was found. 
In 1982, working with much higher mobility sample, Tsui et al.\
\cite{Tsui} have
discovered the fractional quantization of the Hall conductivity. 
The fractional quantizations have been observed around characteristic  
partial fillings in the lowest LL. 
Around these fillings, 
electron-electron interaction plays a predominant role, which  
results in a unique collective phenomenon. 
Among a number of theoretical studies done to solve this problem, 
it was Laughlin \cite{Laughlin} who has succeeded in discovering a 
{\it Laughlin's variational wave function} 
which explains the fractional quantization at the characteristic fillings. 
The fact that 
the Laughlin wave function is a good variational wave function 
implies that the ground state at these fillings is a
translationally invariant liquid state. 
Now, various studies support this conclusion. 
After the discovery of the Laughlin's wave function, 
his novel work has evolved into 
a concept of a {\it composite Fermion} which is an electron bound to an
integer number of magnetic flux quanta. 
Composite Fermion is based on the observation by Jain \cite{Jain1} that 
the Laughlin's wave function could be viewed as integer quantum Hall
states of composite Fermions with an even number of flux quanta, 
in a very natural way. 
%Based on 
On the basis of 
this prominent idea, most experimental results for 
the FQHE can be explained naturally, which supports the idea of the
composite particles experimentally. 
The composite Fermion has provided exciting fields of theoretical
studies since its discovery by Jain. 
For more details, see the reviews by Jain and Kamilla \cite{Jain2}. 

There are many interesting topics other than the IQHE and the FQHE, 
especially in higher LLs where various electron solid phases exist. 
In 1996, Koulakov et al.\ \cite{Koulakov} have studied a possibility of 
charge density wave (CDW) states in higher LLs, which
include bubble states (CDW states with two or more electrons per unit cell) 
and striped Hall states (unidirectional charge density wave states), 
within the Hartree-Fock (HF) theory. 
They have found that CDW and bubble states are energetically favorable 
in higher LLs, 
and 
in particular 
%especially 
around half fillings, striped Hall states are
energetically favorable. 
Indeed, in 1999, working with ultra-high mobility samples at low
temperatures, Lilly et al.\ \cite{Lilly} and Du et al.\ \cite{Du} 
have reported the highly anisotropy in the longitudinal resistivity, 
which is interpreted as due to a formation of stripe states or 
anisotropic charge density wave (ACDW) states. 
Away from half-fillings, more precisely around $\nu=4+1/4$, $4+3/4$, 
reentrant integral quantizations of the Hall 
conductivity have been observed \cite{Du,Cooper}. 
This reentrant integer quantum Hall effect (RIQHE) is interpreted as due
to a collective insulating property of a bubble state pinned by impurities. 
Although these intriguing phenomena in higher LLs have been studied
intensively over ten years, many issues still remain to be solved. 
Among them, we focus on the highly anisotropic states. 
While experimental features of the anisotropic states suggest that the
anisotropic states are the striped Hall states, 
the ACDW state has a slightly lower energy than the striped Hall state 
within the HF theory with no electric current. 
This has been an enigma as to the anisotropic states.  
In experiments of the anisotropic states, current is injected.
This effect has not been taken into account in
the previous calculations. 
We focus on this fact and 
investigate the effect of the injected current on the striped Hall states 
and the ACDW states by calculating response functions against the
injected current \cite{Tsuda1}. 
This is one of main topics of this thesis. 
The details are given in Chapter \ref{chapter4}. 

\vskip 0.7cm %insert vertical space

Another main topic of this thesis is a calculational method for 
periodic states in the
quantum Hall system by means of a {\it von Neumann lattice} (vNL) basis 
 \cite{vNL_original,vNL_review}. 
Most theoretical studies of the quantum Hall system start by expansion
of 
the Hamiltonian using a complete set of eigenstates of free electrons in
a magnetic field. 
There are mainly three different complete sets to be used according to
the purpose: 
%
%item
\begin{enumerate}
 \item Eigenstates of a magnetic angular momentum operator 
 \item Eigenstates of a magnetic translation operator in one-direction  
 \item A complete set of a vNL basis which we use aggressively in this thesis
\end{enumerate}
All of them are closely related to a magnetic flux $\phi_0=h/e$, and 
$\phi_0$ is closely related to a magnetic translation group as follows. 
As has been mentioned, eigenenergies of electrons in a uniform 
magnetic field $B$ 
are quantized to LLs and each LL has a large number of degeneracy 
given by $N_\phi=SB/\phi_0$, where $S$ is a total area of the 2D system. 
This number of degeneracy is equal to the number of magnetic flux quanta
penetrating the system, which 
means that one state in each LL occupies an area penetrated by one
magnetic flux. 
This fact can be viewed from a point of view of a magnetic translation. 
In the case of no magnetic field, 
translation operators commute with each other and they commute with 
a free Hamiltonian, too. 
Therefore, a momentum is well-defined as a good quantum number. 
On the other hand, 
in the presence of a uniform magnetic field, 
previous translation operators are not commutative with 
the free Hamiltonian anymore. 
This is because 
the vector potential generating the magnetic field is not uniform in
coordinate space even if a magnetic field is uniform. 
To obtain the proper translation operator, 
the previous translation operator %in the absence of the magnetic field
should be modified to a {\it magnetic translation operator} which is
commutative with the free Hamiltonian in the uniform magnetic field. 
After we obtain the proper translation operators, 
two linearly independent magnetic translation operators are still 
non-commutative with each other for an arbitrary translation. 
Only discrete translation operators 
are commutative with each other, which span a 
{\it magnetic translation group} \cite{Zak}. 
For these discrete operators, 
a momentum is well-defined as a good quantum number, and 
the momentum is  reduced to a magnetic Brillouin zone 
%due to 
owing to 
the
discrete translation symmetry. 
Actually, the vNL basis is an eigenstate of these discrete magnetic
translation operators so that it has a momentum as a good quantum
number and periodicities in two linearly independent directions. 
The area spanned by the two different unit magnetic translation operators 
is given by $S_a=\phi_0/B=h/eB$ which corresponds to an area penetrated by
$\phi_0$. 
Hence, the area occupied by one electron and a magnetic translation
group are closely related to each other. 

%Whereas, 
Three complete sets mentioned above are discretized as 
one state occupies an area $S_a$ as follows:  
%
%item
\begin{enumerate}
 \item For the complete set of eigenstates of a magnetic angular momentum
       operator, 
       the eigenstate is characterized by the eigenvalue of a magnetic 
       angular momentum operator $\hbar(l-m)$, where $l$ is the LL index and
       $m$ is a nonnegative integer. 
       The probability density of the eigenstate with $m$ is 
       localized on 
       the circumference with a radius $r_m=\sqrt{m/\pi}\, a$ with $a=\sqrt{h/eB}$, 
       which gives $S_a$ per eigenstate. 
 \item For the complete set of eigenstates of a translation operator in one
       direction (which is set to $y$), while the probability density 
       in the $y$-direction is uniform, 
       the density in the $x$-direction is localized at 
       discrete coordinates with  
       the interval $\Delta x=\phi_0/B L_y$, where $L_y$ is the system size in
       the $y$-direction, which gives $S_a$ per eigenstate, too. 
 \item For the complete set of the vNL basis, 
       the probability density is 
       localized at lattice sites with an area of the unit
       cell given by $S_a$, which naturally gives $S_a$ per eigenstate. 
\end{enumerate}
Thus, all of three basis sets are closely related to $\phi_0$ which is 
closely related to 
the magnetic translation group. 
As for the vNL basis, the periodicity of the vNL can be deformed 
as long as the area of the unit cell is kept to $S_a$. 
From this point of view, we develop the vNL formalism to analyze 
periodic states, which include striped Hall states, ACDW states, and Bubble
states. The vNL basis is the most suitable basis among 
three sets of basis mentioned above since 
it can be deformed as the periodicity of the vNL 
coincides with the periodicity of the periodic state. 
The details are given in Chapter \ref{chapter3}. 
The study of the highly anisotropic states in this thesis is 
one of applications of this formalism. 
There will be some other applications of this formalism 
to study periodic states in 
the quantum Hall system, which is one of our future plans. 

\vskip 0.7cm

This thesis is organized as follows. 
In Chapter \ref{chapter2}, a von Neumann lattice formalism is reviewed. 
Some useful relations and features peculiar to the von Neumann lattice
formalism are also reviewed. 
In Chapter \ref{chapter3}, 
a calculational method by means of the von Neumann lattice formalism is 
developed to study periodic states in the quantum Hall system.  
%methods to study 
%periodic states in the quantum Hall system by means of the von Neumann
%lattice basis are developed. 
In Chapter \ref{chapter4}, effects of an injected current on highly
anisotropic states are studied using the von Neumann lattice formalism
developed in Chapter \ref{chapter3}. 
Some concrete calculations are presented in Appendixes.

\chapter{Review of the von Neumann lattice formalism}
\label{chapter2}
%%%%%%%%%%%%%%%%%%%%%%%%%%%%%%%%%%%%%%%%%%%%%%%%%%%%%%%%%%%%%
%%% Chapter 2 Review of the von Neumann lattice formalism %%%
%%%%%%%%%%%%%%%%%%%%%%%%%%%%%%%%%%%%%%%%%%%%%%%%%%%%%%%%%%%%%
In this chapter, we review the von Neumann lattice formalism which is 
a powerful tool to study periodic states in the quantum Hall
system as has been mentioned in the previous chapter. 
The von Neumann lattice basis is an eigenstate of commutative 
magnetic translation operators, which spans a Hilbert space in the
quantum Hall system in conjunction with a Landau level space. 
The von Neumann lattice basis is first introduced as a complete set of
coherent states. Since they are coherent states, 
two different states are not orthogonalized. 
However, by Fourier transforming and normalizing them, 
we obtain 
a discrete set of orthonormal von Neumann lattice basis 
as a complete set of one-particle eigenstates. 
We review this procedure and show that the obtained basis is indeed 
the eigenstate of commutative magnetic translation operators. 
Some useful relations and features peculiar to the von Neumann lattice basis 
are also reviewed. 

Note that in this thesis, we only consider the rectangular von Neumann
lattice basis for simplicity. However, the oblique von Neumann lattice
can be obtained by a similar procedure. For details, 
see Ref.~\cite{vNL_review}.

%%%%%%%%%%%%%%%%%%%%%%%%%%%%%%%%%%%%%%%%%%%%%%%%%%%%%%%
%% section 2-1 Orthonormal von Neumann lattice basis %%
%%%%%%%%%%%%%%%%%%%%%%%%%%%%%%%%%%%%%%%%%%%%%%%%%%%%%%% 
\section{Orthonormal von Neumann lattice basis}

%%%%%%%%%%%%%%%%%%%%%%%%%%%%%%%%%%%%
% subsection 2-1-1 Coherent states %
%%%%%%%%%%%%%%%%%%%%%%%%%%%%%%%%%%%% 
\subsection{Landau level quantization} 
Let us consider a two-dimensional (2D) electron system in the $x$-$y$ plane
subjected to a 
perpendicular magnetic field ${\bf B}=(0,0,B)$. 
In the absence of electron-electron interactions, 
electrons just perform a cyclotron
motion with a cyclotron frequency $\omega_c=eB/m_e$. 
Since this cyclotron motion corresponds to a harmonic oscillator in quantum
mechanics, one expects that the energy level is quantized to 
$E_l=\hbar\omega_c(l+1/2)$ with $l=0,1,2,\dots$. 
We confirm this quantization by a microscopic analysis below.

The cyclotron motion is represented by a guiding center coordinates 
${\bf X}=(X,Y)$ and the relative coordinates ${\bm \xi}=(\xi,\eta)$ as 
% 
%== Eq ==%
%guiding center coordinate and relative coordinate
\begin{equation}
 x=X+\xi, \quad y=Y+\eta, 
\end{equation}
\begin{equation}
 \xi=\frac{1}{eB}(-i\hbar\partial_y+eA_y),
  \quad \eta=-\frac{1}{eB}(-i\hbar\partial_x+eA_x).  
\end{equation}
where 
${\bf B}=\nabla\times{\bf A}({\bf x})$ and $e>0$ is the electron charge. 
Here, ${\bf X}$ and ${\bm \xi}$ satisfy the following commutation relations,
%
%== Eq ==% 
%commutation relations of the guiding center coordinates 
\begin{gather}
 [X,Y]=-[\xi,\eta]=i\frac{\hbar}{eB},\quad 
 [X,\xi]=[X,\eta]=[Y,\xi]=[Y,\eta]=0.
 \label{eq2:commutation}
\end{gather} 
By using these coordinates, the one-particle free Hamiltonian 
$H_0=(-i\hbar\nabla+e{\bf A}({\bf x}))^2/2m_e$ is expressed by 
%
%== Eq ==%
%one-particle Hamiltonian in a perpendicular magnetic field
\begin{equation}
 H_0=\frac{m_e\omega_c^2}{2}(\xi^2+\eta^2). 
 \label{eq2:Free_Hamiltonian1}
\end{equation}

Next, we introduce new operators $L_A,L_A^\dag,L_B,L_B^\dag$ as 
%== Eq ==%
%New operators L_A,B
\begin{gather}
 L_A=-\frac{1}{\sqrt{2}\,l_B}(\eta+i\xi),\quad 
 L_A^\dag=-\frac{1}{\sqrt{2}\,l_B}(\eta-i\xi)\notag\\
 L_B=\frac{1}{\sqrt{2}\,l_B}(X+iY),\quad 
 L_B^\dag=\frac{1}{\sqrt{2}\,l_B}(X-iY),
 \label{eq2:LALB} 
\end{gather}
where $l_B=\sqrt{\hbar/eB}$ is the magnetic length. 
Since these operators satisfy the following commutation relations, 
%
%== Eq ==%
%Commutation relations for L_A,B
\begin{gather}
 [L_A,L_A^\dag]=[L_B,L_B^\dag]=1,\quad 
 [L_A,L_B]=[L_A^\dag,L_B^\dag]=[L_A,L_B^\dag]=[L_A^\dag,L_B]=0. 
\end{gather}
new operators can be regarded as independent two kinds of 
creation and annihilation operators. 
Using these operators, 
we rewrite the one-particle free Hamiltonian $H_0$ as 
%
%== Eq ==%
%Hamiltonian by L_A
\begin{equation}
 H_0=\hbar\omega_c(L_A^\dag L_A + \frac{1}{2}). 
 \label{eq2:Free_Hamiltonian2}
\end{equation}
This is a Hamiltonian of a harmonic oscillator with a frequency $\omega_c$. 
The eigenstate and eigenvalue of $H_0$ are given by 
%
%== Eq =%
%Landau level eigenstates
\begin{gather}
 |f_l\rangle=\frac{(L_A^\dag)^l}{\sqrt{l!}}|f_0\rangle,\quad 
 H_0|f_l\rangle=E_l|f_l\rangle, 
\end{gather} 
where $|f_0\rangle$ satisfies $L_A|f_0\rangle=0$.
Hence, the naive expectation with respect to the quantization of the energy level is confirmed. 
The energy level labeled by $l$, $E_l=\hbar\omega_c(l+1/2)$, is called 
the $l$th {\it Landau level} (LL).

%%%%%%%%%%%%%%%%%%%%%%%%%%%%%%%%%%%%%%%%%%%%%%%%%%%%%%%%%%%%%%%%%%%%%%%%%%
% subsection 2-1-2 Eigenstates of the magnetic angular momentum operator %
%%%%%%%%%%%%%%%%%%%%%%%%%%%%%%%%%%%%%%%%%%%%%%%%%%%%%%%%%%%%%%%%%%%%%%%%%%
\subsection{Eigenstate of the magnetic angular momentum operator}
In the above discussion, the guiding center coordinates are not used to
derive the LL quantization. Since the free Hamiltonian $H_0$ is
commutative with the guiding center coordinates, each LL has degeneracy 
coming from the eigenstate related to the guiding center coordinates. 
The most simple one is the eigenstate of the operator $L^\dag_B L_B$. 
This eigenstate is given by
%
%== Eq ==%
%Magnetic angular momentum eigenstates 
\begin{equation}
|J_M\rangle=\frac{(L^\dag_B)^M}{\sqrt{M!}}|J_0\rangle, \quad 
L^\dag_B L_B |J_M\rangle=M |J_M\rangle \quad (M=0,1,2,\dots), 
\end{equation}
where $L_B|J_0\rangle=0$. 
The direct product 
$|f_l\rangle \otimes |J_M\rangle \equiv |l,M\rangle$ is a simultaneous
eigenstate of the free Hamiltonian $H_0$ and the operator 
$J=(eB/2)(\xi^2+\eta^2-X^2-Y^2)=\hbar(L^\dag_A L_A-L^\dag_B L_B)$, and 
spans a Hilbert space of a one-particle state. 
Actually, the operator $J$ is a magnetic angular momentum operator 
since it commutes with the free Hamiltonian, and when the magnetic field
is set to zero, 
it takes $J(B=0)=-i\hbar(x\partial_y-y\partial_x)$ which is an angular momentum
operator in the absence of a magnetic field. 
Here, the fact that ${\bf A}({\bf x})$ is proportional to $B$ is used. 
The electron density of the eigenstate with $M$ is 
localized on 
the circumference with a radius $r_M=\sqrt{M/\pi}\, a$ with $a=\sqrt{h/eB}$, 
which gives the area $S_a=h/eB=\phi_0/B$ ($\phi_0=h/e$ is a magnetic flux
quantum) 
per eigenstate as has been mentioned in
the previous chapter (see also Appendix \ref{appA:wave_function}). 
Note that each LL has the degeneracy factor $N_\phi=BS/\phi_0$ 
($S$ is a total area of the 2D system), which means that the 
number of degeneracy in each LL is equal to the number of magnetic flux
quanta penetrating the 2D system.

%%%%%%%%%%%%%%%%%%%%%%%%%%%%%%%%%%%%
% subsection 2-1-3 Coherent states %
%%%%%%%%%%%%%%%%%%%%%%%%%%%%%%%%%%%% 
\subsection{Coherent state}
There is another eigenstate constructed by the guiding center
coordinate, that is, a coherent states of the guiding center
coordinates,  
%
%== Eq ==%
%von Neumann lattice eigenstates
\begin{equation}
(X+iY)|\alpha_{m,n}\rangle = z_{m,n}|\alpha_{m,n}\rangle,\quad 
z_{m,n}=a(r_s m+i\frac{n}{r_s}),  
\end{equation}
where $m$ and $n$ are integers. 
In coordinate space, these coherent states are localized at the
rectangular lattice point $a(mr_s, n/r_s)$, where 
$a=\sqrt{h/eB}$ is a lattice constant, 
and $r_s$ is an asymmetry parameter of the unit cell 
(see Appendix \ref{appA:wave_function}). 
The magnitude of $a$ is of the order of tens of nano meter for a few
Tesla of magnetic field. 
Since the number of these coherent states in each Landau level is equal to 
$N_\phi$, the direct product 
$|f_l\rangle \otimes |\alpha_{m,n}\rangle \equiv |l,{\bf N}\rangle$ 
with ${\bf N}=(m,n)$ forms the complete set 
of a one-particle state. 
Note that the completeness of this coherent set is ensured
mathematically \cite{Perelomov,Bargmann,Bacry} and this set is called {\it von Neumann lattice} (vNL)
basis \cite{vNL_original,vNL_review}.

Since $[L_B,L_B^\dag]=1$, we can take $L_B=\partial/\partial L_B^\dag$. 
Thus, 
%
%== Eq ==%
%Wave function of vNL1
\begin{equation}
 L_B|\alpha_{m,n}\rangle=\frac{\partial}{\partial
  L_B^\dag}|\alpha_{m,n}\rangle
  =\frac{z_{m,n}}{\sqrt{2}\, l_B}|\alpha_{m,n}\rangle,  
\end{equation}
and $|\alpha_{m,n}\rangle$ is given by 
%
%== Eq ==%
%Wave function of vNL2
\begin{equation}
 |\alpha_{m,n}\rangle=
  C_{m,n}e^{(z_{m,n}/\sqrt{2}\, l_B)L_B^\dag}|\alpha_{0,0}\rangle, 
\end{equation}
where $C_{m,n}$ is a normalization constant. 
If we set $\langle \alpha_{m,n}|\alpha_{m,n}\rangle=1$, 
then $|C_{m,n}|=e^{-\pi|z_{m,n}|^2/2a^2}$ with an arbitrary phase
factor. 
We use $e^{i\pi(m+n+mn)}$ as a phase factor of $C_{m,n}$
throughout this thesis and set $a=1$ unless otherwise stated. 
%The natural unit $\hbar=c=1$ is also used. 

The eigenstate $|\alpha_{m,n}\rangle$ is not an orthogonal basis since it is a coherent
state. 
However, the inner product, 
%
%== Eq ==%
%Inner product of alpha
\begin{equation}
\langle \alpha_{m,n} | \alpha_{m',n'}\rangle =
 e^{i\pi\{(m-m')+(n-n')+(m-m')(n-n')\}}
 e^{-(\pi/2)\{(m-m')^2r_s^2+(n-n')^2/r_s^2\}},  
\end{equation}
is a function of the difference, $(m-m')$ and $(n-n')$, so that 
the Fourier representation of $|\alpha_{m,n}\rangle$ becomes an orthogonal
basis. 
Indeed, by using the momentum representation of $|\alpha_{m,n}\rangle$, 
%
%== Eq ==%
%Fourier transformation of alpha
\begin{equation}
 |\alpha_{\bf p}\rangle=\sum_{m,n}e^{imp_x+inp_y}|\alpha_{m,n}\rangle, 
\end{equation}
the inner product of $|\alpha_{\bf p}\rangle$ is calculated as (see
Appendix \ref{appA:inner_product})
%
%== Eq ==%
%Inner product of alpha2
\begin{gather}
 \langle \alpha_{\bf p}|\alpha_{{\bf p}'}\rangle =
  \beta^\ast({\bf p}) \beta({\bf p}')(2\pi)^2
  \sum_{\bf N}\delta^2({\bf p}-{\bf p}'-2\pi{\bf N})e^{i\phi({\bf
 p},{\bf N})},\notag \\
 \beta({\bf p})=(\sqrt{2}\, r_s)^{1/2}e^{-(r_sp_y)^2/4\pi}\theta_1
 \left(\frac{p_x+ir_s^2p_y}{2\pi}\Biggm| ir_s^2\right), 
 \label{eq2:inner_product}
\end{gather}
where
$\theta_1(v|\tau)=i\sum^\infty_{n=-\infty}
e^{i\pi n+i\pi\tau(n-1/2)^2+i\pi v(2n-1)}$ 
is a Jacobi's theta function of the first kind,  
$\phi({\bf p},{\bf N})=\pi(m+n)-p_x n$, 
and we set $a=1$.
Thus, we obtain the orthonormal vNL basis by normalizing 
$|\alpha_{\bf p}\rangle$ as 
%
%== Eq ==%
%Normalization of alpha
\begin{equation}
 |\beta_{\bf p}\rangle=\frac{1}{\beta({\bf p})}|\alpha_{\bf p}\rangle .
\end{equation}
From the property of the theta function, $\beta({\bf p})$ and
$|\beta_{\bf p}\rangle$ obeys the following 
nontrivial boundary condition, 
%
%== Eq ==%
%Boundary condition of beta
\begin{equation}
 \beta({\bf p}+2\pi{\bf N})=e^{i\phi({\bf p},{\bf N})}\beta({\bf
  p}),\quad 
 |\beta_{{\bf p}+2\pi{\bf N}}\rangle=e^{-i\phi({\bf p},{\bf N})}
 |\beta_{\bf p}\rangle.
 \label{eq2:boundary}
\end{equation} 
Using this boundary condition, we obtain the momentum space reduced to 
the Brillouin zone (BZ), $-\pi<p_x,p_y<\pi$, hence, the Hilbert space of
a one-particle state is spanned by the state 
$|l,{\bf p}\rangle=|f_l\rangle\otimes |\beta_{\bf p}\rangle$. 
Owing to 
%Due to 
the boundary condition (\ref{eq2:boundary}), 
$|\beta_{\bf p}\rangle$ satisfies 
%
%== Eq ==%
%Orthonormal equation for beta
\begin{equation}
 \langle \beta_{\bf p}|\beta_{{\bf p}'}\rangle=(2\pi)^2\sum_{\bf N}
  \delta^2({\bf p}-{\bf p}'-2\pi{\bf N})e^{i\phi({\bf p}, {\bf N})}. 
\label{eq2:orthonormal}
\end{equation}
The left hand side and the right hand side of Eq.~(\ref{eq2:orthonormal}) obey the
same boundary condition. If we restrict the momentum range to 
the first Brillouin zone, then Eq.~(\ref{eq2:orthonormal}) becomes 
$\langle \beta_{\bf p}|\beta_{{\bf p}'}\rangle=(2\pi)^2\delta({\bf
p}-{\bf p}')$ which expresses a usual orthonormal relation.

%%%%%%%%%%%%%%%%%%%%%%%%%%%%%%%%%%%%%%%%%%%%%%%%%%%%%%%%%%%%%%%%%%%%%%%%%
%% Section2-2 Magnetic translation group and von Neumann lattice basis %%
%%%%%%%%%%%%%%%%%%%%%%%%%%%%%%%%%%%%%%%%%%%%%%%%%%%%%%%%%%%%%%%%%%%%%%%%%
\section{Magnetic translation group and von Neumann Lattice basis}

%%%%%%%%%%%%%%%%%%%%%%%%%%%%%%%%%%%%%%%%%%%%%%%
% Subsection 2-2-1 Magnetic translation group %
%%%%%%%%%%%%%%%%%%%%%%%%%%%%%%%%%%%%%%%%%%%%%%%
\subsection{Magnetic translation group}
The translation operator in the system with zero magnetic field is 
given by 
%
%== Eq ==%
%normal translation operator
\begin{equation}
 T({\bm \delta})=e^{i{\bm \delta}\cdot \hat{\bf p}/\hbar}
\end{equation} 
where $\hat{\bf p}=-i\hbar\nabla$. 
Indeed, if $T({\bm \delta})$ acts on a one-particle wave function 
$\psi({\bf x})$, 
then $\psi({\bf x})$ is translated to $\psi({\bf x}+{\bm \delta})$. 
The translation operator $T({\bm \delta})$ commutes with the
one-particle free Hamiltonian $H_0(B=0)=\hat{\bf p}^2/2m_e$, 
and $T({\bm \delta}_x)$ and $T({\bm \delta}_y)$ are commutative 
for an arbitrary $\delta_x$ and $\delta_y$, 
where ${\bm \delta}_x=\delta_x\hat{\bf e}_x$, 
${\bm \delta}_y=\delta_y\hat{\bf e}_y$, and $\hat{\bf e}_x$ and
$\hat{\bf e}_y$ are unit vectors in the $x$ and $y$ directions, respectively. 
Hence, we can take 
$\{H_0,T({\bm \delta}_x),T({\bm \delta}_y)\}$ as 
a set which can be diagonalized simultaneously. 
The situation becomes different when a magnetic field is applied. 
Let us consider the 2D electron system in a uniform magnetic field ${\bf B}=(0,0,B)$. 
In this case, the one-particle Hamiltonian is given by 
$H_0=(\hat{\bf p}+e{\bf A}({\bf x}))^2/2m_e$ 
and the electron performs a cyclotron
motion. Clearly, $T({\bm \delta})$ does not commute with $H_0$ 
since the vector potential ${\bf A}({\bf x})$ depends on ${\bf x}$ even
if the magnetic field is uniform. 
In order to obtain the proper translation operator, 
{\it magnetic translation operator}, which commutes with $H_0$, 
the gauge transformation which brings ${\bf A}({\bf x}+{\bm \delta})$ back to
${\bf A}({\bf x})$ is required in conjunction with $T({\bf \delta})$. 
This magnetic translation operator is given by 
%
%== Eq ==%
%magnetic translation operator
\begin{equation}
 \hat{T}({\bm \delta})=e^{i{\bm \delta}\cdot {\bf K}/\hbar},\quad 
 {\bf K}=\hat{\bf p}+e{\bf A}({\bf x})-e{\bf B}\times {\bf x}. 
\end{equation}
Since ${\bf K}$ is rewritten as ${\bf K}=eB(Y,-X)$ and 
$H_0$ is rewritten as Eq.~(\ref{eq2:Free_Hamiltonian1}), it is clear that 
$\hat{T}({\bm \delta})$ indeed commutes with $H_0$.

The magnetic translation operator $\hat{T}({\bm \delta})$ includes the
spatial translation and the gauge transformation. 
To make this point clear, we divide $\hat{T}({\bm \delta})$ into two
parts. 
Using the Cambell-Hausdorff formula, 
%
%== Eq ==%
%Cambell-Hausdorff formula
\begin{equation}
 \exp{A}\exp{B}=\exp(A+B+\frac{1}{2}[A,B]+\frac{1}{12}[A-B,[A,B]]+\dots), 
  \label{eq2:CH_formula}
\end{equation}
we rewrite $\hat{T}({\bm \delta})$ as 
%
%== Eq ==%
%Divide the magnetic translation operator
\begin{equation}
 \hat{T}({\bm \delta})=
e^{i\theta/2}
e^{i{\bm \delta}\cdot e({\bf A}-{\bf B}\times {\bf x})/\hbar}
e^{i{\bm \delta}\cdot \hat{\bf p}/\hbar},
\end{equation}
where 
$i\theta=
[i{\bm \delta}\cdot e({\bf A}({\bf x})-{\bf B}\times {\bf x})/\hbar, 
i{\bm \delta}\cdot \hat{\bf p}/\hbar]$ 
is constant and 
the vector potential ${\bf A}({\bf x})$ is assumed to 
be a linear function of {\bf x}. 
This expression clearly shows that $\hat{T}(\bm \delta)$ consists of 
the spatial translation operator $T({\bm \delta})$ and the gauge
transformation operator 
$e^{i\theta/2}e^{i{\bm \delta}\cdot e({\bf A}-{\bf B}\times {\bf
x})/\hbar}$.

The magnetic translation operators do not commute with each other for
arbitrary two translations owing to 
%due to 
the non-commutativity of $K_x$ and
$K_y$, i.e., $[K_x,K_y]=i\hbar eB$. 
For simplicity, we consider two linearly
independent magnetic translation operators, 
$\hat{T}({\bm \delta}_x)=e^{i\delta_xK_x/\hbar}$ 
and $\hat{T}({\bm \delta}_y)=e^{i\delta_yK_y/\hbar}$. 
Using the commutation relation of $K_x$ and $K_y$, we obtain 
%
%== Eq ==%
%commutation relation of magnetic translation operator
\begin{equation}
 \hat{T}({\bm \delta}_x)\hat{T}({\bm \delta}_y)=
  \hat{T}({\bm \delta}_y)\hat{T}({\bm \delta}_x)
  e^{-ieB\delta_x\delta_y/\hbar}.
\end{equation}
Clearly, $\hat{T}({\bm \delta}_x)$ does not commute with 
$\hat{T}({\bm \delta}_y)$ for arbitrary $\delta_x$ and $\delta_y$. 
However, if $\Phi\equiv B\delta_x\delta_y$ is equal to $2n\pi\hbar/e$
with an integer $n$, $\hat{T}({\bm \delta}_x)$ commutes with 
$\hat{T}({\bm \delta}_y)$. Here, notice that $\Phi$ 
represents the magnetic flux penetrating 
the rectangular area which is spanned by 
${\bm \delta}_x$ and ${\bm \delta}_y$, and 
$\phi_0=2\pi\hbar/e=h/e$ represents a magnetic flux quantum. 
This implies that $\hat{T}({\bm \delta_x})$ and 
$\hat{T}({\bm \delta}_y)$ are commutative 
only when the magnetic flux penetrating 
the area spanned by ${\bm \delta}_x$ and ${\bm \delta}_y$ is equal to 
integral multiples of a magnetic flux quantum $\phi_0$. 
This statement can be easily generalized for arbitrary 
two linearly-independent magnetic translation operators. 
Hence, in the system with a uniform magnetic field, 
we can take $\{H_0,\hat{T}({\bm \delta}_1),\hat{T}({\bm \delta}_2)\}$
with ${\bm \delta}_1$ and ${\bm \delta}_2$ which satisfy the relation 
${\bf B}\cdot ({\bm \delta}_1\times {\bm \delta}_2)=\phi_0 n$, 
as the simultaneously diagonalizable set. 
$\hat{T}({\bm \delta}_1)$ and $\hat{T}({\bm \delta}_2)$ form a 
group called {\it  magnetic translation group} \cite{Zak}.

%%%%%%%%%%%%%%%%%%%%%%%%%%%%%%%%%%%%%%%%%%%%%%%%%%%%%%%%%%%%%%%%%%%%%%%%
% Subsection 2-2-2 Eigenstate of the commutative magnetic translation  %
% operators                                                            %   
%%%%%%%%%%%%%%%%%%%%%%%%%%%%%%%%%%%%%%%%%%%%%%%%%%%%%%%%%%%%%%%%%%%%%%%%
\subsection{Eigenstate of the commutative magnetic translation operators}
The vNL basis is a simultaneous eigenstate of the set 
$\{H_0, \hat{T}(r_s a, 0), \hat{T}(0, a/r_s)\}$,  
so that it is an irreducible representation of magnetic translation
group. 
Let us check this statement in what follows.

First, $\hat{T}(r_s a, 0)$ and $\hat{T}(0, a/r_s)$ are 
the magnetic translation operators along the $x$ and $y$ direction of 
a rectangular unit cell, respectively. 
Since the magnetic flux penetrating the unit cell is equal to $\phi_0$, 
$\hat{T}(r_s a, 0)$ and $\hat{T}(0, a/r_s)$ are 
commutative. 
Next, when $\hat{T}(r_s a,0)$ and $\hat{T}(0, a/r_s)$ act on
$|\alpha_{m,n}\rangle$, 
$|\alpha_{m,n}\rangle$ is translated to the nearest neighbor states, 
$|\alpha_{m-1,n}\rangle$ and $|\alpha_{m,n-1}\rangle$ up to a phase factor,
respectively, that is, 
%
%== Eq ==%
%Action of the magnetic translation operators on the vNL basis
\begin{equation}
 \hat{T}(r_s a,0)|\alpha_{m,n}\rangle = -|\alpha_{m-1,n}\rangle, \quad 
\hat{T}(0,\frac{a}{r_s})|\alpha_{m,n}\rangle = -|\alpha_{m,n-1}\rangle. 
\label{eq2:action_on_alpha}
\end{equation}
Equation (\ref{eq2:action_on_alpha}) gives the action of 
$\hat{T}(r_s a, 0)$ and $\hat{T}(0, a/r_s)$ on 
$|\beta_{\bf p}\rangle$ by 
\begin{equation}
 \hat{T}(r_s a, 0)|\beta_{\bf p}\rangle=-e^{ip_x}|\beta_{\bf p}\rangle,
  \quad 
\hat{T}(0, \frac{a}{r_s})|\beta_{\bf p}\rangle=-e^{ip_y}|\beta_{\bf
p}\rangle,  
\label{eq2:convention}
\end{equation}
hence, the vNL basis $|\beta_{\bf p}\rangle$ is a simultaneous eigenstate
with eigenvalues $-e^{ip_x}$ and $-e^{ip_y}$ for $\hat{T}(r_s a, 0)$
and $\hat{T}(0, a/r_s)$, respectively. 
Note that the minus sign in Eq.~(\ref{eq2:convention}) is a convention
and 
we can take a different phase factor instead of the minus sign
in Eq.~(\ref{eq2:convention}), 
in which case the vNL basis $|\beta_{\bf p}\rangle$ has a different
normalization factor \cite{Ferrari,Rashba}.

%%%%%%%%%%%%%%%%%%%%%%%%%%%%%%%%%%%%%%%%%%%%%%%
%% Section 2-3 von Neumann lattice formalism %%
%%%%%%%%%%%%%%%%%%%%%%%%%%%%%%%%%%%%%%%%%%%%%%%
\section{von Neumann lattice formalism}
In this section, we derive some useful relations in the quantum Hall
system in the vNL formalism. We use the natural unit $(\hbar=c=1)$, 
set $a=1$, and ignore the spin degree of freedom, unless otherwise
stated.

%%%%%%%%%%%%%%%%%%%%%%%%%%%%%%%%%%%%%%%%%%%%%%%%%%%%%%%%%%%%
% Subsection 2-3-1  Hamiltonian of the quantum Hall system %  
%%%%%%%%%%%%%%%%%%%%%%%%%%%%%%%%%%%%%%%%%%%%%%%%%%%%%%%%%%%%
\subsection{Hamiltonian of the quantum Hall system}
In the second quantized form, 
the Hamiltonian of the quantum Hall system is given by 
%
%== Eq ==%
%Hamiltonian in the second quantization 
\begin{gather}
  H=\mathcal{K}+\mathcal{V}, 
 \label{eq2:Hamiltonian}\\
  \mathcal{K}=\int d^2x 
 \Psi^\dag({\bf x})\frac{(-i\nabla+
 e{\bf A}({\bf x}))^2}{2m_e}\Psi({\bf x}),\quad 
 \mathcal{V}=\frac{1}{2}\int d^2x d^2x':\rho({\bf x})
  V({\bf x}-{\bf x}')\rho({\bf x}'):,
\end{gather}
where $\Psi({\bf x})$ is an electron field operator, 
$\rho({\bf x})=\Psi^\dag({\bf x})\Psi({\bf x})$ is a density operator, 
colons represent a normal ordering with respect to creation and
annihilation operators, 
$V({\bf x})=4\pi e^2/\epsilon |{\bf x}|$ is a Coulomb potential, and 
$\epsilon$ is the dielectric constant.
The electron field operator is expanded by the vNL basis as
%
%== Eq ==%
%Electron field operator
\begin{equation}
\Psi({\bf x})=\sum^\infty_{l=0}
 \int_{{\rm BZ}}\frac{d^2p}{(2\pi)^2}b_l({\bf p})\langle {\bf
 x}|l,{\bf p}\rangle, 
\label{eq2:electron_field_operator}
\end{equation}
where 
$b_l({\bf p})$ obeys the boundary condition, 
%
%== Eq ==%
%boundary condition
\begin{equation}
 b_l({\bf p}+2\pi{\bf N})=e^{i\phi({\bf p},{\bf N})}b_l({\bf p}),  
\label{eq2:boundary_condition_for_b}
\end{equation} 
and satisfies the following anti-commutation relation,  
%
%== Eq ==%
%Commutation relation for b
\begin{equation}
 \{b_l({\bf p}), b^\dag_{l'}({\bf p}')\}=\delta_{l,l'}\sum_{\bf N}(2\pi)^2
  \delta^2({\bf p}-{\bf p}'-2\pi{\bf N})e^{i\phi({\bf p},{\bf N})}. 
\label{eq2:commutation_for_b}
\end{equation} 
The Fourier transform of the density operator 
$\rho({\bf k})
=\int d^2x\, \Psi^\dag({\bf x})\Psi({\bf x})e^{i{\bf k}\cdot {\bf x}}$ 
is written as (Appendix \ref{appA:density_operator})
%
%== Eq ==%
%Density operator
\begin{align}
 \rho({\bf k})=&\sum_{l,l'}
  \int_{\rm BZ}\frac{d^2p}{(2\pi)^2}b_l^\dag({\bf p})
  b_{l'}({\bf p}-\hat{\bf k})
  f^0_{l,l'}({\bf k}) e^{-(ir_s/4\pi)k_x(2p_y-k_y/r_s)},
\label{eq2:density_operator}
\end{align}
where 
$f_{l,l'}^0({\bf k})=\langle f_l| e^{i{\bf k}\cdot{\bm \xi}}|f_{l'}\rangle$
(the explicit form is given in Appendix \ref{appA:LL_matrix}) and 
$\hat{\bf k}=(r_sk_x,k_y/r_s)$. 
Note that the integrand of $\rho({\bf k})$ is invariant under the
transformation ${\bf p} \to {\bf p}+2\pi{\bf N}$. 
Substituting Eqs.~(\ref{eq2:electron_field_operator}) and
(\ref{eq2:density_operator}) into Eq.~(\ref{eq2:Hamiltonian}), 
we obtain the kinetic term $\mathcal{K}$ given by 
%
%== Eq ==%
%Kinetic term
\begin{equation}
 \mathcal{K}=\sum_l E_l
  \int_{\rm BZ}\frac{d^2 p}{(2\pi)^2}b_l^\dag({\bf p})b_l({\bf p}),
  \quad 
  E_l=\omega_c(l+\frac{1}{2}), 
\end{equation}
and the Coulomb interaction term given by 
%
%== Eq ==%
%Potential term 
\begin{equation}
 \mathcal{V}=\frac{1}{2}\int\frac{d^2k}{(2\pi)^2}
  :\rho({\bf k})V({\bf k})\rho(-{\bf k}):,\quad
  V({\bf k})=\frac{2\pi q^2}{|{\bf k}|}\quad 
  \hbox{(${\bf k}\neq 0$)},\quad  q^2=\frac{e^2}{4\pi\epsilon}, 
\end{equation}
where $V(0)=0$ owing to %due to 
the charge neutrality of the system. 
Using the new density operator defined by 
%
%== Eq ==%
%bar density operator
\begin{equation}
 \bar{\rho}_{l,l'}({\bf k})\equiv \int_{\rm BZ} \frac{d^2p}{(2\pi)^2}
  b^\dag_l({\bf p})b_{l'}({\bf p}-\hat{\bf k})
  e^{-(ir_s/4\pi)k_x(2p_y-k_y/r_s)}, 
\end{equation}
we write $\rho({\bf k})$ as 
$\rho({\bf k})=\sum_{l,l'}f^0_{l,l'}({\bf k})\bar{\rho}_{l,l'}({\bf k})$ 
and obtain  
%
%== Eq ==%
%Potential term 2 
\begin{equation}
\label{eq2:Coulomb_int_term}
 \mathcal{V}=\frac{1}{2}\int\frac{d^2k}{(2\pi)^2}\sum_{l_1,l_2,l_3,l_4}
  V({\bf k})f^0_{l_1,l_2}({\bf k})f^0_{l_3,l_4}(-{\bf k})
  :\bar{\rho}_{l_1,l_2}({\bf k})\bar{\rho}_{l_3,l_4}(-{\bf k}):. 
\end{equation}

%%%%%%%%%%%%%%%%%%%%%%%%%%%%%%%%%%%%%%%%%%%%%%
% Subsection 2-3-2  Hartree-Fock Hamiltonian %  
%%%%%%%%%%%%%%%%%%%%%%%%%%%%%%%%%%%%%%%%%%%%%%
\subsection{Hartree-Fock Hamiltonian}
The Hamiltonian of the quantum Hall system takes a simple form in the 
Hartree-Fock (HF) approximation by virtue of the magnetic field. 
Both the Hartree term and the Fock term are proportional to the
density operator $\bar{\rho}_{l,l'}({\bf k})$ 
(Appendix \ref{appA:HF_Hamiltonian}). 
The Coulomb interaction term $\mathcal{V}$ is approximated by 
$\mathcal{V}_{\rm HF}-\langle \mathcal{V}_{\rm HF}\rangle /2$, 
where $\mathcal{V}_{\rm HF}$ is given by 
%
%== Eq ==%
%HF Coulomb interaction
\begin{equation}
\label{eq2:Coulomb_int_term_HF}
 \mathcal{V}_{\rm HF}=\sum_{l_1,l_2,l_3,l_4}
  \int\frac{d^2k}{(2\pi)^2}v^{\rm HF}_{l_1,l_2,l_3,l_4}(\tilde{\bf k})
  \langle\bar{\rho}_{l_1,l_2}(-\tilde{\bf k})\rangle
  \bar{\rho}_{l_3,l_4}(\tilde{\bf k}), 
\end{equation}
with the HF potential 
%
%== Eq ==%
%HF potential
\begin{equation}
 v^{\rm HF}_{l_1,l_2,l_3,l_4}({\bf k})=
 V({\bf k})
 f_{l_1,l_2}^0(-{\bf k})f_{l_3,l_4}^0({\bf k})-
 \int\frac{d^2k'}{(2\pi)^2}V({\bf k}')f_{l_1,l_4}^0(-{\bf k}')
 f_{l_3,l_2}^0({\bf k}')e^{-(i/2\pi)(k'_xk_y-k'_yk_x)},  
\label{eq2:HF_potential}
\end{equation}
where $\tilde{\bf k}=(k_x/r_s,r_sk_y)$. 
%and $f^0_{l_1,l_2}({\bf k})$ is given in Appendix \ref{appA:LL_matrix}. 
In Eq.~(\ref{eq2:HF_potential}), 
the first term and the 
second term in the right hand side represent 
the Hartree term and the Fock term, 
respectively.

If the Hamiltonian is projected to the $l$th LL, 
the kinetic term is quenched and the Hamiltonian is given only by the 
Coulomb interaction term projected to the $l$th LL. 
In this case, the $l$th LL projected Hamiltonian is given by 
%
%== Eq ==%
%Hamiltonian in the lth LL
\begin{equation}
 \mathcal{H}^{(l)}=\frac{1}{2}\int\frac{d^2k}{(2\pi)^2}
  v_l({\bf k})
  :\bar{\rho}_{l}({\bf k})\bar{\rho}_{l}(-{\bf k}):. 
\end{equation}
with 
%
%== Eq ==%
%Notation 
\begin{gather}
 v_l({\bf k})=V({\bf k})[F_l({\bf k})]^2, 
 \quad F_l({\bf k})=f^0_{l,l}({\bf k}) \\
 \bar{\rho}_l({\bf k})=\int_{\rm BZ}\frac{d^2p}{(2\pi)^2}
 b^\dag_l({\bf p})b_l({\bf p}-\hat{\bf k})
 e^{-(ir_s/4\pi)k_x(2p_y-k_y/r_s)}
 \label{eq2:projected_dens}.
\end{gather}
We call $\bar{\rho}({\bf k})$ a 
{\it projected density operator}.

The projected HF Hamiltonian is given by 
$H_{\rm HF}^{(l)}=\mathcal{H}_{\rm HF}^{(l)}-\langle \mathcal{H}_{\rm HF}^{(l)}\rangle/2$,
where 
%
%== Eq ==%
%projected HF Hamiltonian 
\begin{equation}
 \mathcal{H}_{\rm HF}^{(l)}=\int\frac{d^2k}{(2\pi)^2}
  v_l^{\rm HF}(\tilde{\bf k})\langle \bar{\rho}_l(-\tilde{\bf k})\rangle 
  \bar{\rho}_l(\tilde{\bf k}), 
  \label{eq2:HF_Hamiltonian}
\end{equation}
with 
%
%== Eq ==%
%Notation 2
\begin{equation}
 v^{\rm HF}_l({\bf k})=v_l({\bf k})-
  \int\frac{d^2k'}{(2\pi)^2}v_l({\bf k}')
  e^{(i/2\pi)(k'_xk_y-k'_yk_x)}. 
\end{equation}
These notations will be used in the following chapters.

%%%%%%%%%%%%%%%%%%%%%%%%%%%%%%%%%%%%%%%%
%% Subsection 2-3-3 Density operator  %%  
%%%%%%%%%%%%%%%%%%%%%%%%%%%%%%%%%%%%%%%%
\subsection{Density operator}
The projected density operator (\ref{eq2:projected_dens}) is 
noncommutative. Indeed, the commutation relation of projected
density operators is calculated using Eq.~(\ref{eq2:commutation_for_b}) 
as 
%
%== Eq ==%
%commutation relation of projected density operators
\begin{equation}
 [\bar{\rho}_l({\bf k}),\bar{\rho}_l({\bf k}')]=
  -2i\sin\left(
	  \frac{({\bf k}\times {\bf k}')_z}{4\pi}
	  \right)
  \bar{\rho}_l({\bf k}+{\bf k}'). 
\label{eq2:commutation_of_density}
\end{equation}
Owing to %Due to 
this noncommutativity, various phases are realized 
in the quantum Hall system.

The operator of the total number of electrons is given by 
$\hat{N}_{\rm total}=\int d^2x\rho({\bf x})=\rho({\bf k}=0)$. 
Substituting Eq.~(\ref{eq2:density_operator}) into this expression, 
we obtain %$\hat{N}_{\rm total}$ as 
%
%== Eq ==%
%operator of the total number of electrons
\begin{equation}
 \hat{N}_{\rm total}=\sum_{l,l'}\int_{\rm BZ}\frac{d^2 p}{(2\pi)^2}
  b^\dag_l({\bf p})b_{l'}({\bf p})\langle f_l|f_{l'}\rangle 
  =\sum_{l}\int_{\rm BZ}\frac{d^2p}{(2\pi)^2}b^\dag_{l}({\bf p})b_l({\bf p}). 
\end{equation}
Hence, in the LL projected space, the density operator in the momentum
space is given by $b^\dag_l({\bf p})b_l({\bf p})$.

%%%%%%%%%%%%%%%%%%%%%%%%%%%%%%%%%%%%%%%%%%%%%%%%%%%%%%%%
%% Subsection 2-3-4  Magnetic field in momentum space %%  
%%%%%%%%%%%%%%%%%%%%%%%%%%%%%%%%%%%%%%%%%%%%%%%%%%%%%%%%
\subsection{Magnetic field in momentum space}
\begin{figure}[tb]
\begin{center}
\includegraphics[width=6cm]{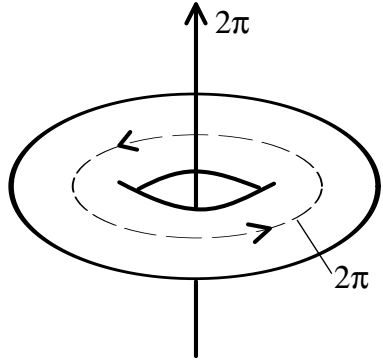}
\end{center}
\caption{\label{fig2:2Dtorus}
Two magnetic fluxes in the 2D torus in momentum space. 
}
\end{figure}

In the vNL formalism, nontrivial phase factors appear in 
expressions such as 
the commutation relation of the field operator 
Eq.~(\ref{eq2:commutation_for_b}) and the density operator 
Eq.~(\ref{eq2:density_operator}). 
This nontrivial phase factor can be interpreted as due to 
the magnetic field in momentum space. 
On the vNL, the momentum is defined in the Brillouin zone and 
has a periodicity $2\pi$ in both $p_x$ and $p_y$ directions. 
Hence, the momentum is defined on a 2D torus. 
Since the field operator $b_l({\bf p})$ obeys a boundary condition 
$b_l({\bf p}+2\pi{\bf N})=e^{i\phi({\bf p},{\bf N})}b_l({\bf p})$, 
for one period, 
the phases $e^{i\pi}$ in the $p_x$-direction and 
$e^{i\pi-ip_x}$ in the $p_y$-direction are obtained. 
If these phases are interpreted as due to 
the Aharonov-Bohm phase caused by the magnetic field in momentum space, 
the phase $e^{i\pi}$ can be regarded as due to two 
magnetic fluxes with the magnitude $2\pi$ shown in Fig.~\ref{fig2:2Dtorus} 
and the phase $e^{-ip_x}$ can be regarded as
due to the magnetic field perpendicular to the surface of the torus 
with the magnitude $-1/2\pi$. 
%
%== Fig ==%
%magnetic field on the torus. 
%== end ==%
%
Indeed, the expression of the density operator 
Eq.~(\ref{eq2:density_operator}) is rewritten as 
%
%== Eq ==%
%Density operator
\begin{align}
 \rho({\bf k})=&\sum_{l,l'}
  \int_{\rm BZ}\frac{d^2p}{(2\pi)^2}b_l^\dag({\bf p})
  b_{l'}({\bf p}-\hat{\bf k})
  f^0_{l,l'}({\bf k}) 
 \exp\left\{
 i\int^{{\bf p}-\hat{\bf k}}_{\bf p}{\bf A}({\bf p}')
 \cdot d{\bf p}'
 \right\},
\label{eq2:density_operator2}
\end{align}
where the ${\bf p}'$ integral is a line integral on the straight line
from ${\bf p}$ to ${\bf p}-\hat{\bf k}$ and ${\bf A}({\bf
p})=-(a^2/2\pi)(-p_y,0,0)$ which corresponds to 
a vector potential in momentum in Landau gauge. 
The magnetic field obtained from this vector potential is 
${\bf B}=(0,0,-1/2\pi)$. 
The phase factor in 
Eq.~(\ref{eq2:density_operator2}) is a gauge connection form 
${\bf p}$ to ${\bf p}-\hat{\bf k}$ in order to keep the integrand of 
$\rho({\bf k})$ 
invariant against the gauge transformation in momentum space 
%
%== Eq ==%
%gauge transformation
\begin{equation}
 b_l({\bf p})\to e^{i\lambda({\bf p})}b_l({\bf p}),\quad 
  {\bf A}({\bf p})\to {\bf A}({\bf p})-\nabla\lambda({\bf p}). 
\end{equation}
Hence, the nontrivial phase factor in the expression of the density
operator Eq.~(\ref{eq2:density_operator}) can be interpreted as due to 
the magnetic field in momentum space consistently. 
While the vector potential takes the form in Landau gauge, 
other gauges can be taken by the gauge transformation in momentum
space.

\chapter{Periodic states in the quantum Hall system}
\label{chapter3}
 %%%%%%%%%%%%%%%%%%%%%%%%%%%%%%%%%%%%%%%%%%%%%%%%%%%%%%%%
%% Periodic states in the quantum Hall system  states %%
%%%%%%%%%%%%%%%%%%%%%%%%%%%%%%%%%%%%%%%%%%%%%%%%%%%%%%%%
In this chapter, we develop a formalism based on the von Neumann lattice
(vNL) basis to treat 
periodic states in the quantum Hall system, which include charge
density wave (CDW) states or Wigner crystal states, bubble states, and striped
Hall states. Wigner crystal states or CDW states 
have been observed at low partial fillings 
and e.g., the Landau level filling factor $\nu=1/5$ 
in the lowest Landau level (LL) \cite{Jiang}, 
whereas bubble 
states and anisotropic charge density wave (ACDW) or striped Hall states 
have been observed at, e.g., partial filling factor $\nu^\ast\sim$ 1/4, 3/4 
in higher LLs for the bubble state \cite{Du,Cooper} and $\nu^\ast\sim$ 1/2 in the higher
LLs for the ACDW or striped Hall state \cite{Lilly,Du}. 
The vNL basis is the most suitable basis to study these periodic states
consistently. We first review the background of this issue, then 
we develop the vNL formalism for periodic states 
in the quantum Hall system.

%%%%%%%%%%%%%%%%%%%%%%%%%%%%
%% section 3-1 Background %%
%%%%%%%%%%%%%%%%%%%%%%%%%%%%
\section{Background} 
In the quantum Hall system, energy levels split into LLs owing to %due to 
the
cyclotron motion of electrons and the energy difference between the
nearest LLs is given by $\hbar\omega_c$. 
If the magnetic field is so strong that the energy difference 
$\hbar\omega_c$ is much larger than the typical order of the
Coulomb interaction $e^2/4\pi\epsilon l_B$, the LL mixing effects can be
neglected. 
In this case, it is enough to consider only the system projected to 
the uppermost partially-filled LL 
where the kinetic term is quenched and the Hamiltonian 
is given by only the projected Coulomb interaction term. 
This projected Coulomb interaction term is totally different from the 
unprojected Coulomb interaction term in that the projected density
operators are noncommutative with each other as seen in 
Eq.~(\ref{eq2:commutation_of_density}). 
If we neglect this noncommutativity, 
an expected ground state in this system is a {\it Wigner crystal}
state. 
A Wigner crystal state is a state where electrons form a
triangular lattice through the Coulomb interaction so that the Coulomb
energy becomes minimum. 
When the partial filling factor is small enough, 
the triangular Wigner crystal is indeed the ground state. 
This is because for low partial fillings, 
the superposition of wave functions is negligible and 
the problem can be treated as a classical one. 
The situation totally changes 
when the superposition of wave functions becomes
large and the problem cannot be treated as a classical one anymore. 
In this case, if we exclude a possibility of liquid states or fractional
quantum Hall states, 
naively expected ground states are 
``{\it CDW states}''. 
The CDW states are obtained as a self-consistent 
solution within the Hartree-Fock (HF) approximation by 
proper treatment of the noncommutativity of the density operators 
\cite{Yoshioka_Fukuyama}. 
In the low partial filling limit, the self-consistently obtained 
CDW state coincides with the triangular
Wigner crystal state. However, as the partial filling factor approaches
the half-filling, 
bubble states which are CDW states with two or
more electrons per unit cell, ACDW states,
and striped Hall states which are unidirectional charge density wave
states, become ground states. 
To treat these states, some theoretical approaches have been developed 
in early studies.

%%%%%%%%%%%%%%%%%%%%%%%%%%%%%%%%%%
% subsection 3-1-1 Early studies %
%%%%%%%%%%%%%%%%%%%%%%%%%%%%%%%%%% 
\subsection{Early studies}
Among early studies of CDW states in the quantum Hall system 
\cite{Kuramoto,Fukuyama_Platzman_Anderson,Yoshioka_Fukuyama,Gerhardts}, 
the self-consistent HF solution was 
first studied by Yoshioka and Fukuyama \cite{Yoshioka_Fukuyama}. 
They focused on the triangular and square CDW state in the 
lowest LL and calculated the energy dispersion and the total energy of
the ground state, neglecting the higher harmonics of the CDW. 
The calculation was done by means of the Green function. 
In their calculation, first, the Hamiltonian is expanded by the 
eigenstate of the magnetic translation operator in one direction, and then, 
the Green function was ingeniously introduced to solve the
problem self-consistently. 
Although their formalism is a bit complicated, the self-consistent
calculation including the noncommutativity of the projected density
operator is performed properly. 
Later, Yoshioka and Lee \cite{Yoshioka_Lee} developed a simpler 
formalism and calculated the same problem including higher harmonics
of the CDW, in a different way. 
Using the same basis as the previous one, 
they more directly obtained the solution by 
%In their calculation, using the same basis as
%the previous one, the solution is more directly obtained by 
diagonalizing the HF Hamiltonian self-consistently. 
Their results are 
essentially the same as the previous one except for small
corrections. 

There is a totally different approach to obtain the same results 
first developed by C\^{o}t\'{e} and MacDonald \cite{Cote_MacDonald}, 
which is an approach using an equation of motion. 
They calculated the total energy and the mean value of the density
operator for the CDW ground state by solving 
the HF equation of motion numerically, and obtained the same results as
the previous one. As pointed out in their paper \cite{Cote_MacDonald}, 
it is not possible to obtain the one-particle energy dispersion with
this approach in contrast to the previous studies %works
\cite{Yoshioka_Fukuyama,Yoshioka_Lee}. 
However, their main topic was to investigate collective modes of the
CDW state, and for this purpose, it was enough to know 
the mean value of the density operator for the CDW state since 
the collective modes are associated with poles of the density-density
response function which can be derived in the time-dependent Hartree-Fock
approximation (TDHFA). 
Later, the TDHFA has been applied for anisotropic charge density wave
(ACDW) states \cite{Cote_Fertig} 
and bubble states \cite{Cote_etal} to investigate the
collective modes.

%%%%%%%%%%%%%%%%%%%%%%%%%%%%%%%%%%%%%%%%%%%%%%%%%%%%%%%%%%%%%%%%%%%%%%%%%%%
% subsection 3-1-2 vNL formalism for striped Hall, CDW, and bubble states %
%%%%%%%%%%%%%%%%%%%%%%%%%%%%%%%%%%%%%%%%%%%%%%%%%%%%%%%%%%%%%%%%%%%%%%%%%%%
\subsection{vNL formalism for striped Hall, CDW, and bubble states}
Our vNL formalism for the CDW state is a similar one to the 
diagonalization technique developed by Yoshioka and Lee
\cite{Yoshioka_Lee}, except for the use of the vNL basis instead of the 
eigenstate of the magnetic translation operator in one direction. 
Expanding the HF Hamiltonian by the vNL basis and adjusting the
periodicity of the vNL to an integer multiple of the periodicity 
of the CDW, we can easily diagonalize the HF Hamiltonian 
self-consistently. 
The essential difference is that a physical picture is more clear in
the vNL formalism. 
In the vNL formalism, in contrast to other basis,
momenta are clearly defined as eigenvalues of the commutative magnetic
translation operators and 
the one-particle energy dispersion is obtained as a function of the
momenta. 
The one-particle energy dispersion of the CDW states 
resembles that of
noninteracting electrons in an external periodic potential
\cite{Hofstader,TKNN}. In the latter case, when the filling factor $\nu$
(not the LL filling factor) is
rational, i.e., for 
$\nu=\phi_0/BS_0=q/p$, where $p$ and $q$ have no
factors in common and $S_0=n^{-1}$ with an electron density $n$ 
is the unit-cell area of the crystal, 
the Landau level splits into $p$ non-overlapping subbands. 
On the other hand, in the former case, while there is no periodic
potential or external lattice structure, the von Neumann lattice 
plays a similar role and the one-particle energy dispersion of the CDW
state has $p$ bands for the partial LL filling factor $\nu^\ast=q/p$. 
In both cases, the momenta are defined in the Brillouin zone 
(or the magnetic Brillouin zone) and 
the one-particle energy dispersion has 
the $p$-band structure owing to %due to 
the $p$-fold
reduction of the Brillouin zone. 
In this sense, the physical picture is more clear for the vNL
formalism. 

The vNL formalism was first developed for the striped Hall state by
Ishikawa and others \cite{Ishikawa_stripe,Maeda_stripe,Aoyama_stripe}. 
Recently, this formalism has been developed for CDW states \cite{Tsuda1} 
and bubble states \cite{Tsuda2}. 
We review the details in the following sections.

%%%%%%%%%%%%%%%%%%%%%%%%%%%%%%%%%%%%
%% section 3-2 Striped Hall state %%
%%%%%%%%%%%%%%%%%%%%%%%%%%%%%%%%%%%%
\section{Striped Hall state}

%%%%%%%%%%%%%%%%%%%%%%%%%%%%%%%%%%%%%%%%%%%%%%%%%%%%%%%%%%%%%
% subsection 3-2-1 Assumption of the unidirectional density %
%%%%%%%%%%%%%%%%%%%%%%%%%%%%%%%%%%%%%%%%%%%%%%%%%%%%%%%%%%%%%
\subsection{Assumption of the unidirectional density}
Let us consider the case of the partial filling factor $\nu^\ast$ 
($0< \nu^\ast < 1$) in the $l$th LL. 
The striped Hall state is a unidirectional CDW state in the quantum Hall 
system, which has the following unidirectional density
(Fig.~\ref{fig3:stripe_dens}): 
%
%== Fig ==%
%Schematic view of the density of the striped Hall state. 
\begin{figure}[tb]
\begin{center}
\includegraphics[width=6cm]{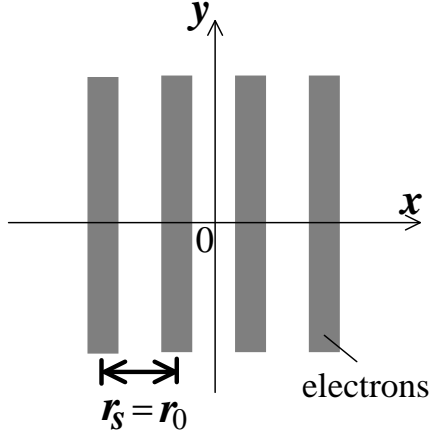}
\end{center}
\caption{\label{fig3:stripe_dens}
Schematic view of the density profile of the striped Hall state. 
The density of the striped Hall state 
is uniform in one direction and periodic with the period $r_0$ in the 
other direction. If we take the vNL asymmetry parameter $r_s=r_0$, 
the HF Hamiltonian is diagonalized on the vNL basis.
}
\end{figure}
%== end ==%
%
%== Eq ==%
%Mean value of the density operator for the striped Hall state
\begin{equation}
 \langle \rho_l({\bf x}) \rangle_{\rm
  stripe}=\sum_{N_x}\Delta_l(N_x)F_l(\frac{2\pi N_x}{r_0}, 0)
  e^{i(2\pi N_x /r_0)x},
  \label{eq3:stripe_dens}
\end{equation}
where $r_0$ is the period of the density in the $x$-direction, 
$\Delta_l(N_x)$ is an order parameter
determined self-consistently, and $\Delta_l(0)=\nu^\ast$. 
Equation (\ref{eq3:stripe_dens}) gives the following form of 
the mean value of the projected density operator, 
%
%== Eq ==%
%Mean value of the projected density operator 
\begin{equation}
\label{eq3:assumption_stripe0}
 \langle \bar{\rho}_l({\bf k}) \rangle _{\rm stripe}=
  \sum_{N_x}\Delta_l(N_x)(2\pi)^2
  \delta(k_x+\frac{2\pi N_x}{r_0})\delta(k_y).  
\end{equation}
If we take the vNL asymmetry parameter $r_s=r_0$, 
Eq.~(\ref{eq3:assumption_stripe0}) with 
$\tilde{\bf k}=(k_x/r_s,r_s k_y)$ takes the simple form, 
%
%== Eq ==%
%Mean value of the projected density operator for \tilde{\bf k}
\begin{equation}
 \label{eq3:assumption_stripe}
 \langle \bar{\rho}_l(\tilde{\bf k}) \rangle _{\rm stripe}=
  \sum_{N_x}\Delta_l(N_x)(2\pi)^2
  \delta(k_x+2\pi N_x)\delta(k_y).  
\end{equation}
%

%%%%%%%%%%%%%%%%%%%%%%%%%%%%%%%%%%%%%%%%%%%%%%%%%%%%%%%%%%%%%%%%%
% subsection 3-2-2 Diagonalization, eigenstates and eigenvalues %
%%%%%%%%%%%%%%%%%%%%%%%%%%%%%%%%%%%%%%%%%%%%%%%%%%%%%%%%%%%%%%%%%
\subsection{Diagonalization, eigenstates and eigenvalues}
The HF Hamiltonian of the striped Hall state is already diagonalized 
on the vNL basis. 
Substitution of Eq.~(\ref{eq3:assumption_stripe}) into the HF Hamiltonian 
Eq.~(\ref{eq2:HF_Hamiltonian}) gives 
the HF Hamiltonian for the striped Hall state by 
%
%== Eq ==%
%HF Hamiltonian for the striped Hall state
\begin{equation}
 \mathcal{H}_{\rm HF-stripe}^{(l)}=\int _{\rm BZ}\frac{d^2 p}{(2\pi)^2}
  \epsilon_l({\bf p})b^\dag_l({\bf p})b_l({\bf p}), 
  \label{eq3:Hamiltonian_stripe}
\end{equation}
where $\epsilon_l({\bf p})$ is a one-particle energy given by 
%
%== Eq ==%
%eigenenergy of the striped Hall state
\begin{equation}
 \epsilon_l({\bf p})=\epsilon^{(0)}_l+\sum_{N_x\neq 0}\Delta_l(N_x)
  v_l^{\rm HF}(\frac{2\pi N_x}{r_s}, 0)(-1)^{N_x}e^{-iN_xp_y}. 
  \label{eq3:one-particle_ene_stripe}
\end{equation} 
In Eq.~(\ref{eq3:one-particle_ene_stripe}), $\epsilon^{(0)}_l$ is a uniform
Fock energy given by $\nu^\ast v^{\rm HF}_l(0)$. 
The values of $v^{\rm HF}_l(0)$ are shown in Table
\ref{table3:Fock_energy}. 
%
%== Table ==%
%rs vs total energy
\begin{table}[tb]
\setlength{\extrarowheight}{1.05mm}\newcolumntype{Y}{>{\centering\arraybackslash}X}
\begin{tabularx}{\linewidth}{YY}\hline\hline
$l$ & $v^{\rm HF}_l(0)/(q^2/l_{\rm B})$\\ \hline
0 & $-1.25331$ \\ 
1 & $-0.93999$ \\ 
2 & $-0.80290$ \\ 
3 & $-0.71968$ \\ \hline\hline
\end{tabularx}
\caption{Values of $v^{\rm HF}_l(0)$ for each
 LL. 
}
\label{table3:Fock_energy}
\end{table}
%== end ==%
%

For the ground state of the striped Hall state, 
the two-point function of the operator $b_l({\bf p})$ is 
given by \cite{Ishikawa_stripe,Maeda_stripe} 
%
%== Eq ==%
%two-point function for the striped Hall state
\begin{align}
 \langle b^\dag_l({\bf p})b_{l'}({\bf p}')\rangle_{\rm stripe}
 =&\sum_{\bf N}\delta_{l,l'}\theta[\epsilon_{\rm F}-\epsilon_l({\bf p})]
 (2\pi)^2\delta^2({\bf p}-{\bf p}'-2\pi{\bf N})
 e^{-i\phi({\bf p},{\bf N})}, 
 \label{eq3:two_point_stripe}
\end{align}
where $\epsilon_{\rm F}$ is a Fermi energy and $\theta$ is a step function.

%%%%%%%%%%%%%%%%%%%%%%%%%%%%%%%%%%%%%%%%%%%%%%%%%%%%%%%%%%%%
% subsection 3-2-3 Self-consistency condition and solution %
%%%%%%%%%%%%%%%%%%%%%%%%%%%%%%%%%%%%%%%%%%%%%%%%%%%%%%%%%%%%
\subsection{Self-consistency condition and solution}
\label{chap3:stripe_solutions}
\begin{figure}[tb]
\begin{center}
 \includegraphics[width=6cm]{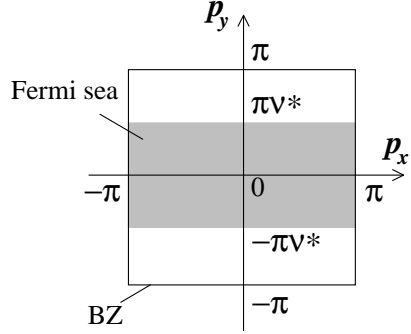}
\end{center}
\caption{\label{fig3:Fermi_sea}
Fermi sea of the striped Hall state at $\nu^\ast$. 
The occupied state is represented by the dark region. 
When the stripe faces the $y$-direction, the $p_x$-direction 
of the Brillouin zone is fully occupied. 
In this case, the Fermi sea has the inter-LL energy gap in the
 $p_x$-direction and is gapless in the $p_y$-direction.
} 
\end{figure}

The self-consistent equation for $\Delta_l(N_x)$ is obtained by 
substitution of Eq.~(\ref{eq3:two_point_stripe}) into the left hand side of
Eq.~(\ref{eq3:assumption_stripe}).
$\Delta_l(N_x)=(-1)^{N_x}\sin(\nu^\ast\pi N_x)/\pi N_x$ 
is a solution of the self-consistent equation. 
This solution has the Fermi sea, $|p_y|<\pi\nu^\ast$  
(shown in Fig.~\ref{fig3:Fermi_sea}) 
and gives the one-particle energy as 
%
%== Eq ==%
%One-particle energy dispersion
\begin{equation}
 \epsilon_l({\bf p})=\epsilon^{(0)}_l+\sum_{N_x\neq 0}v_l^{\rm HF}(\frac{2\pi N_x}{r_s},
  0)\frac{\sin(\nu^\ast\pi N_x)}{\pi N_x}e^{-iN_x p_y}. 
\end{equation}
%
%== Fig ==%
%Fermi sea of the striped Hall state at half-filling
%== end ==%
%
%== Fig ==%
%one-particle energy of the striped Hall state
\begin{figure}[tbhp]
\vskip 4cm
\begin{center}
 \includegraphics[height=5cm,clip]{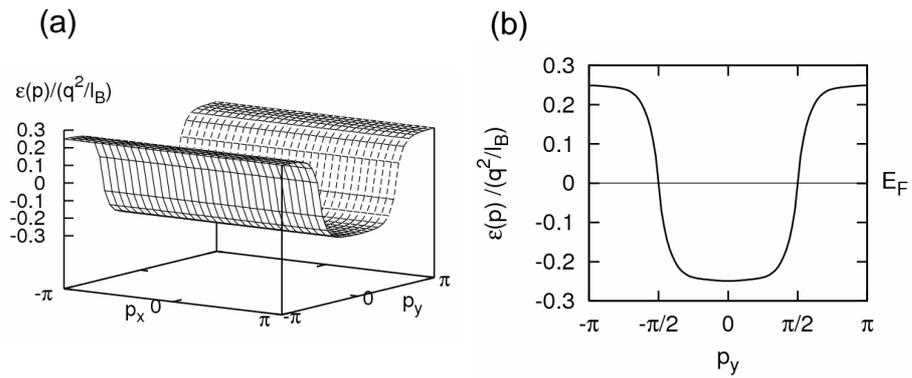}
\end{center}
\caption{\label{fig3:ene_stripeL2}
(a) 3D plot of the one-particle energy of the $l=2$ 
striped Hall state at 
half-filling. 
When the stripe faces the $y$-direction, the one-particle energy is 
uniform in the $p_x$-direction. 
(b) 2D plot of the one-particle energy as a function of $p_y$. 
The Fermi velocity is logarithmically divergent.} 
\end{figure}
%== end ==%
%
The HF energy per particle is given as a function of $r_s$ by 
%
%== Eq ==%
%total energy of the striped Hall state
\begin{align}
 E_{\rm stripe}^{(l)}(r_s)=
 \frac{\langle H_{\rm HF}^{(l)}\rangle_{\rm
 stripe}}{N_e^{(l)}}
 =\frac{\langle \mathcal{H}_{\rm HF-stripe}^{(l)}\rangle_{\rm
 stripe}}{2 N_e^{(l)}}
 =
 \frac{1}{2}\epsilon^{(0)}_l
 +\frac{1}{2}\sum_{N_x\neq 0}\nu^\ast 
 v_l^{\rm HF}(\frac{2\pi N_x}{r_s}, 0) 
 \left(\frac{\sin(\nu^\ast\pi N_x)}{\nu^\ast\pi N_x}\right)^2. 
\end{align}
where $N_e^{(l)}$ is the total number of particles within the $l$th LL. 
We determine the optimal value of $r_s$ by minimizing $E_{\rm
stripe}^{(l)}(r_s)$ with respect to $r_s$. 
The optimal value of $r_s$ and the minimum energy at each LL are
shown in Table \ref{table3:stripe_table} \cite{Maeda_stripe}. 
%
%== Table ==%
%rs vs total energy
\begin{table}[tb]
\setlength{\extrarowheight}{1.05mm}\newcolumntype{Y}{>{\centering\arraybackslash}X}
\begin{tabularx}{\linewidth}{YYY}\hline\hline
$l$ & $r_s^{\rm stripe}$ & $E_{\rm stripe}/(q^2/l_{\rm B})$\\ \hline
0 & 1.636 & $-0.4331$ \\ 
1 & 2.021 & $-0.3490$ \\ 
2 & 2.474 & $-0.3074$ \\ 
3 & 2.875 & $-0.2800$ \\ \hline\hline
\end{tabularx}
\caption{Minimum energy and corresponding parameter $r_s$ of the 
striped Hall states at $\nu^\ast=1/2$.}
\label{table3:stripe_table}
\end{table}
%== end ==%
%

\begin{figure}[tb]
\begin{center}
\includegraphics[width=8cm]{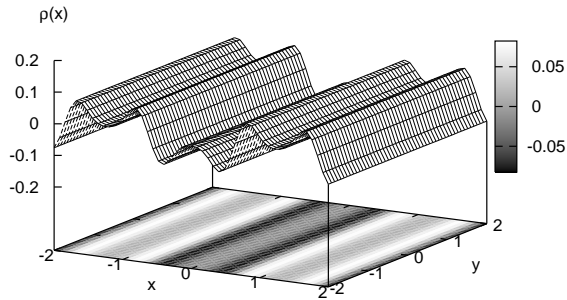}
\end{center}
\caption{\label{fig3:stripe_densL2}Density profile of the 
$l=2$ striped Hall state at half-filling. 
The uniform part $\rho_0=\nu^\ast$ is subtracted. 
The density of the striped Hall state 
is uniform in the $y$-direction 
and periodic with a period $r_s$ 
in the $x$-direction.}
\end{figure}

For the optimal value of $r_s$, the one-particle energy and the density
profile of the striped Hall state at $\nu^\ast=1/2$ in the $l=2$ LL are
plotted in Fig.~\ref{fig3:ene_stripeL2} and
Fig.~\ref{fig3:stripe_densL2}, respectively. 
%
%
%== Fig ==%
%Density of the striped Hall state at \nu^\ast=1/2 in the l=2 LL. 
%== end ==%
%
When the density is uniform in the $y$-direction, 
the one-particle energy is a function of only $p_y$ 
and uniform in the $p_x$-direction. 
The energy dispersion becomes gapless in the $p_y$-direction and has a
inter-LL gap in the $p_x$-direction. 
The Fermi velocity 
$v_F=\partial \epsilon_l({\bf p})/\partial p_y|_{p_y=\pi\nu^\ast}$ is 
logarithmically divergent \cite{Ishikawa_stripe}. 

%%%%%%%%%%%%%%%%%%%%%%%%%%%%%%%%%%%%%%%%%%%
%% section 3-3 Charge density wave state %%
%%%%%%%%%%%%%%%%%%%%%%%%%%%%%%%%%%%%%%%%%%%
\section{Charge density wave state}
\label{chap3:CDW}

%%%%%%%%%%%%%%%%%%%%%%%%%%%%%%%%%%%%%%%%%%%%%%%%%%%%%%%%%%%%%%%%%%
% subsection 3-3-1 Assumption of the charge density wave density %
%%%%%%%%%%%%%%%%%%%%%%%%%%%%%%%%%%%%%%%%%%%%%%%%%%%%%%%%%%%%%%%%%%
\subsection{Assumption of the CDW density}
\begin{figure}[tb]
\begin{center}
\includegraphics[width=5cm]{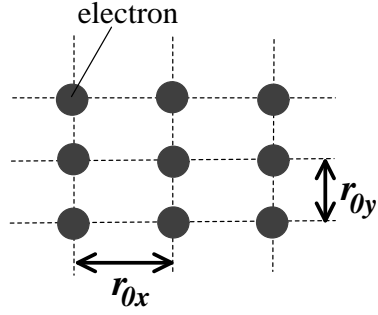}
\end{center}
\caption{\label{fig3:CDW_dens}
Schematic view of the density of the rectangular CDW state. 
The density is periodic with the periodicity $r_{0x}$ and $r_{0y}$ in the  
$x$-direction and $y$-direction, respectively. 
}
\end{figure}

\begin{figure}[tb]
\begin{center}
\includegraphics[width=7cm]{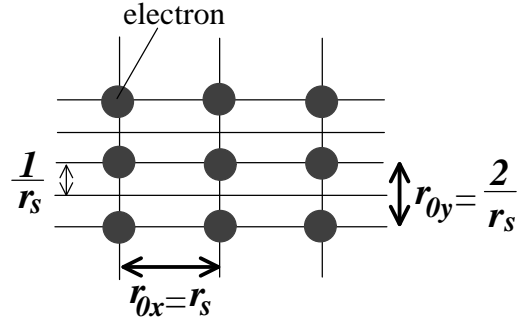}
\end{center}
\caption{\label{fig3:CDW_vs_vNL}
The vNL unit cell and the CDW unit cell in the case of $\nu^\ast=1/2$. 
The thin lines represent the vNL. 
For $\nu^\ast=1/2$, the CDW unit cell is just twice as large as the vNL
 unit cell. If we take $r_s=r_{0x}$, 
the HF Hamiltonian can be block-diagonalized.
}
\end{figure}

Let us consider the case of $\nu^\ast$ in the $l$th LL. Here, we
consider only the rectangular CDW state for simplicity. 
The CDW state has the following periodic density (Fig.~\ref{fig3:CDW_dens}), 
%
%== Fig ==%
%Schematic view of the density of the striped Hall state. 
%== end ==%
%
%== Eq ==%
%Assumption of the CDW density
\begin{equation}
 \langle \rho_l ({\bf x})\rangle_{\rm CDW}=
  \sum_{\bf N}\Delta_l({\bf Q}_{\bf N})F_l({\bf Q}_{\bf N})
  e^{-i{\bf Q}_{\bf N}\cdot {\bf x}},
 \label{eq3:CDW_dens}
\end{equation}
where ${\bf Q}_{\bf N}=(mQ_{0x}, nQ_{0y})$ and ${\bf N}=(m,n)$ with
integers $m$,$n$. 
Equation (\ref{eq3:CDW_dens}) gives the following form of 
the mean value of the projected density operator, 
%
%== Eq ==%
%projected density operator of the CDW state 
\begin{equation}
 \langle \bar{\rho}_l({\bf k})\rangle_{\rm CDW} =\sum_{\bf N}
  \Delta_l({\bf Q}_{\bf N})(2\pi)^2\delta^2({\bf k}-{\bf Q}_{\bf N}). 
\end{equation}
The reciprocal vector ${\bf R}_{{\bf N}'}$ is given by 
${\bf R}_{{\bf N}'}=2\pi(m'/Q_{0x},n'/Q_{0y})\equiv(m'r_{0x},n'r_{0y})$, 
which satisfies ${\bf Q}_{\bf N}\cdot {\bf R}_{{\bf N}'}=2\pi M$ 
with an integer $M$. 
Since one electron in the CDW state with the density 
Eq.~(\ref{eq3:CDW_dens}) occupies the area $S_0=r_{0x}r_{0y}$ and 
one state in each LL occupies the area $S_\phi=a^2$ with the vNL
constant $a=\sqrt{h/eB}$, 
the partial filling factor $\nu^\ast$ is given by 
\begin{equation}
 \nu^\ast=\frac{\hbox{Number of electrons in the $l$th LL}}
  {\hbox{\rm Number of states in one LL}}=
  \frac{a^2}{r_{0x}r_{0y}}. 
\end{equation}
Thus, the relation $r_{0x}=a^2/\nu^\ast r_{0y}$ holds. 
%Especially, 
In particular, 
in the case of $\nu^\ast=N/M$, this relation is given by 
$r_{0x}=M/Nr_{0y}$, where we set $a=1$ and the integers $M$,$N$ have no
factors in common. 
By using this relation, ${\bf Q}_{\bf N}$ is rewritten as 
\begin{equation}
 {\bf Q}_{\bf N}=(\frac{2\pi m}{r_{0x}}, 2\pi nr_{0x}\frac{N}{M}). 
\end{equation}
Hence, $\langle \bar{\rho}_l({\bf k})\rangle_{\rm CDW}$ is given by 
\begin{equation}
\langle \bar{\rho}_l({\bf k})\rangle_{\rm CDW}=\sum_{m,n}
\Delta_l({\frac{2\pi m}{r_{0x}},2\pi nr_{0x}\nu^\ast})
 (2\pi)^2\delta(k_x-\frac{2\pi m}{r_{0x}})\delta(k_y-2\pi nr_{0x} \nu^\ast), 
\end{equation}
with $\nu^\ast=M/N$. 
If we take $r_s=r_{0x}$ (Fig.~\ref{fig3:CDW_vs_vNL}), 
$\langle \bar{\rho}_l(\tilde{\bf k})\rangle_{\rm CDW}$ is
given by 
\begin{align}
 \langle \bar{\rho}_l(\tilde{\bf k})\rangle_{\rm CDW}
  =&\sum_{m,n}\Delta_l({\frac{2\pi m}{r_s}, 2\pi n r_s\nu^\ast})
  (2\pi)^2\delta(k_x-2\pi m)\delta(k_y-2\pi n \nu^\ast) \nonumber \\
 =&\sum_{m,j=-\infty}^\infty \sum_{k=0}^{M-1}
 \Delta_l(\frac{2\pi m}{r_s}, 2\pi r_s(jM+k)\frac{N}{M})
 (2\pi)^2\delta(k_x-2\pi m)\delta(k_y-2\pi(jM+k)\frac{N}{M}), 
\label{eq3:CDW_dens2}
\end{align}
where $n=jM+k$ with integers $j,k$ ($k=0,1,2,\dots,M-1$). 
%
%== Fig ==%
%vNL unit cell and the CDW unit cell. 
%== end ==%
%

%%%%%%%%%%%%%%%%%%%%%%%%%%%%%%%%%%%%%%%%%%%%%%%%%%%%%%%%%%%%%%%%%
% subsection 3-3-2 Diagonalization, eigenstates and eigenvalues %
%%%%%%%%%%%%%%%%%%%%%%%%%%%%%%%%%%%%%%%%%%%%%%%%%%%%%%%%%%%%%%%%%
\subsection{Diagonalization, eigenstates and eigenvalues} 
Substituting Eq.~(\ref{eq3:CDW_dens2}) into the HF Hamiltonian 
Eq.~(\ref{eq2:HF_Hamiltonian}), we can block-diagonalized 
the HF Hamiltonian 
as follows. 
Under the assumption Eq.~(\ref{eq3:CDW_dens}) at $\nu^\ast=N/M$, 
the HF Hamiltonian is written as 
%
%== Eq ==%
%HF Hamiltonian for the CDW state
\begin{align}
 \mathcal{H}_{\rm HF-CDW}^{(l)}=&\int \frac{d^2k}{(2\pi)^2}
 v_l^{\rm HF}(\tilde{\bf k}) \nonumber \\
 &\times \sum^{\infty}_{m,j=-\infty}\sum^{M-1}_{k=0}
 \Delta_l(\frac{2\pi m}{r_s},2\pi r_s(jM+k)\frac{N}{M})
 (2\pi)^2\delta(k_x+2\pi m)
 \delta(k_y+2\pi (jM+k)\frac{N}{M})\bar{\rho}_l(\tilde{\bf k}), \nonumber \\
=&\epsilon_l^{(0)}N_e^{(l)}+\sum^{\infty}_{m,j=-\infty}\sum^{M-1}_{k=0}
 \Delta_l^{\rm HF}(m;j,k)
 \bar{\rho}_l(-\frac{2\pi m}{r_s}, -2\pi r_s(jM+k)\frac{N}{M}),
 \label{eq3:HF_CDW}  
\end{align}
where 
%
%== Eq ==%
%order parameter of the CDW state
\begin{equation}
 \Delta_l^{\rm HF}(m;j,k)\equiv 
  \Delta_l(\frac{2\pi m}{r_s}, 2\pi r_s (jM+k)\frac{N}{M})
  v_l^{\rm HF}(-\frac{2\pi m}{r_s},-2\pi r_s(jM+k)\frac{N}{M}),  
\end{equation}
except for $m=j=k=0$ and $\Delta_l^{{\rm HF}}(0;0,0)=0$. 
Substituting Eq.~(\ref{eq2:projected_dens}) 
into Eq.~(\ref{eq3:HF_CDW}) 
and using the boundary condition of $b_l({\bf p})$ given in 
Eq.~(\ref{eq2:boundary_condition_for_b}), 
we rewrite the HF Hamiltonian as 
%
%== Eq ==%
%HF Hamiltonian for the CDW state 2
\begin{align}
 \mathcal{H}_{\rm HF-CDW}^{(l)}-\epsilon_l^{(0)}N_e^{(l)}=&
 \int_{\rm BZ}\frac{d^2p}{(2\pi)^2}
 \sum^{\infty}_{m,j=-\infty}\sum^{M-1}_{k=0}
 \Delta_l^{\rm HF}(m;j,k)e^{i\pi(m+jN+m(j+k/M)N)}\nonumber \\
 &\times b_l^\dag({\bf p})b_l(p_x,p_y+2\pi k\frac{N}{M})e^{imp_y-i(jN)p_x}. 
\end{align}
Here, if we divide the interval of the $p_y$-integral into $M$ same
intervals and 
reduce each interval to the range $[0,2\pi/M]$ 
using the boundary condition Eq.(\ref{eq2:boundary_condition_for_b}), 
$\mathcal{H}_{\rm HF-CDW}^{(l)}$ is given by 
%
%== Eq ==%
%HF Hamiltonian for the CDW state 3
\begin{align}
 \mathcal{H}_{\rm HF-CDW}^{(l)}-\epsilon_l^{(0)}N_e^{(l)}=&\int^{2\pi}_{0}\frac{dp_x}{2\pi}
  \int^{2\pi/M}_{0}\frac{dp_y}{2\pi}
  \sum^\infty_{m,j=-\infty}\sum^{M-1}_{k,k'=0}
  \Delta_l^{\rm HF}(m;j,k-k')e^{i\pi(m+jN+m(j+(k-k')/M)N)}\nonumber \\
  &\times b_l^\dag(p_x,p_y+2\pi k' \frac{N}{M})b_l(p_x,p_y+2\pi k \frac{N}{M})
  e^{imp_y-i(jN)p_x+i2\pi m k' N/M} \nonumber \\
 =&\int^{2\pi}_{0}\frac{dp_x}{2\pi}\int^{2\pi/M}_{0}\frac{dp_y}{2\pi}
 \sum^{M-1}_{k,k'=0}b_l^\dag({\bf p},k')
 \Delta_l^{\rm HF}({\bf p},k',k)b_l({\bf p},k), 
\end{align}
where 
%
%== Eq ==%
%Notation
\begin{gather}
 b({\bf p},k)\equiv b(p_x,p_y+2\pi k \frac{N}{M}), \\
 \Delta_{\rm HF}({\bf p},k',k)\equiv 
 \sum^\infty_{m=-\infty}\sum^\infty_{j=-\infty}
 \Delta_{\rm
 HF}(m;j,k-k')e^{i\pi(m+jN+m(j+(k+k')/M)N)}e^{imp_y-i(jN)p_x}.
\end{gather}
Hence, the HF Hamiltonian $\mathcal{H}_{\rm HF-CDW}^{(l)}$ is
block-diagonalized and the problem is reduced to the diagonalization
problem of the $M\times M$ Hermite matrix $\Delta_l^{\rm HF}({\bf p},k',k)$. 
Note that the fact that $\langle \rho_l({\bf r})\rangle$ is real gives  
the relation $\Delta_l^\ast(-{\bf Q}_N)=\Delta_l({\bf Q}_N)$, and by using
this relation, it is proved that the $M\times M$ matrix 
${\bf \Delta}_l^{\rm HF}({\bf p})$ with 
$\{{\bf \Delta}^{\rm HF}_l({\bf p})\}_{k',k}\equiv \Delta_{\rm HF}({\bf
p},k',k)$ is Hermite.

The Hermite matrix ${\bf \Delta}_l^{\rm HF}({\bf p})$ is diagonalized
using the unitary matrix $U({\bf p})$ as 
%
%== Eq ==%
%Diagonalization of the HF Hamiltonian for the CDW state
\begin{equation}
 \Lambda({\bf p})
  =U^\dag({\bf p}){\bf \Delta}_l^{\rm HF}({\bf p})U({\bf p})
  =\left(
    \begin{array}{cccc}
     \epsilon_0({\bf p}) & & & \\
     & \epsilon_1({\bf p}) & & \\
     & & \ \ddots \  & \\
     & & & \epsilon_{M-1}({\bf p})\\
    \end{array}
   \right), 
\end{equation}
here, $\epsilon_s({\bf p})$ represents 
the eigenvalue of the $(s+1)$th energy band, 
and the eigenvector for $\epsilon_s$ is given by 
$\{{\bf v}_s({\bf p})\}_{s'}=U_{s,s'}({\bf p})$.
${\rm RBZ}$ represents the reduced Brillouin zone, $0<p_x<2\pi$, 
$0<p_y<2\pi/M$. 
Hence, the HF Hamiltonian $\mathcal{H}_{\rm HF-CDW}^{(l)}$ is diagonalized as 
%
%== Eq ==%
%HF Hamiltonian for the CDW state final
\begin{equation}
 \mathcal{H}_{\rm HF-CDW}^{(l)}-\epsilon_l^{(0)}N_e^{(l)}=\int_{\rm RBZ}\frac{d^2 p}{(2\pi)^2}
  {\bf c}_l^\dag({\bf p})\Lambda({\bf p}){\bf c}_l({\bf p})
  =\int_{\rm RBZ}\frac{d^2 p}{(2\pi)^2}\sum_{s=0}^{M-1}
  \epsilon_s({\bf p})c^\dag_s({\bf p})c_s({\bf p}),  
\end{equation}
where 
\begin{gather}
 {\bf c}_l({\bf p})=(c_0({\bf p}), c_1({\bf p}), \cdots,
  c_{M-1}({\bf p}))^t=U^\dag({\bf p}){\bf b}_l({\bf p})\quad 
  {\rm with}\notag \\ \{{\bf b}_l({\bf p})\}_k=b_l({\bf p},k), \quad 
  c_s({\bf p})=\sum_{k=0}^{M-1}U^\dag_{s,k}({\bf p})b_l({\bf p},k). 
\label{eq3:b_to_c} 
\end{gather}
In the present case of $\nu^\ast=N/M$, 
the lower N bands (from $s=0$ to $s=N-1$) 
are fully occupied for the ground state.

%%%%%%%%%%%%%%%%%%%%%%%%%%%%%%%%%%%%%%%%%%%%%%%%
% subsection 3-3-3 Self-consistency condition  %
%%%%%%%%%%%%%%%%%%%%%%%%%%%%%%%%%%%%%%%%%%%%%%%%
\subsection{Self-consistency condition}
The HF Hamiltonian should be diagonalized self-consistently. 
The ground state is given by 
%
%== Eq ==%
%Ground state of the CDW state
\begin{equation}
 |\Omega\rangle =N_c\prod_{s=0}^{N-1}\prod_{{\bf p}\in {\rm RBZ}}
  c_s^\dag({\bf p})|0\rangle, 
\end{equation}
where $N_c$ is a normalization constant and $|0\rangle$ is a vacuum
state in which the $(l-1)$th and lower Landau levels are fully
occupied. 
Using this ground state, we obtain the self-consistency condition 
%
%== Eq ==%
%Self-consistency condition
\begin{equation}
 \langle \Omega| \bar{\rho}_l({\bf q})|\Omega\rangle
  =\sum_{\bf N}\Delta_l({\bf Q}_{\bf N})(2\pi)^2\delta^2({\bf q}-{\bf
  Q}_{\bf N}).
\label{eq3:CDW_dens3}  
\end{equation}
In what follows, we derive the self-consistent equation for 
$\Delta({\bf Q}_{\bf N})$ from Eq.(\ref{eq3:CDW_dens3}).

The field operators $b_l^\dag({\bf p})$ and $b_l({\bf p})$ satisfy the
anti-commutation relation Eq.~(\ref{eq2:commutation_for_b}). 
For ${\bf p}$ and ${\bf p}'$ within the RBZ, 
Eq.~(\ref{eq2:commutation_for_b}) becomes 
%
%== Eq ==%
%Commutation relation for b
\begin{equation}
 \{b_l^\dag({\bf p},k),b_{l'}({\bf p}',k')\}=(2\pi)^2
  \delta_{l,l'}\delta_{k,k'}
  \delta^2({\bf p}-{\bf p}'). 
\end{equation}
For $c_s^\dag({\bf p})$ and $c_s({\bf p})$, the corresponding relation is given by 
%
%== Eq ==%
%Commutation relation for c
\begin{equation}
 \{c^\dag_s({\bf p}),c_{s'}({\bf
  p}')\}=\delta_{s,s'}(2\pi)^2\delta^2({\bf p}-{\bf p}').
\label{eq3:commutation_for_c}
\end{equation}
Using Eq.~(\ref{eq2:projected_dens}), 
we rewrite 
the self-consistency condition Eq.~(\ref{eq3:CDW_dens3}) as %is rewritten as 
%
%= Eq ==%
%Self-consistency condition 2
\begin{equation}
 \langle \Omega|
  b_l^\dag({\bf p})b_l({\bf p}')|\Omega \rangle
  =\sum^{M-1}_{k=0}F({\bf p},k)
  \sum_{\bf N}(2\pi)^2\delta(p_x-p_x'-2\pi N_x)
  \delta(p_y-p_y'-2\pi N_y-2\pi k\frac{N}{M})e^{-i\phi({\bf p},{\bf N})}, 
\label{eq3:self_bb}
\end{equation}
where 
%
%== Eq ==%
%Notation for F
\begin{equation}
 F({\bf p},k)
  =\sum_{m,j=-\infty}^\infty 
  \Delta_l(\frac{2\pi m}{r_s}, 2\pi r_s(jM+k)\frac{N}{M})
  e^{-i\pi(m+jN)-i\pi m(j+k/M)N}. 
\end{equation}
The left hand side of Eq.~(\ref{eq3:self_bb}) is calculated using 
Eqs.~(\ref{eq3:b_to_c}) and (\ref{eq3:commutation_for_c}), and the
result is 
%
%== Eq ==%
%Self-consistency condition 3
\begin{equation}
 \langle \Omega|b_l^\dag({\bf p},k)b_l({\bf p}',0)|\Omega\rangle
  =\left(\sum_{s=0}^{N-1}U^\dag_{s,k}({\bf p})U_{0,s}({\bf p})\right)
  (2\pi)^2\delta^2({\bf p}-{\bf p}'), 
\label{eq3:self_bb2}
\end{equation}
for ${\bf p}$ and ${\bf p}'$ within the RBZ. 
From Eqs.~(\ref{eq3:CDW_dens3}) and (\ref{eq3:self_bb2}), 
the self-consistent equation for 
$\Delta_l({\bf Q}_N)$ is given by 
%
%== Eq ==%
%Self-consistent equation for D
\begin{equation}
 \Delta_l(\frac{2\pi m}{r_s}, 2\pi r_s (jM+k)\frac{N}{M})
  =\int_{BZ}\frac{d^2p}{(2\pi)^2}
  \left(
   \sum_{s=0}^{N-1}U^\dag_{s,k}({\bf p})U_{0,s}({\bf p})
  \right)
  e^{ip_x(jN)-imp_y}e^{i\pi(m+jN)+i\pi m(j+k/M)N}.
\label{eq3:self_Delta}
\end{equation}
If the right hand side of Eq.~(\ref{eq3:self_Delta}) coincides 
with the left hand side, 
the problem is self-consistently solved.

%%%%%%%%%%%%%%%%%%%%%%%%%%%%%%%%%%%%%%%%%%%%%%%%%%%%%%%%%%%%%%%%%%%%%%%
% subsection 3-3-4 Density profile and Energy dispersion of CDW states %
%%%%%%%%%%%%%%%%%%%%%%%%%%%%%%%%%%%%%%%%%%%%%%%%%%%%%%%%%%%%%%%%%%%%%%%
\subsection{Density profile and Energy dispersion of CDW states}
As in the case of the striped Hall state, the HF energy per particle is
given as a function of $r_s$ by 
%
%== Eq ==%
%HF energy for CDW
\begin{equation}
 E_{\rm CDW}^{(l)}(r_s)=
  \frac{\langle H_{\rm HF}^{(l)}\rangle_{\rm CDW}}{N_e^{(l)}}
  =\frac{1}{2}\epsilon_l^{(0)}+\frac{1}{2}\int^{2\pi}_{0}\frac{dp_x}{2\pi}
  \int^{2\pi/M}_{0}\frac{dp_y}{\left(\frac{2\pi}{M}\right)}\sum^{N-1}_{s=0}
  \epsilon_s({\bf p}). 
\end{equation}
The optimal value of $r_s$ is determined by minimization of 
$E_{\rm CDW}^{(l)}(r_s)$ to $r_s$. 
The optimal value of $r_s$ and the minimum energy per particle 
at several fillings in
$l=0, 2$ are shown
in Table \ref{table3:CDW_table} and Fig.~\ref{fig3:cohesive_rs}, 
where cohesive energy, which is defined as 
$\hbox{(Total energy)}-\hbox{(Uniform Fock energy)}$, 
is shown for the minimum energy. 
As for $l=0$ LL, the optimal value of $r_s$ is approximately given by 
$\sqrt{1/\nu^\ast}$, which gives $r_{0x}=r_{0y}$. 
Hence, the CDW states in the $l=0$ LL become isotropic for the $x$ and
$y$-directions. 
On the other hand, as for the $l=2$ LL, 
the optimal value of $r_s$ takes a different value from $1/\nu^\ast$
especially near $\nu^\ast=1/2$, which means that 
anisotropic states are energetically favorable at these fillings.  
Hence, the CDW states in the $l=2$ LL become anisotropic near
$\nu^\ast=1/2$. 
Since the original Hamiltonian has a rotational
symmetry, the direction of the anisotropy can take any direction. 
In the present case of the rectangular vNL, two solutions whose 
anisotropy direction in density face the $x$ or $y$-direction are 
degenerate. 
This degeneracy yields two optimal
values of $r_s$ for anisotropic states. When we set the smaller one to
$r_s^{(1)}$ and the larger one to $r_s^{(2)}$, they are related by 
%
%== Eq ==%
%relation between two rs
\begin{equation}
 r_s^{(1)}=\frac{1}{\nu^\ast r_s^{(2)}}. 
\end{equation}
Only the smaller value $r_s^{(1)}$ is shown in Table \ref{table3:CDW_table} 
as the optimal value of $r_s$ for the CDW states in the $l=2$ LL.

The density profile of the CDW state is given by 
%
%== Eq ==%
%Density of the CDW state 
\begin{equation}
 \langle \rho_l({\bf x})\rangle _{\rm CDW}=\sum^{\infty}_{m,n=-\infty} 
  \Delta_l(\frac{2\pi m}{r_s},2\pi r_s n\frac{N}{M})
  F_l(\frac{2\pi m}{r_s}, 2\pi r_s n\frac{N}{M})
  \cos(\frac{2\pi m}{r_s}x+2\pi r_s n\frac{N}{M}y), 
\end{equation}
where the $x$ and $y$-inversion symmetry is assumed. 
In Figs.~\ref{fig3:2DdensL0} and \ref{fig3:2DdensL2}, 
the density profile is plotted 
at several fillings in $l=0,2$ LLs for the optimal value of $r_s$. 
As for the $l=0$ CDW density, 
the density becomes isotropic for all partial
fillings. On the other hand, as for the $l=2$ CDW density, 
the density becomes highly anisotropic near $\nu^\ast=1/2$, and as 
the partial filling decreases, the density approaches the isotropic
one gradually. 
This difference between the lowest LL and the higher LLs is mainly due to 
the difference of the form of the HF potential $v_l^{\rm HF}({\bf k})$. 
We refer to the direction with a narrower periodicity in density as 
``{\it stripe direction}'' in this thesis.

In the present vNL formalism, the energy dispersion of the CDW state is
obtained as a function of momenta $p_x$ and $p_y$ which are conjugate to 
the vNL. 
As for the $l=0$ LL, the energy dispersion is isotropic in the $p_x$
and $p_y$-directions and  
the band width becomes narrower as the density becomes lower. 
In Fig.~\ref{fig3:2DeneL0}, 
the energy dispersion at $p_y=0$ is plotted as a function of $p_x$. 
As for the $l=2$ LL, the energy dispersion becomes almost flat in 
the direction perpendicular to the stripe direction. 
In Fig.~\ref{fig3:2DeneL2}, the energy dispersion at $p_y=0$ is plotted as a
function of $p_x$. 
In both cases, there is a large energy gap
between the filled bands and the empty bands. 
This large gap coincides with the previously obtained results 
by Yosioka and Lee
\cite{Yoshioka_Lee}.

%%%%%%%%%%%%%%%%%%%%%%%%%%%%%%
%% section 3-4 Bubble state %%
%%%%%%%%%%%%%%%%%%%%%%%%%%%%%%
\section{Bubble state}

%%%%%%%%%%%%%%%%%%%%%%%%%%%%%%%%%%%%%%%%%%%%%%%%%%%%%
% subsection 3-4-1 Assumption of the bubble density %
%%%%%%%%%%%%%%%%%%%%%%%%%%%%%%%%%%%%%%%%%%%%%%%%%%%%%
\subsection{Assumption of the bubble density}
The bubble state is a CDW state with two or more electrons per unit
cell. 
We call the bubble state with $N$ electrons per unit cell the 
``{\it $N$-electron bubble state}''. 
Let us consider the $N$-electron rectangular bubble state at 
$\nu^\ast=1/M$ in the $l$th LL, 
with a period $r_{0x}$ and $r_{0y}$ in the $x$ and $y$
direction, respectively. 
In this case, since $N$ electrons are 
in the area of the unit cell $S_0=r_{0x}r_{0y}$ and 
the density of states is given by $1/a^2$, 
the partial filling factor is given by 
%
%== Eq ==%
%Filling factor 
\begin{equation}
 \nu^\ast=\frac{\hbox{Number of electrons in the $l$th LL}}
  {\hbox{Number of states in one LL}}
  =\frac{\frac{L_xL_y}{r_{0x}r_{0y}}N}{\frac{L_xL_y}{a^2}}
  =\frac{a^2}{r_{0x}r_{0y}}N, 
\end{equation} 
where $L_x$ and $L_y$ are the size of the system in the $x$ and $y$
directions, respectively. 
%Especially, 
In particular, 
in the case of $\nu^\ast=1/M$, this relation becomes 
$r_{0x}=NM/r_{0y}$, where we set $a=1$. 
By using this relation, 
the mean value of the projected density operator of 
the $N$-electron bubble state is given by 
%
%== Eq ==%
%assumption of the Bubble density 
\begin{equation}
 \langle \bar{\rho}_l(\tilde{\bf k})\rangle_{\rm bubble}=
  \sum_{m,n=-\infty}^\infty 
  \Delta_l(\frac{2\pi m}{r_s},2\pi nr_s\frac{1}{NM})(2\pi)^2
  \delta(k_x-2\pi m)\delta(k_y-2\pi n\frac{1}{NM}), 
\label{eq3:bubble_dens}
\end{equation}
where we set $r_s=r_0$. 
Hence, the diagonalization procedure of the HF Hamiltonian is the same as
in the case of the CDW state at $\nu^\ast=1/K$ with $K=NM$.

%%%%%%%%%%%%%%%%%%%%%%%%%%%%%%%%%%%%%%%%%%%%%%%%%%%%%%%%%%%%
% subsection 3-4-2 Self-consistency condition and solution %
%%%%%%%%%%%%%%%%%%%%%%%%%%%%%%%%%%%%%%%%%%%%%%%%%%%%%%%%%%%%
\subsection{Self-consistency condition and solution}
The important difference from the CDW state is a self-consistency condition. 
The one-particle energy of the $N$-electron bubble state at $\nu^\ast=1/M$ 
has $K=NM$ energy bands. Since the filling factor is $1/M$, 
the lowest $N$ bands are fully occupied for the ground state. 
As a result, the self-consistency condition is given by 
%
%== Eq ==%
%Self-consistency condition
\begin{equation}
 \Delta_l(\frac{2\pi m}{r_s}, 2\pi(n+\frac{j}{K})r_s)=
  \int_{\rm BZ}\frac{d^2 p}{(2\pi)^2} 
  \left(
   \sum^{N-1}_{s=0}U^\ast_{j,s}({\bf p})U_{0,s}({\bf p})
  \right)(-1)^{m+n+mn}
  e^{ip_x n-ip_y m -i\pi m(j/K)}.
\end{equation} 

For the $l=0$ LL, no self-consistent bubble solution has been obtained. 
For the $l=2$ LL, a self-consistent bubble solution is obtained for
$\nu^\ast=1/4$. The obtained solution is a 2-electron bubble state. 
The optimal value of $r_s$ is $r_s=2.8$ and the 
corresponding cohesive energy is $-0.1576$ $(q^2/l_{\rm B})$ which is
slightly lower than that of the CDW state at the same filling. 
Hence, for the $\nu^\ast=1/4$ in the $l=2$ LL, the 
2-electron bubble state is more stable. 
The density profile of this state is shown in
Fig.~\ref{fig3:2DdensL2_bubble}, which is almost isotropic. 
The energy dispersion consists of eight bands and the lower two bands are
fully occupied. As in the case of the CDW state, there is a large gap
between the filled bands and the empty bands.

Since electrons of this 2-electron bubble state are well-localized in
density, it is expected that the state is easily pinned by impurities,
which results in a quantization of the Hall conductivity owing to %due to 
its collective insulating property. Indeed, this quantization 
has been observed around $\nu^\ast=1/4, 3/4$ in the $l=2$ LL 
\cite{Du,Cooper}, and it is
believed that this observed quantization is an evidence of the bubble
state.

%%%%%%%%%%%%%%%%%%%%%%%%%%%%%%%%%%%%%%%%%%%%%%%%%%%%%%%%%%%%%%%%%%%%%%%%
% subsection 3-4-3 Relation between the bubble state and the CDW state %
%%%%%%%%%%%%%%%%%%%%%%%%%%%%%%%%%%%%%%%%%%%%%%%%%%%%%%%%%%%%%%%%%%%%%%%%
\subsection{Relation between the $N$-electron bubble state and 
the CDW state at $\nu^\ast=1/M$} 
The diagonalization formalism for the $N$-electron bubble state at
$\nu^\ast=1/M$ includes the CDW solution at $\nu^\ast=1/M$. 
To make this point clear, let us compare the mean value of the
density of the CDW state and the bubble state. 
For the CDW state at $\nu^\ast=1/M$, 
the mean values of the density operators are given by 
%
%== Eq ==%
%CDW density for $\nu=1/M$
\begin{align}
 &\langle \rho_l({\bf x})\rangle_{\rm CDW}=\sum^\infty_{m,n=-\infty}
  \Delta_l^{\rm CDW}(\frac{2\pi m}{r_{\rm C}},2\pi r_{\rm C}n\frac{1}{M})
  F_l(\frac{2\pi m}{r_{\rm C}},2\pi r_{\rm C}n\frac{1}{M})
  e^{-i(2\pi m/r_{\rm C})x-i(2\pi r_{\rm C}/M)y}, \\
 &\langle \bar{\rho}_l(\tilde{\bf k})\rangle _{\rm CDW}=
 \sum^\infty_{m,n=-\infty}
 \Delta_l^{\rm CDW}(\frac{2\pi m}{r_{\rm C}},\frac{2\pi r_{\rm C} n}{M})
 (2\pi)^2\delta(k_x-2\pi m)\delta(k_y-2\pi n\frac{1}{M}), 
\end{align}
and for the $N$-electron bubble state at $\nu^\ast=1/M$, they are given
by 
%
%== Eq ==%
%Bubble density for $\nu=1/M$
\begin{align}
 &\langle \rho_l({\bf x})\rangle _{\rm bubble}=\sum^\infty_{m,n=-\infty}
  \Delta_l^{\rm bubble}(\frac{2\pi m}{r_{\rm B}},2\pi r_{\rm B}n\frac{1}{K})
  F_l(\frac{2\pi m}{r_{\rm B}},2\pi r_{\rm B} n\frac{1}{K})
  e^{-i(2\pi m/r_{\rm B})x-i(2\pi r_{\rm B}n/K)y}, \\
 &\langle \bar{\rho}_l(\tilde{\bf k})\rangle _{\rm bubble}=
 \sum^\infty_{m,n=-\infty}
 \Delta_l^{\rm bubble}(\frac{2\pi m}{r_{\rm B}},\frac{2\pi r_{\rm B} n}{K})
 (2\pi)^2\delta(k_x-2\pi m)\delta(k_y-2\pi n\frac{1}{K}), 
\end{align}
with $K=MN$, where $r_s=r_{\rm C}$ for the CDW state and 
$r_s=r_{\rm B}$ for the bubble state. 
%
%== Fig ==%
%Schematics view of the 2-electron bubble density 
\begin{figure}[tbp]
\begin{center}
\includegraphics[width=5cm]{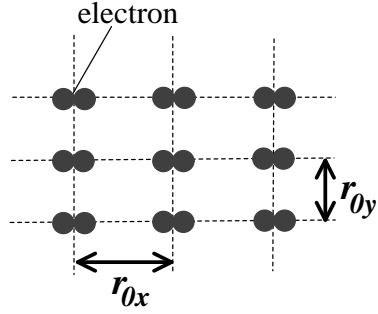}
\end{center}
\caption{\label{fig3:2eBubble_dens}
Schematic view of the density of the 2-electron bubble state.
}
\end{figure}
%== end ==%
%
If we set 
$r_{\rm B}=N r_{\rm C}$, 
$\Delta_l^{\rm bubble}(2\pi N m/r_{\rm B},2\pi r_{\rm B}n/K)=
\Delta_l^{\rm CDW}(2\pi m/r_{\rm C},2\pi r_{\rm C}n/M)$ 
and other $\Delta_l^{\rm bubble}$s is equal to zero, 
$\langle \rho_l({\bf x})\rangle_{\rm bubble}$ becomes the same as 
$\langle \rho_l({\bf x})\rangle _{\rm CDW}$. 
Hence, the HF Hamiltonian with the assumption (\ref{eq3:bubble_dens}) becomes the same 
as the HF Hamiltonian with the assumption (\ref{eq3:CDW_dens2}). 

%\newpage

%%%%%%%%%%%%%%%%%%%%%% Figures %%%%%%%%%%%%%%%%%%%%%%%%%%%%%%%%%%%%%%%%
%
%== Table ==%
%rs vs total energy
\begin{table}[tbhp]
\setlength{\extrarowheight}{1.05mm}\newcolumntype{Y}{>{\centering\arraybackslash}X}
\begin{tabularx}{\linewidth}{YYY}\hline\hline
$\nu^\ast(l=0)$ & $r_s^{\rm CDW}$ & $E_{\rm CDW}/(q^2/l_{\rm B})$\\ \hline
1/2  & 1.41 & $-0.1303$ \\ 
3/7  & 1.52 & $-0.1516$ \\ 
2/5  & 1.58 & $-0.1597$ \\ 
1/3  & 1.73 & $-0.1768$ \\ 
2/7  & 1.87 & $-0.1866$ \\ 
1/4  & 2.00 & $-0.1918$ \\ 
1/5  & 2.24 & $-0.1943$ \\ 
1/6  & 2.45 & $-0.1920$ \\ 
1/7  & 2.65 & $-0.1880$ \\ 
1/10 & 3.16 & $-0.1739$ \\ 
\hline\hline
$\nu^\ast(l=2)$ & $r_s^{\rm CDW}$ & $E_{\rm CDW}/(q^2/l_{\rm B})$\\ \hline
1/2  & 0.82 & $-0.1090$ \\ 
3/7  & 0.96 & $-0.1245$ \\ 
2/5  & 1.03 & $-0.1304$ \\ 
1/3  & 1.22 & $-0.1426$ \\ 
2/7  & 1.44 & $-0.1498$ \\ 
1/4  & 1.67 & $-0.1550$ \\ 
1/5  & 2.24 & $-0.1672$ \\ 
1/6  & 2.45 & $-0.1757$ \\ 
1/7  & 2.65 & $-0.1791$ \\ 
1/10 & 3.16 & $-0.1748$ \\ 
\hline\hline
\end{tabularx}
\caption{Minimum energy and corresponding parameter $r_s$ of the 
CDW states at several fillings in the $l=0,2$ LL.
As for the minimum energy, the uniform Fock energy is subtracted. 
The optimal value of $r_s$ for the $l=2$ LL takes two different values 
which give the same energy. 
Only the smaller one is shown in this table. 
These values are plotted in Fig.~\ref{fig3:cohesive_rs}
}
\label{table3:CDW_table}
\end{table}
%== end ==%
%
%%%%%%%%%%%%%%%%%%% l=0 and l=2 %%%%%%%%%%%%%%%%%%%%%%%%%%%%%
%
%== Fig ==%
%Optimal values of rs for the CDW density at l=0 and l=2. 
\begin{figure}[tbhp]
\begin{center}
\includegraphics[width=12cm]{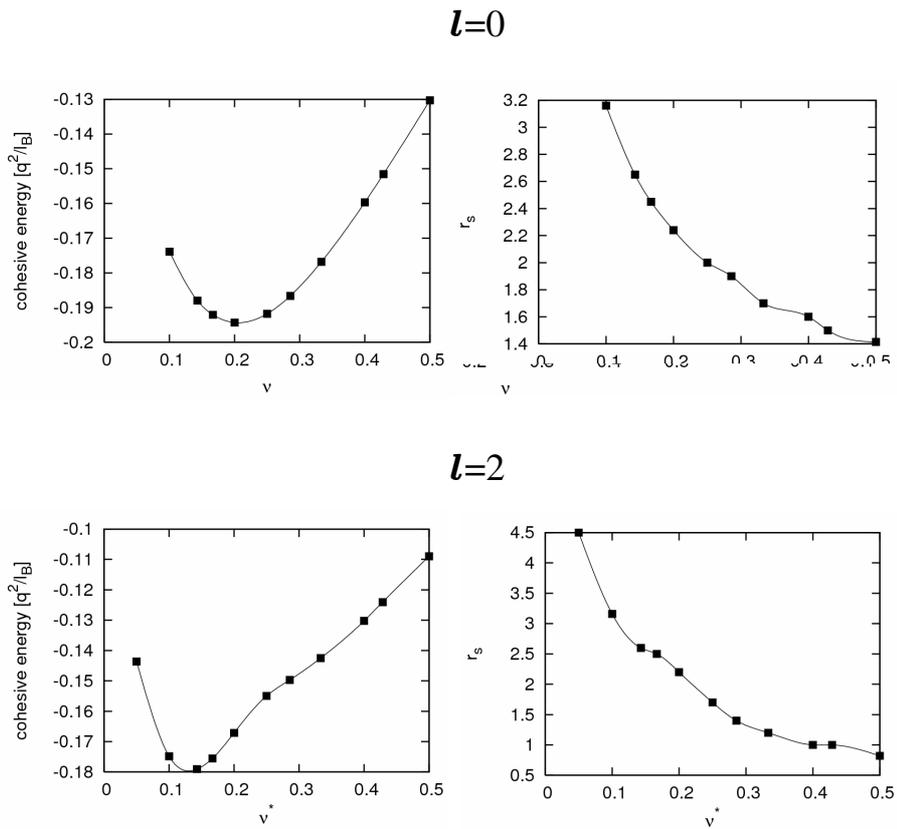}
\end{center}
\caption{\label{fig3:cohesive_rs}
The optimal value of $r_s$ is plotted for $l=2$ CDW states. 
Cohesive energy is plotted for $l=0$ CDW states with the optimal value
of $r_s$. 
}
\end{figure}
%== end ==%
%
%%%%%%%%%%%%%%% Density %%%%%%%%%%%%%%%%%%%%%%%%%%
%
%== Fig ==%
%2D plot of the CDW density at l=0. 
\begin{figure}[tbhp]
\begin{center}
\includegraphics[height=17cm,clip]{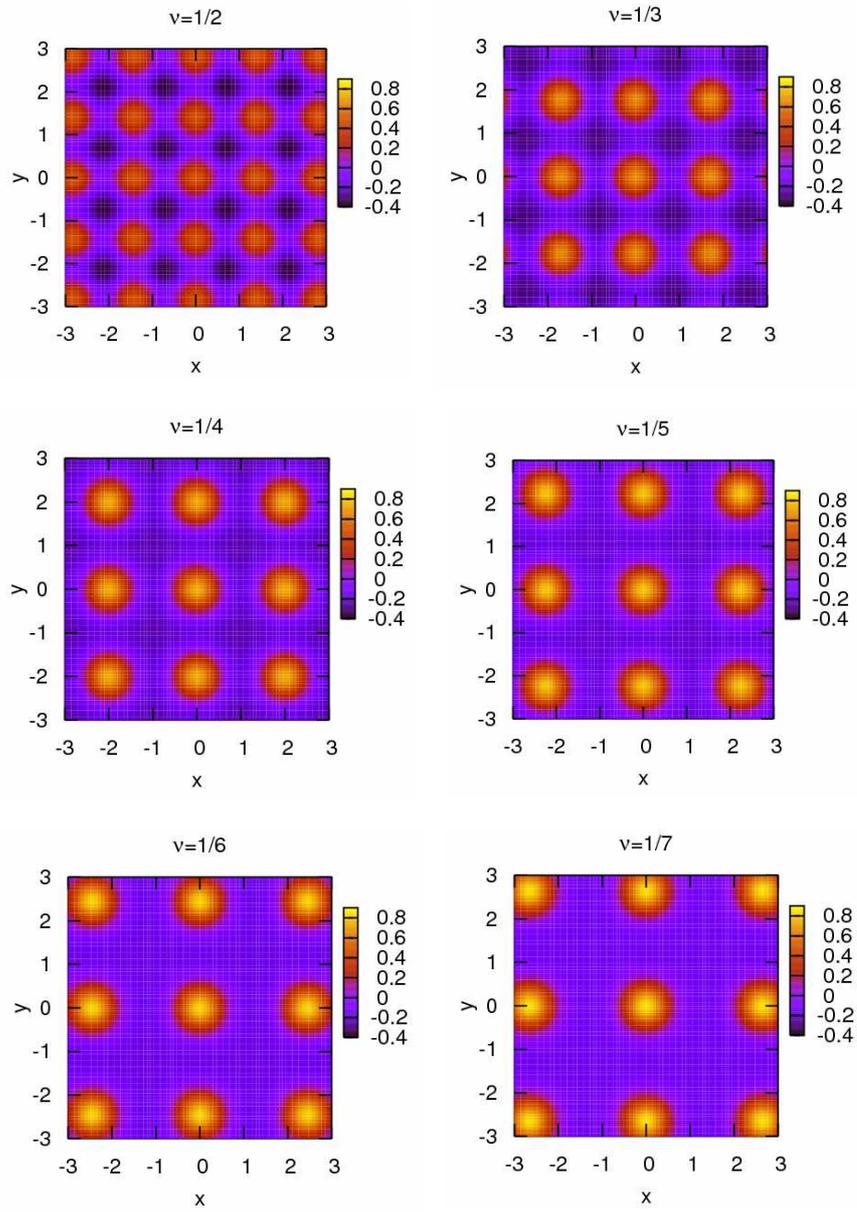}
\end{center}
\caption{\label{fig3:2DdensL0}
2D plot of the density of the $l=0$ CDW state. 
As the filling decreases, the distance between localized electrons 
increases isotropically.
}
\end{figure}
%== end ==%
%
%
%== Fig ==%
%2D plot of the CDW density at l=2. 
\begin{figure}[tbhp]
\begin{center}
\includegraphics[height=18cm,clip]{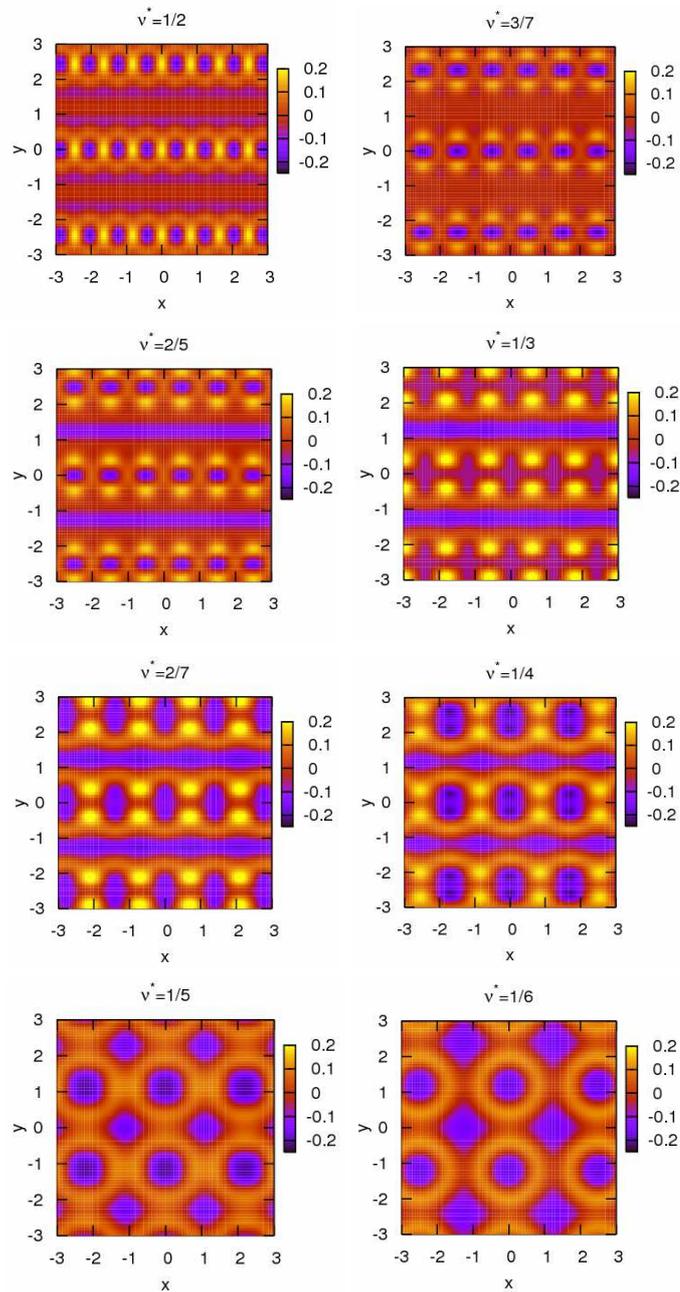}
\end{center}
\caption{\label{fig3:2DdensL2}
2D plot of the density of the $l=2$ CDW state. 
Near $\nu^\ast=1/2$, the density is highly anisotropic. 
As $\nu^\ast$ decreases, the density becomes isotropic gradually. 
}
\end{figure}
%== end ==%
%
%%%%%%%%%%%%%%% Energy %%%%%%%%%%%%%%%%%%%%%%%%%%%%%%%%5
%
%== Fig ==%
%2D plot of the energy dispersion at l=0. 
\begin{figure}[tbhp]
\vskip 2cm
\begin{center}
\includegraphics[height=7cm,clip]{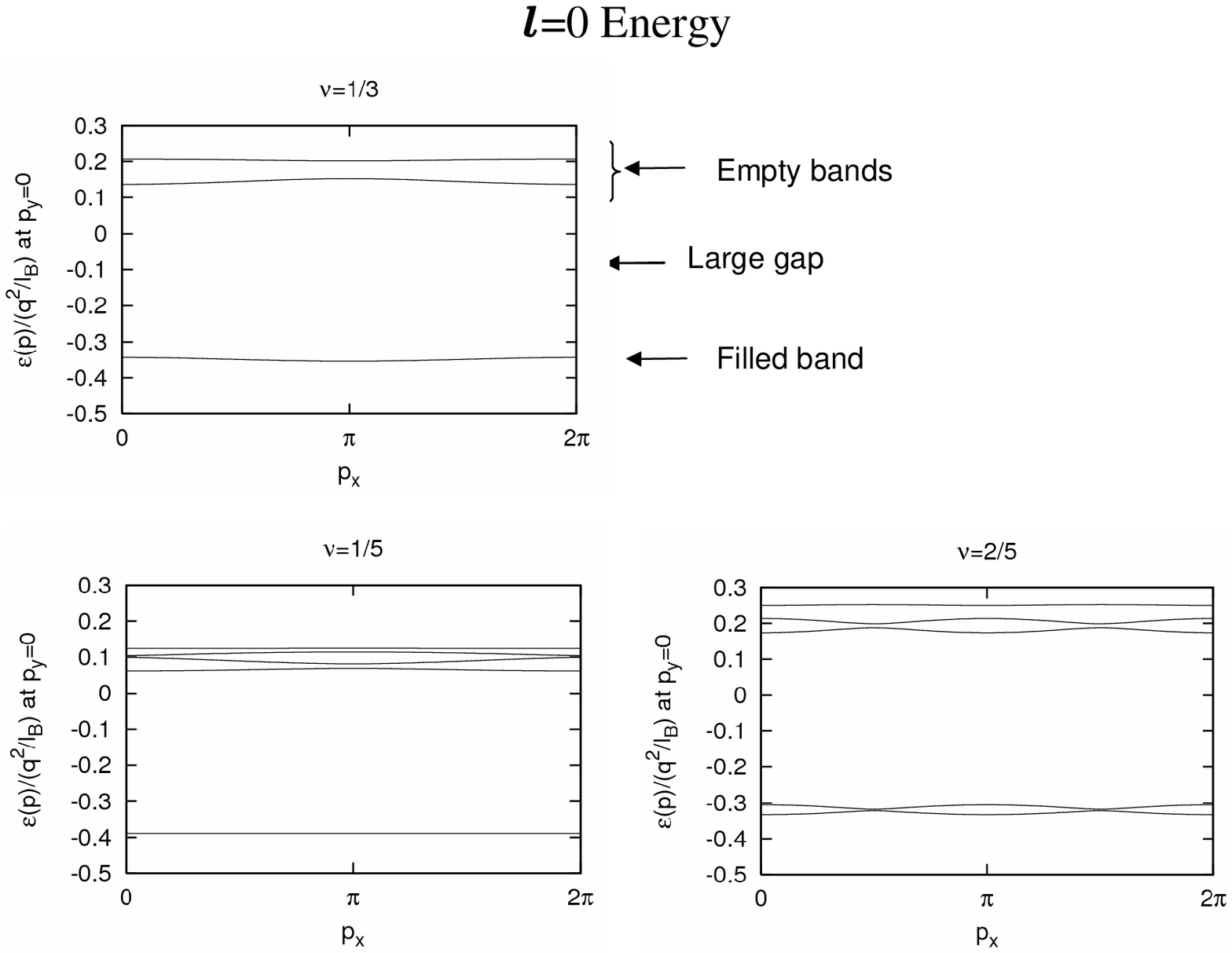}
\end{center}
\caption{\label{fig3:2DeneL0}
2D plot of the one-particle energy dispersion of the $l=0$ CDW state. 
The energy dispersion at $p_y=0$ is plotted as a function of 
$p_x$. 
There is a large gap between the filled bands and the empty bands. 
}
\end{figure}
%== end ==%
%
%
%== Fig ==%
%2D plot of the energy dispersion at l=0. 
\begin{figure}[tbhp]
\vskip 2cm
\begin{center}
\includegraphics[height=7cm,clip]{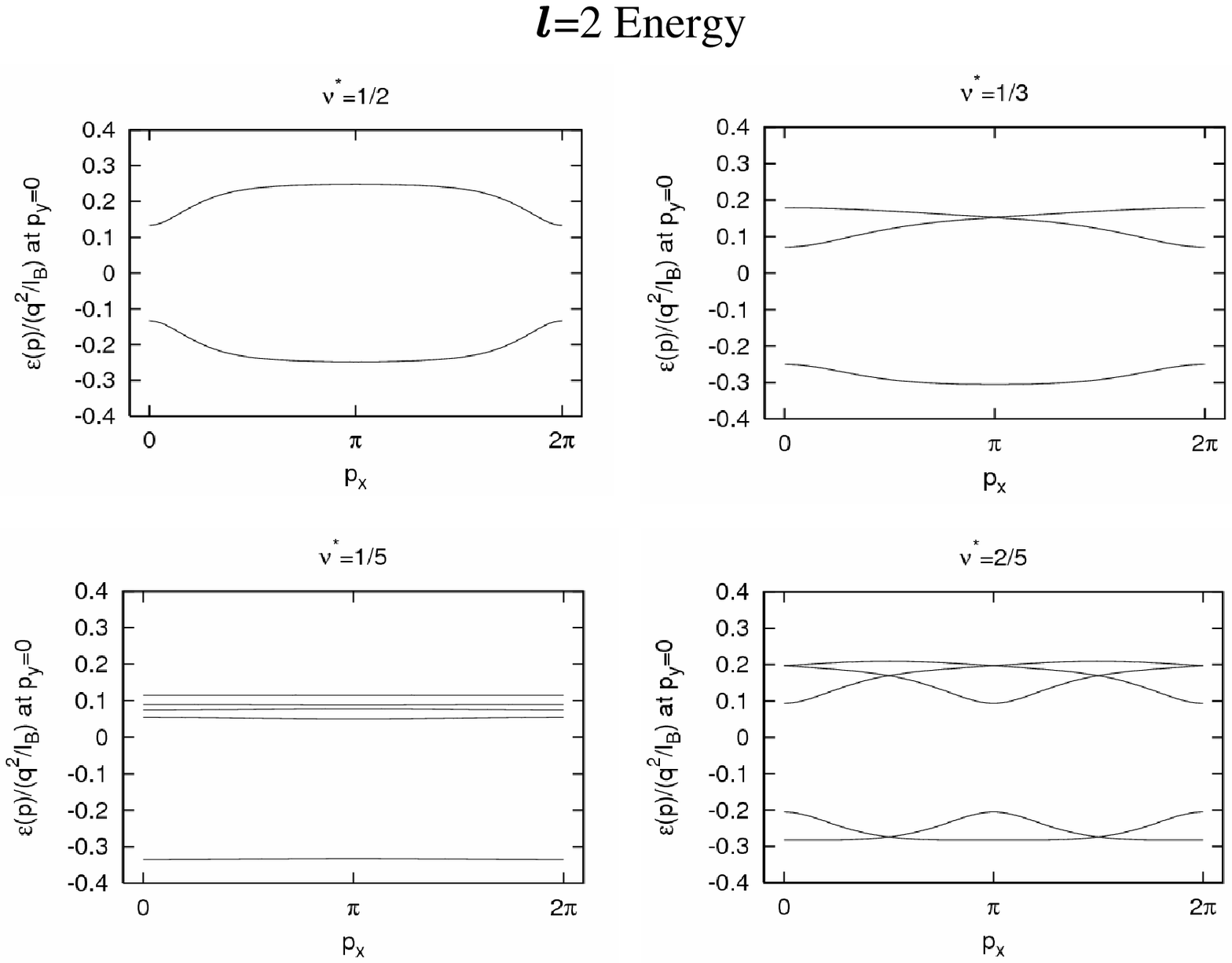}
\end{center}
\caption{\label{fig3:2DeneL2}
2D plot of the one-particle energy dispersion of the $l=2$ CDW state. 
The energy dispersion at $p_y=0$ is plotted as a function of $p_x$. 
There is a large gap between the filled bands and the empty bands. 
}
\end{figure}
%== end ==%
%
%%%%%%%%%%%%%%%%% Bubble %%%%%%%%%%%%%%%%%%%%%%%%%%%%%%
%
%== Fig ==%
%2D plot of the bubble density at l=2. 
\begin{figure}[tbhp]
\vskip 1cm
\begin{center}
\includegraphics[height=7cm,clip]{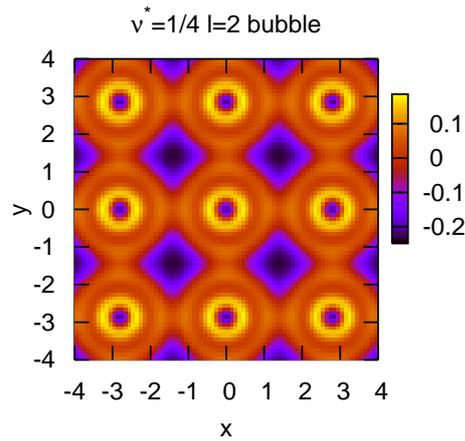}
\end{center}
\caption{\label{fig3:2DdensL2_bubble}
2D plot of the density of the $l=2$ 2-electron bubble state at 
$\nu^\ast=1/4$. 
The density is almost isotropic in the $x$ and $y$ directions.
}
\end{figure}
%== end ==%
%
%== Fig ==%
%2D plot of the bubble density at l=2. 
\begin{figure}[tbhp]
\vskip 4cm
\begin{center}
\includegraphics[height=5cm,clip]{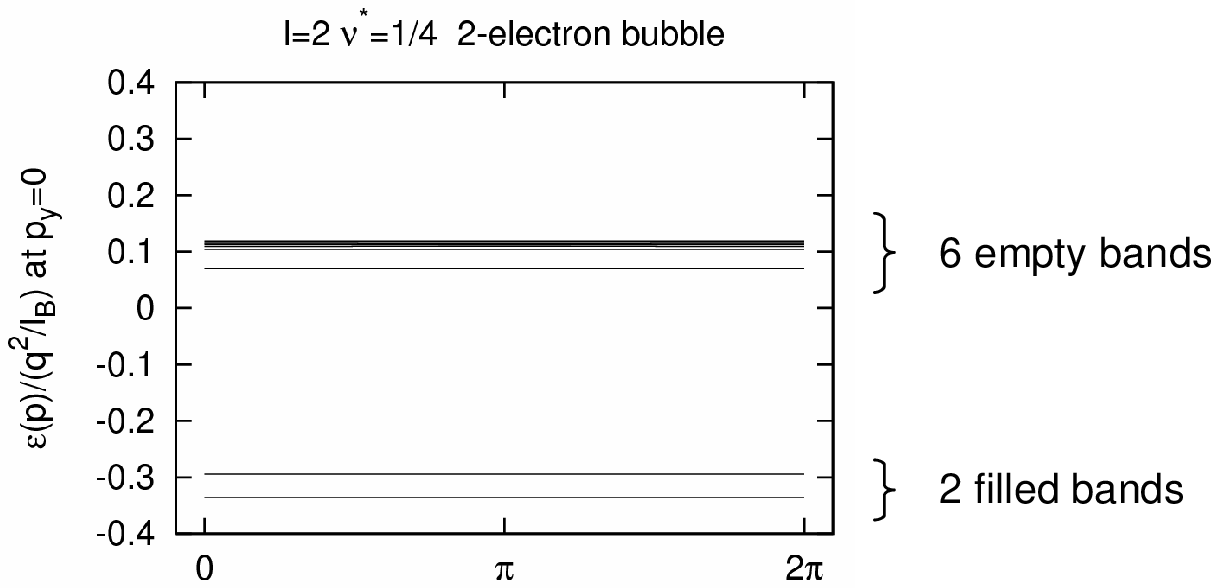}
\end{center}
\caption{\label{fig3:2DeneL2_bubble}
2D plot of the one-particle 
energy dispersion of the $l=2$ 2-electron bubble state at
 $\nu^\ast=1/4$. The energy dispersion at $p_y=0$ is plotted as a
 function of $p_x$. 
There is a large gap between the filled bands and the empty bands. 
}
\end{figure}
%== end ==%

\chapter{Effects of an injected current on highly anisotropic states}
\label{chapter4}
%%%%%%%%%%%%%%%%%%%%%%%%%%%%%%%%%%%%%%%%%%%%%%%%%%%%%%%%%%%%%%%%%%%%%%%%%%%
%% Chapter 4 Effects of an injected current on highly anisotropic ground %%
%%%%%%%%%%%%%%%%%%%%%%%%%%%%%%%%%%%%%%%%%%%%%%%%%%%%%%%%%%%%%%%%%%%%%%%%%%%
In this chapter, we study effects of an injected current on highly 
anisotropic states which have been discovered 
around half-filled third and higher Landau levels (LLs) 
in experiments with ultra-high mobility samples at low temperatures. 
While several states have been proposed 
to theoretically explain the experimental results so far, 
it is still an open problem which state is realized in
the experiments. 
Most previous studies have been done within the Hartree-Fock (HF) 
approximation with no injected current. 
In experiments of anisotropic states, however, 
a current is injected to investigate properties of the system. 
This effect has not been taken into account 
in the previous studies. 
In this chapter, we focus on two different anisotropic HF states, i.e., 
a striped Hall state and an anisotropic charge density wave (ACDW) state, 
and investigate 
the effect of the injected current on these two HF states using 
response functions against the injected current. 
The calculations are done by means of a von Neumann lattice (vNL) 
formalism developed in Chapter \ref{chapter2}. 
With no injected current, the ACDW state has a lower energy. 
We find that  
the striped Hall state becomes lower energy state when the
injected current exceeds a critical value. 
The critical value is estimated at 
about $0.04$-$0.05$ nA, which is much smaller than the current used in
experiments of anisotropic states.

%%%%%%%%%%%%%%%%%%%%%%%%%%%%%%
%% section 4-1 Introduction %%
%%%%%%%%%%%%%%%%%%%%%%%%%%%%%%
\section{Introduction}
In the quantum Hall system, 
half-filled states at each Landau level (LL) 
exhibit much attractive features. 
Around the half-filled lowest LL, 
isotropic compressible states have been observed \cite{Willett,HLR}, 
which are widely believed to be the Fermi liquid of composite fermions. 
Around the half-filled second LL, 
the $5/2$ fractionally quantized Hall conductance has been observed 
\cite{5/2}. 
The p-wave Cooper pairing state of composite fermions, 
which is called the Pfaffian state, 
has been proposed to explain this state 
\cite{Moore_Read,Greiter_Wen_Wilczek}. 
Around half-filled third and higher LLs, 
highly anisotropic states, which have extremely anisotropic longitudinal 
resistivities and un-quantized Hall resistivities, 
have been found in ultra-high mobility samples at low
temperatures \cite{Lilly,Du}. 
%Many theoretical works have 
Much theoretical work has 
been done to study the anisotropic states 
\cite{Koulakov,stripe00,stripe01,stripe02,stripe03,stripe04,stripe05,
stripe06,stripe07,stripe08,Aoyama_stripe,Cote_Fertig} and 
several states have been proposed so far. 
In this chapter, we focus on 
two different HF states among them, i.e., 
a striped Hall state \cite{Koulakov,stripe00} 
and an anisotropic charge density wave (ACDW) state \cite{Cote_Fertig}.

\begin{figure}[tbhp]
\begin{center}
 \includegraphics[width=10cm]{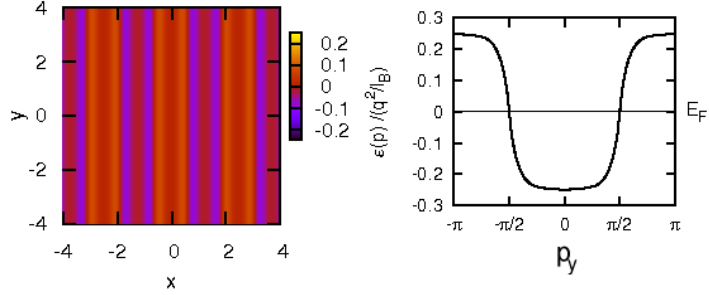}
\end{center}
\caption{\label{fig4:striped_Hall_state}Left: Density modulation of 
the striped Hall state at the half-filled third LL. 
The uniform part is subtracted. 
The density is periodic in the $x$-direction and uniform in the
$y$-direction (stripe direction). 
The injected current flows easily in the stripe direction. 
Right: Energy spectrum 
of the striped Hall state at the same filling. 
It is uniform in the $p_x$-direction. 
The BZ is $|p_i|<\pi$ and 
$E_F$ is a Fermi energy. 
}
\end{figure}

\begin{figure}[tbhp]
\begin{center}
 \includegraphics[width=10cm]{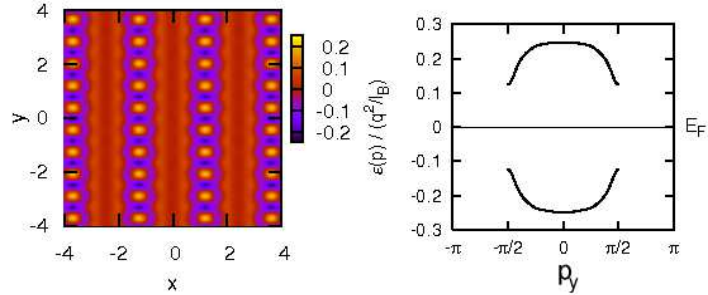}
\end{center}
\caption{\label{fig4:ACDW_state}Left: Density modulation of the ACDW
state at the half-filled third LL. The uniform part is subtracted. 
The density is modulated along stripes. 
Right: Energy spectrum of the ACDW state at the same filling. 
It depends on $p_x$ weakly and the spectrum at $p_x=0$ is plotted. 
The BZ is reduced to $|p_y|<\pi/2$ owing to %due to 
the modulations along stripes
and the two bands are formed. 
}
\end{figure}

The striped Hall state is a unidirectional charge density wave state 
which is a gapless state with an anisotropic Fermi surface
\cite{Ishikawa_stripe,Maeda_stripe}. 
The Fermi surface  has an energy gap in one
direction and is gapless in the other direction 
(Figs.~\ref{fig3:Fermi_sea} and \ref{fig4:striped_Hall_state}). 
The ACDW state is a CDW state which has a similar periodic density 
in a direction perpendicular to stripes, and in addition, has 
a density modulation along stripes (Fig.~\ref{fig4:ACDW_state}). 
The density modulation along stripes results in energy gaps in both directions. 
These features suggest that the anisotropic state found in
experiments is the striped Hall state since the anisotropic longitudinal
resistivity and the un-quantized Hall resistivity are naturally
explained by the anisotropic Fermi surface
\cite{Ishikawa_stripe,Maeda_stripe},  
while it is difficult to explain these experimental features with the
ACDW state because of the energy gap. 
However, it has been pointed out that the striped Hall state is unstable
within the HF approximation to formation of modulations along stripes so
that the ACDW state is the lower energy state \cite{Cote_Fertig}. 
This has been an enigma as to the anisotropic states.

%
%== Fig ==%
%%% figure %%%
%
%
%== Fig ==% 
%%% figure %%%

In the experiments of the anisotropic states, 
current is injected. 
This effect has not been taken into account 
in the previous calculations of the total energy. 
About two decades ago, 
MacDonald et al.~have studied 
the effect of the injected current on the integer quantum Hall system 
\cite{MacDonald}. 
They calculated the current and charge distributions 
and found that charges accumulate around both edges 
of the sample with the opposite sign, 
as expected from the classical Hall effect 
\cite{MacDonald,Thouless,Beenakker1,Beenakker2}. 
The charge accumulation causes the energy enhancement via the 
Coulomb interaction between charged particles. 
The same type of energy corrections may exist even in 
the present highly correlated quantum Hall states. 
However, the effect of the injected current on the anisotropic state 
has not been studied.

In this chapter, we calculate the correlation energies of 
the striped Hall state and the ACDW state in the system with the
injected current flowing in the direction along stripes, and with 
no impurities and no metallic contacts. 
It is important to know if the ACDW state has a lower energy even in the
system with the injected current. 
The density modulation along stripes shown in Fig.~\ref{fig4:ACDW_state}
suggest that the effect of the injected current becomes larger in the
ACDW state. 
To verify whether this naive expectation is true or not, 
the dependence of the correlation energies on the
injected current is studied in detail. 
Effects of impurities and metallic contacts are ignored in our
calculations of correlation energies since these effects are 
expected to be small in experiments of anisotropic states in 
ultra-high mobility samples and are outside the scope of this work. 
Effects of the injected current are investigated 
using response functions for electromagnetic fields. 
The current and charge distributions are determined and  
energies of the two states are calculated from these distributions. 
It is found that the energy of the ACDW state  
increases faster than that of the striped Hall state 
as the injected current increases. 
Hence, the striped Hall state becomes the lower energy state 
when the current exceeds the critical value. 
The critical value is estimated at about $0.04$ - $0.05$ nA, 
which is much smaller than the current used in the experiments. 
Our result suggests that the anisotropic states observed in 
experiments are the striped Hall states. 
Hence, the enigma as to the anisotropic states is resolved.

This chapter is organized as follows. 
In Sec.~\ref{chap4:Experiments}, we review the experiments of the 
anisotropic states and discuss the problem briefly. 
In Sec.~\ref{chap4:Hartree_Fock}, properties of 
the striped Hall state and the ACDW state 
are discussed within the HF approximation with no injected current. 
In Sec.~\ref{chap4:response_functions}, 
electromagnetic response functions 
of the two HF states are calculated in the long wavelength limit. 
Using these response functions, 
we determine the current and charge distributions and calculate 
the energy corrections due to injected currents 
in Sec.~\ref{chap4:energy_corrections}. 
A summary is given in Sec.~\ref{chap4:summary}.

%%%%%%%%%%%%%%%%%%%%%%%%%%%%%%%%%%%%%%%%%%%%%%%%%%%%%%%%
% section 4-2 Experiments of highly anisotropic states %
%%%%%%%%%%%%%%%%%%%%%%%%%%%%%%%%%%%%%%%%%%%%%%%%%%%%%%%%
\section{Experiments of highly anisotropic states}
\label{chap4:Experiments}
\begin{figure}[tb]
\begin{center}
 \includegraphics[width=9cm]{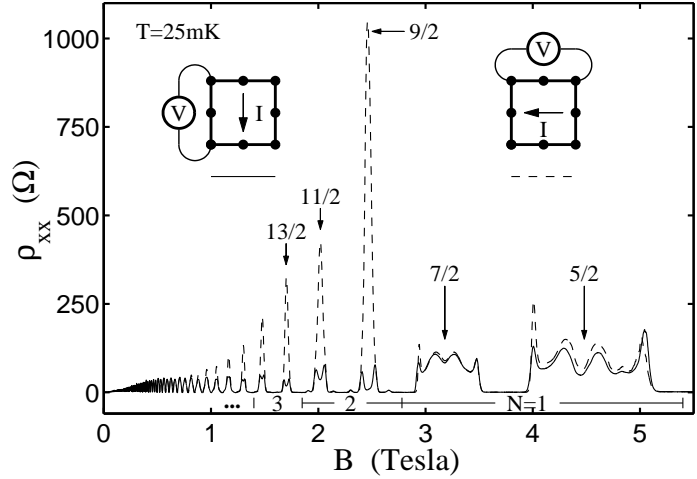}
\end{center}
\caption{\label{fig4:Lilly}
Longitudinal resistivity $\rho_{xx}$ in sample A at temperature $T=25$
 mK for two perpendicular directions of the current flow through the
 sample. 
 The two traces result from simply change of the direction of current
 through the sample; the sample itself is {\it not} rotated. 
 The diagrams in the figure depict the difference between two
 configurations. From Ref.~\cite{Lilly}. 
}
\end{figure}

In 1999, working with 2D ultra-high mobility samples at low temperatures, 
Lilly et al.~\cite{Lilly} and Du et al.~\cite{Du} have reported 
strong anisotropy in longitudinal resistivities $\rho_{xx}$
around $\nu^\ast=1/2$ in $l\geq2$ LLs. 
The experimental result by Lilly et al.~is shown in
Fig.~\ref{fig4:Lilly}. 
%
%== Fig ==% 
%%% figure %%%
In this figure, the sample A is a GaAs/AlGaAs heterojunction grown by
molecular beam epitaxy, where the density of the sample 
is close to $n_s=2.67\times 10^{11}$ ${\rm cm}^{-2}$, the low
temperature mobility is $\mu\geq 9\times 10^{6} {\rm cm}^2/V s$, 
and the sample size is a $5\times 5$ mm square. 
The electrical transport measurements have been performed using $2$-$20$
nA, $13$ Hz excitation. For more details about experimental parameters 
used by Lilly et al.~, see Ref.~\cite{Lilly}. 
We refer to these experimental parameters as {\it Lilly's parameters} in
this thesis. 

In this experiment, longitudinal resistivities $\rho_{xx}$ and Hall
resistivities $\rho_{xy}$ have been measured for two different
configurations. 
For the configuration shown in the left inset in Fig.~\ref{fig4:Lilly}, 
$\rho_{xx}$ is almost zero around half-fillings in $l\geq2$ LLs, such as 
$\nu=9/2$, $11/2$, $13/2$. 
On the other hand, for the other configuration shown in the right inset
in Fig.~\ref{fig4:Lilly}, $\rho_{xx}$ revels very large values 
around the same fillings, where the sample itself is not rotated and only 
the direction of current is changed. 
Here, no quantized plateau in the Hall resistivity 
$\rho_{xy}$ have been observed around these fillings. 
These experimental results indicate that 
the ground state around these fillings is not an isotropic state such as
a fractional quantum Hall state but an anisotropic state. 
Since the original Hamiltonian of the quantum Hall system has a
rotational symmetry, it is considered that the anisotropic state is
energetically favorable and a spontaneous symmetry breaking occurs with 
respect to 
the direction of the anisotropy which would be determined by 
some small residual inhomogeneity of the sample. 

%The longitudinal resistivities $\rho_{yy}$, $\rho_{yy}$ and the Hall
%resistivities $\rho_{xy}$, $\rho_{yx}$ are determined by measuring the
%longitudinal voltage $V_{x}$ or $V_{y}$ and the Hall voltage $V_{x}$ or
%$V_{y}$ under a small injected current. 

%%%%%%%%%%%%%%%%%%%%%%%%%%%%%%%%%%%%%%%%%%%%%%%%%%%%%%%%%%%%%%%%%%%%
% section 4-3 Highly anisotropic states in the Hartree-Fock theory %
%%%%%%%%%%%%%%%%%%%%%%%%%%%%%%%%%%%%%%%%%%%%%%%%%%%%%%%%%%%%%%%%%%%%
\section{Highly anisotropic states in the Hartree-Fock theory}
\label{chap4:Hartree_Fock}
In this section, the striped Hall state and the ACDW state at the
partial filling factor $\nu^\ast=1/2$ in higher LLs are investigated. 
Calculations are performed by means of a vNL basis 
within the HF approximation with no 
injected current. 
As has been mentioned in Chapter \ref{chapter2}, 
the vNL basis is a suitable basis to study spatially periodic
states.

%%%%%%%%%%%%%%%%%%%%%%%%%%%%%%%%%%%%%%%
% subsection 4-3-1 Striped Hall state %
%%%%%%%%%%%%%%%%%%%%%%%%%%%%%%%%%%%%%%%
\subsection{Striped Hall state at $\nu^\ast=1/2$}
The striped Hall state is obtained as a HF solution under the
assumption of the unidirectional charge density wave 
Eq.~(\ref{eq3:stripe_dens}). The solutions at $\nu^\ast=1/2$ are 
given in subsection \ref{chap3:stripe_solutions} in Chapter
\ref{chapter3} in detail. 
Let us consider the striped Hall state in which stripes 
face the $y$-direction. 
In this case, the anisotropic Fermi surface of the striped Hall state 
has the inter-LL energy gap in the $p_x$-direction and is gapless in the
$p_y$-direction. When a current flows in the $y$-direction, an electric
field is generated in the $x$-direction owing to %due to 
the Hall effect so that 
the Fermi sea slides in the $p_x$-direction. 
Since the Brillouin zone in the $p_x$-direction is fully occupied, 
there is no dissipation in this case and the longitudinal resistivity in
the $y$-direction is expected to be zero. 
On the other hand, when the current flows in the $x$-direction, 
the Fermi sea slides in the $p_y$-direction so that the 
energy dissipation occurs owing to %due to 
the gapless structure of the Fermi
surface in this direction, which would result in a large longitudinal
resistivity. 
Hence, it is expected that the striped Hall state revels the
highly anisotropic longitudinal resistivity, which coincides with the
experimental results.

%%%%%%%%%%%%%%%%%%%%%%%%%%%%%%%
% subsection 4-3-2 ACDW state %
%%%%%%%%%%%%%%%%%%%%%%%%%%%%%%%
\subsection{Anisotropic charge density wave state at $\nu^\ast=1/2$}
CDW states studied in Sec.~\ref{chap3:CDW} in Chapter \ref{chapter3} 
become highly anisotropic 
around $\nu^\ast=1/2$ in higher LLs. 
We call these highly anisotropic states at $\nu^\ast=1/2$ 
``{\it anisotropic charge density wave} (ACDW) states'' and 
the direction with a narrower periodicity in density 
``{\it stripe direction}'' in this thesis. 
%
%Since the original Hamiltonian in the quantum Hall system has a 2D rotational
%symmetry, the stripe direction can take any direction. 
%However, due to small residual impurities in the sample, 
In the vNL formalism, the HF Hamiltonian of the ACDW state can be
rewritten in the form of a $2\times 2$ matrix. 
Therefore, eigenvalues and eigenstates for the ACDW state can be easily 
obtained in a more analytical form as a function of order parameters. 
In what follows, we derive them and discuss the features of the
self-consistent ACDW solution. 

\begin{figure}[tb]
\begin{center}
 \includegraphics[width=6cm]{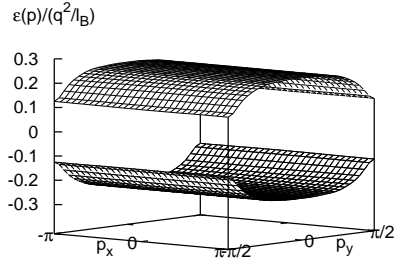}
\end{center}
\caption{\label{fig4:ene_ACDWL2}
One-particle energy of the $l=2$ 
ACDW state at half-filling. 
The uniform Fock energy is subtracted. 
Two bands are formed and  
the lower band is completely filled.}
\end{figure}

Let us start by the HF Hamiltonian of the CDW state at $\nu^\ast=1/2$. 
From Eq.~(\ref{eq3:HF_CDW}), this is given by 
%
%== Eq ==%
%HF Hamiltonian for CDW state at $\nu^\ast=1/2$
\begin{equation}
 \mathcal{H}^{(l)}_{{\rm HF-ACDW}}=\epsilon^{(0)}_l N_e^{(l)}+
  \int_{{\rm RBZ}}\frac{d^2p}{(2\pi)^2}{\bf
  b}_l^\dag({\bf p})D_l({\bf p}){\bf b}_l({\bf p}),
\label{eq4:HF-ACDW}
\end{equation}
%
%
%== Eq ==%
%Notation 1
\begin{gather}
 {\bf b}_l({\bf p})=
 \left(
  \begin{array}{cc}
   b_l(p_x, p_y)\\
   b_l(p_x, p_y+\pi)
  \end{array}
 \right),\quad 
 D_l({\bf p})=
 \left(
 \begin{array}{cc}
  A({\bf p}) & B({\bf p})\\
  B^\ast({\bf p}) & A(p_x, p_y+\pi)
 \end{array}
 \right), 
\end{gather}
where $\epsilon^{(0)}_l$ is a uniform Fock energy given by 
$\nu^\ast v_l^{\rm HF}(0)$, 
the momentum integration is performed over 
the reduced Brillouin zone (RBZ), $|p_x|<\pi$ and $|p_y|<\pi/2$, 
$D_l({\bf p})$ is a $2\times 2$ Hermite matrix, 
and $A({\bf p})$, $B{\bf p}$ are given by 
%
%== Eq ==%
%Notation 2
\begin{align}
 A({\bf p})=&\sum_{{\bf N}\neq 0}v_l^{\rm HF}(\frac{2\pi N_x}{r_s}, 2\pi N_yr_s)
 \Delta_l(\frac{2\pi N_x}{r_s}, 2\pi N_yr_s)
 e^{i\pi(N_x+N_y+N_xN_y)-ip_xN_y+ip_yN_x}\nonumber \\
 B({\bf p})=&\sum_{\bf N}v_l^{\rm HF}
 (\frac{2\pi N_x}{r_s},\pi(2N_y+1)r_s)
 \Delta_l(\frac{2\pi N_x}{r_s},\pi(2N_y+1)r_s)
 e^{i\pi(N_x+N_y+N_x(N_y+1/2))-ip_xN_y+ip_yN_x}. 
\end{align} 
The Hamiltonian Eq.~(\ref{eq4:HF-ACDW}) can be diagonalized 
at each momentum just by unitary transformation of the field operator 
${\bf b}_l({\bf p})$. 
In the present case of $\nu^\ast=1/2$, 
the Brillouin zone is reduced to the half size of 
the original domain and 
two energy bands are formed (Fig.~\ref{fig4:ene_ACDWL2}). 
%
%== Fig ==%
%Energy dispersion of ACDW state
$D_l({\bf p})$ is diagonalized using
the unitary matrix $U({\bf p})$ as
%
%== Eq ==%
%Diagonalization of D_l({\bf p}) 
\begin{equation}
 U^\dag({\bf p}) D_l({\bf p})U({\bf p})=
  \left(
   \begin{array}{cc}
    \epsilon_+({\bf p})& 0           \\
    0                & \epsilon_-({\bf p})
   \end{array}
  \right). 
\end{equation}
where $\epsilon_+({\bf p})$ and $\epsilon_-({\bf p})$ represent 
the upper energy band and the lower energy band respectively. 
If we assume the $x$- and $y$-inversion symmetries for the density, 
the order parameters become real and have the property 
$\Delta_l(Q_x,Q_y)=\Delta_l(Q_x,-Q_y)=\Delta_l(-Q_x,Q_y)$. 
The self-consistent solution is available 
only when the two energy bands are symmetric with 
respect to the energy gap, i.e., 
$\epsilon_+({\bf p})=-\epsilon_-({\bf p})\equiv \epsilon({\bf p})$, as expected from 
the particle-hole symmetry of the original Hamiltonian. 
This gives ${\rm Tr}_{2\times 2}D_l({\bf p})=0$, where 
${\rm Tr}_{2\times 2}$ denotes the trace with respect to the 
$2\times 2$ matrix indices, and 
$A(p_x, p_y+\pi)=-A({\bf p})$. 
In this case, $U({\bf p})$ and $\epsilon({\bf p})$ are given by 
%
%== Eq ==%
%Definition of U({\bf p}) and \epsilon({\bf p})
\begin{equation}
 U({\bf p})=
 \left(
  \begin{array}{cc}
   \frac{B({\bf p})}{N_+({\bf p})} &\frac{B({\bf p})}{N_- ({\bf p})} \\
   \frac{\epsilon_+ ({\bf p})-A({\bf p})}{N_+ ({\bf p})} & 
    \frac{\epsilon_- ({\bf p})-A({\bf p})}{N_- ({\bf p})}
  \end{array}
 \right), \quad 
 \epsilon({\bf p})=\sqrt{(A({\bf p}))^2+|B({\bf p})|^2}, 
\end{equation}
where 
$N_\pm({\bf p})= 2\epsilon_\pm({\bf p})(\epsilon_\pm({\bf p})-A({\bf p}))$. 
Using the basis ${\bf c}_l({\bf p})=U^\dag({\bf p}){\bf b}_l({\bf p})$, 
we obtain the HF Hamiltonian of the ACDW state 
%
%== Eq ==%
%HF Hamiltonian for ACDW
\begin{align}
 \mathcal{H}^{(l)}_{{\rm HF-ACDW}}=&\epsilon_0 N_e^{(l)}+
  \int_{{\rm RBZ}}\frac{d^2p}{(2\pi)^2}
  {\bf c}_l^\dag({\bf p})
  \left(
   \begin{array}{cc}
    \epsilon_+({\bf p})& 0           \\
    0                & \epsilon_-({\bf p})
   \end{array}
  \right) 
  {\bf c}_l({\bf p}),
\end{align}
where
%
%== Eq ==%
%Definition of b
\begin{equation}
 {\bf c}_l({\bf p})=
 \left(
  \begin{array}{cc}
   c_+(\bf p)\\
   c_-(\bf p)
  \end{array}
 \right),
\end{equation}
and $\Delta_l({\bf Q}_N)$ is determined as a numerical solution of 
%by solving 
the self-consistent
equation Eq.~(\ref{eq3:self_Delta}). % numerically. 

The HF energy of the ACDW state depends on 
the asymmetry parameter $r_s$. 
The optimal value of $r_s$, the HF energy per particle, 
and the magnitude of the energy gap are
given in Table \ref{table4:ACDW_table},  
%
%== Table ==%
%table
\begin{table}[tb]
\setlength{\extrarowheight}{1.05mm}
\newcolumntype{Y}{>{\centering\arraybackslash}X}
\begin{tabularx}{\linewidth}{YYYY}\hline\hline
$l$ & $r_s^{\rm ACDW}$ & $E_{\rm ACDW}/(q^2/l_{\rm B})$ & 
$\Delta_{\rm Gap}/(q^2/l_{\rm B})$\\ \hline
0 & $\sqrt{2}$  & -0.4436 &0.3292\\
1 & 1.02, 1.96 & -0.3583 &0.3077\\
2 & 0.82, 2.44 & -0.3097 &0.2470\\
3 & 0.70, 2.86 & -0.2814 &0.1967\\\hline\hline
\end{tabularx}
\caption{\label{table4:ACDW_table}
Minimum energy and corresponding parameter $r_s$ 
of the ACDW states at $\nu^\ast=1/2$. 
$\Delta_{\rm Gap}$ is a magnitude of the energy gap.}
\end{table}
in which there are two values of $r_s$ at each
LL owing to %due to 
the $\pi/2$-rotational symmetry. 
The magnitude of the energy gap is estimated to be of the order of $10$ K 
for a few tesla. 
Therefore, 
the extremely anisotropic longitudinal resistivities 
are not expected at tens of milli Kelvin and 
it is difficult to explain the experiments with the ACDW state. 
On the other hand, 
the HF energy of
the ACDW state is slightly lower than that of the striped Hall state at
each LL. 
This was one of the remaining issues for the anisotropic states.

%%%%%%%%%%%%%%%%%%%%%%%%%%%%%%%%%%
% section 4-4 Response functions %
%%%%%%%%%%%%%%%%%%%%%%%%%%%%%%%%%%
\section{Response functions}
\label{chap4:response_functions}
In this section, electromagnetic response functions 
of the two HF states are calculated 
in the long wavelength limit. 
We consider the quantum Hall system with an 
infinitesimal external gauge field 
$a_\mu(x)=(a_0(x), -{\bm a}(x))$ 
and calculate the response
functions of the striped Hall state and the ACDW state. 
These response functions will be used to determine 
current and charge distributions and energy corrections due to an injected
current in Sec.~\ref{chap4:energy_corrections}.

%%%%%%%%%%%%%%%%%%%%%%%%%%%%%%%%%%%%%%%%%%%%%%%%%%%%%%%%%%%%%%%%%
% subsection 4-4-1 Response functions of the striped Hall state %
%%%%%%%%%%%%%%%%%%%%%%%%%%%%%%%%%%%%%%%%%%%%%%%%%%%%%%%%%%%%%%%%%
\subsection{Response function of the striped Hall state}
The Hamiltonian in the quantum Hall system with $a_\mu(x)$ is given by 
%
%== Eq ==%
%total Hamiltonian 2
\begin{align}
 H=&\int d^2x \Psi^\dag({\bf x})
 \left(
 \frac{(-i\nabla+e{\bf A}({\bf x})+e{\bm a}({\bf x}))^2}
 {2m_e}-ea_0({\bf x}) 
 \right)\Psi({\bf x})+\frac{1}{2}\int d^2x d^2x'
 :\rho({\bf x})V({\bf x}-{\bf x}')\rho({\bf x}'):, 
 \label{eq4:Hamiltonian_with_a}
\end{align}
where $V({\bf x})=q^2/|{\bf x}|$. 
We project the Coulomb interaction part to each LL 
and apply the HF approximation to 
the projected Coulomb interaction. 
Then, by using the vNL basis, the Hamiltonian 
in the HF approximation is given by
%
%== Eq ==%
%HF Hamiltonian
\begin{align}
 H=&
 \sum_l E_l\int_{\rm BZ}
 \frac{d^2 p}{(2\pi)^2}b_l^\dag({\bf p})b_l({\bf p})
 \nonumber \\
 %%%
 &-\int\frac{d^2k}{(2\pi)^2}\sum_{l,l'}
 ef^\mu_{l,l'}(\tilde{\bf k})a_\mu(\tilde{\bf k})
 \int_{\rm BZ}\frac{d^2p}{(2\pi)^2}
 b^\dag_l({\bf p})b_{l'}({\bf p}-{\bf k})
 e^{-(i/4\pi)k_x(2p_y-k_y)}\nonumber\\
 %%%
 &+\int\frac{d^2k d^2k'}{(2\pi)^4}\sum_{l,l'}
 \frac{e^2\omega_c}{4\pi}
 {\bm a}(\tilde{\bf k})\cdot{\bm a}(\tilde{\bf k}')
 f^0_{l,l'}(\tilde{\bf k}+\tilde{\bf k}')\int_{\rm BZ}\frac{d^2p}{(2\pi)^2}b^\dag_l({\bf p})
 b_{l'}({\bf p}-{\bf k}-{\bf k}')
 e^{-(i/4\pi)(k_x+k'_x)(2p_y-k_y-k'_y)}\nonumber \\
 %%%
 &+\sum_{l}\int\frac{d^2k}{(2\pi)^2}v_l^{\rm HF}({\bf k})
 \langle \bar{\rho}_l(-\tilde{\bf k})\rangle \bar{\rho}_l(\tilde{\bf k}), 
\end{align}
where  
$f^\mu_{l_1,l_2}({\bf k})$ is defined by 
(see Appendix \ref{appA:LL_matrix}) 
%
%== Eq ==%
%Definition of f^{\mu\nu}
\begin{equation} 
 f^\mu_{l_1,l_2}({\bf k})=
  \langle f_{l_1}|\frac{1}{2}\{ 
  v^\mu,e^{i{\bf k}\cdot{\bm \xi}} \}|f_{l_2}\rangle
  \label{eq4:def_of_f}
\end{equation}
in which $v^\mu=(1,-\omega_c\eta,\omega_c\xi)$ is the electron
velocity. 
Repeated Greek indices $\mu$ and $\nu$ are summed for 0, 1, 2 
in throughout the following calculations. 
The action is given by
%
%== Eq ==%
%HF action for striped Hall state 
\begin{align}
 S[a_\mu,b,b^\dag]=\int dt 
  \Big[&\int_{\rm
  BZ}\frac{d^2p}{(2\pi)^2}b^\dag_l({\bf p},t)
  (i\partial_t+\mu_{\rm F})b_l({\bf p},t)-H(t)\Big], 
\label{eq4:HF_action}
\end{align}
where $H(t)$ is the Heisenberg representation of $H$.

Let us concentrate on the striped Hall state with the stripe direction 
facing the $y$-direction 
at half-filling. 
Substituting Eq.~(\ref{eq3:assumption_stripe}) into
Eq.~(\ref{eq4:HF_action}),  
we obtain the action of the striped Hall state given by 
%
%== Eq ==%
%Action of the striped Hall state
\begin{equation}
S_{\rm HF}[a_\mu,b,b^\dag]=\sum_{l,l'}\int_{\rm BZ}\frac{d^3p d^3p'}{(2\pi)^6}
b_l^\dag(p)\Bigl[
(p_0-\xi_l({\bf p}))\delta_{l,l'}(2\pi)^3\delta^3(p-p')
-U_{a_1}^{(l,l')}(p,p')-U_{a_2}^{(l,l')}(p,p')\Bigr]b_{l'}(p'), 
\end{equation}
where
%
%== Eq ==%
%definition of U's
\begin{align*}
 %Ua1
 U_{a_1}^{(l,l')}(p,p')=&-\sum_{\bf N}
 ef^\mu_{l,l'}(\tilde{\bf p}-\tilde{\bf p}'-2\pi\tilde{\bf N})
 h({\bf p}+{\bf p}',{\bf N})
 a_\mu(p_0-p'_0, 
 \tilde{\bf p}-\tilde{\bf p}'-2\pi\tilde{\bf N})
 e^{-(i/4\pi)(p_x-p'_x)(p_y+p'_y)},
 \nonumber  \\
 %Ua2
 U_{a_2}^{(l,l')}(p,p')=&
 \sum_{\bf N}\int\frac{d^3k}{(2\pi)^3}\frac{e^2\omega_c}{4\pi}
 f^0_{l,l'}(\tilde{\bf p}-\tilde{\bf p}'-2\pi\tilde{\bf N})
 h({\bf p}+{\bf p}',{\bf N})\nonumber \\
 &\times 
 {\bm a}(k_0, \tilde{\bf k})
 \cdot {\bm a}(p_0-p'_0-k_0, \tilde{\bf p}-\tilde{\bf p}'-\tilde{\bf
 k}-2\pi\tilde{{\bf N}})e^{-(i/4\pi)(p_x-p'_x)(p_y+p'_y)}
\end{align*}
%
%
%== Eq ==%
%Notation
\begin{equation}
 h({\bf p},{\bf N})\equiv (-1)^{N_x+N_y+N_xN_y}
 e^{-(i/2)p_xN_y+(i/2)p_yN_x}. 
\end{equation}
Here, $p$ denotes $(p_0, {\bf p})$, 
$\xi_l({\bf p})=E_l+\epsilon_l({\bf p})-\epsilon_{\rm F}$, 
and $U_{a_1}^{(l,l')}(p,p')$ and $U_{a_2}^{(l,l')}(p,p')$ are the first order term and the second
order term with respect to $a_\mu(p)$, respectively.

\begin{figure}[tbhp]
\begin{center}
 \includegraphics[width=7cm]{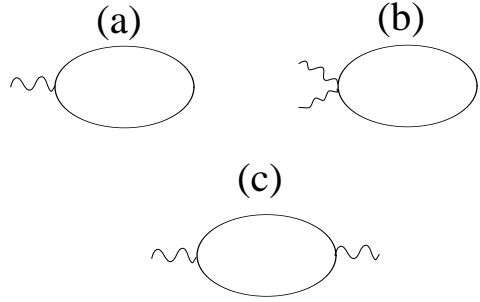}
\end{center}
\caption{\label{fig4:Feynman}(a), (b) and (c) are 
Feynman diagrams for 
$\Delta S_1$, $\Delta S_2$ and $\Delta S_3$, respectively.}
\end{figure}

The partition function $Z[a]$ is calculated using path integrals by
%
%== Eq ==%
%partition function
\begin{align}
 Z[a_\mu]=&\int\mathcal{D}b^\dag\mathcal{D}b\, 
 e^{iS_{\rm HF}[a_\mu,b,b^\dag]} \nonumber \\
 =&\int\mathcal{D}b^\dag\mathcal{D}b\, 
 e^{-(-i)\sum b^\dag(g^{-1}{\bf 1}-U_{a_1}-U_{a_2})b}
 =e^{{\rm Tr}\log[(-i)g^{-1}]}
 e^{{\rm Tr}\log[{\bf 1}-gU_{a_1}-gU_{a_2}]},
\end{align}
where 
the power of the exponent is expressed in the matrix
representation in momentum space and ${\rm Tr}$ denotes the trace of 
the momentum indices and the LL indices. 
$g_l(p)$ is the Green function given by
%
%== Eq ==%
%Green's function
\begin{equation}
 g_l(p)=\frac{\theta(\xi_l({\bf p}))}{p_0-\xi_l({\bf p})+i\delta}
  +\frac{\theta(-\xi_l({\bf p}))}{p_0-\xi_l({\bf p})-i\delta}, 
\end{equation}
where $\delta$ is an infinitesimal positive constant. 
The effective action $S_{\rm eff}[a_\mu]$ is defined as 
$S_{\rm eff}[a_\mu]=-i\log Z[a_\mu]$, which consists of 
the non-perturbed part $S_0={\rm Tr}\log[(-i)g^{-1}]$ and 
the correction part due to the external gauge field 
$\Delta S_{\rm eff}[a_\mu]$. $\Delta S_{\rm eff}[a_\mu]$ is given by 
(Fig. \ref{fig4:Feynman}), 
%
%== Eq ==%
%Effective action
\begin{equation}
 \Delta S_{\rm eff}[a_\mu]=
  \Delta S_1[a_\mu] +\Delta S_2[a_\mu] +\Delta S_3[a_\mu] +\mathcal{O}[a_\mu^3],
  \label{eq4:Seff_stripe}
\end{equation}
where
%
%== Eq ==%
%parts of the effective action
\begin{align}
 %Delta S_1
 \Delta S_1[a_\mu]=&i{\rm Tr}[gU_{a_1}]=
 i\sum_l\int_{\rm BZ}
 \frac{d^3p}{(2\pi)^3}g_l(p)U^{(l,l)}_{a_1}(p,p),\nonumber \\
 %Delta S_2
 \Delta S_2[a_\mu]=&i{\rm Tr}[gU_{a_2}]=
 i\sum_l\int_{\rm BZ}
 \frac{d^3p}{(2\pi)^3}g_l(p)U^{(l,l)}_{a_2}(p,p),\nonumber \\
 %Delta S_3
 \Delta S_3[a_\mu]=&\frac{i}{2}{\rm Tr}[gU_{a_1}gU_{a_1}]
 =\frac{i}{2}\sum_{l,l'}
 \int_{\rm BZ}\frac{d^3pd^3p'}{(2\pi)^6}
 g_l(p)U^{(l,l')}_{a_1}(p,p')
 g_{l'}(p')U^{(l',l)}_{a_1}(p',p).
 \label{eq4:S_stripe}
\end{align}
%
%
%== Fig ==%
%Feynman diagrams
Substituting the expressions for $g_l(p)$, 
$U_{a1}^{(l,l')}(p,p')$ and $U_{a2}^{(l,l')}(p,p')$ into
Eq.~(\ref{eq4:S_stripe}), 
we obtain %$\Delta S_{\rm eff}[a_\mu]$ is given by 
%
%== Eq ==%
%Effective action
\begin{align}
 \Delta S_{\rm eff}[a_\mu]=
 &(e l_0)a_0(0,{\bf 0})+\sum_{N_x} e
 f^\mu_{l_0,l_0}(-2\pi N_x/r_s,0)\frac{\sin(p_FN_x)}{\pi N_x}
 e^{i\pi N_x}
 a_\mu(0,-2\pi N_x/r_s,0)\nonumber \\
 &-\frac{1}{2}\int\frac{d^3 p}{(2\pi)^3}\sum_{\bf N}
 a_\mu(p_0, {\bf p})K^{\mu\nu}(p,{\bf N})
 a_\nu(-p_0, -{\bf p}-2\pi\tilde{\bf N}),
\end{align}
where $l_0$ represents the uppermost partially filled LL 
and $\tilde{\bf N}=(N_x/r_s, r_s N_y)$. 
$K^{\mu\nu}(p,{\bf N})$ is a response function given by 
%
%== Eq ==%
%Response functions
\begin{align}
 \label{eq4:response_stripe}
 K^{\mu\nu}(p,{\bf N})=&\sum_{l,l'}e^2f^\mu_{l,l'}({\bf p})
  f^\nu_{l',l}(-{\bf p}-2\pi\tilde{\bf N})I_{l,l'}(p_0,\hat{\bf p},{\bf N})
  h(\hat{\bf p}, {\bf N})\nonumber \\
 &+\frac{e^2}{2\pi}\omega_c\sum_{\bf N}\left[
 l_0\delta_{{\bf N},0}+f_{l_0,l_0}^0(-2\pi\tilde{\bf N})(-1)^{N_x}
 \frac{\sin(\pi N_x/2)}{\pi N_x}\delta_{N_y,0}\right]
 (\delta_{\mu,1}\delta_{\nu,1}+\delta_{\mu,2}\delta_{\nu,2}), 
\end{align}
and $I_{l,l'}(p_0,{\bf p},{\bf N})$ is the loop integral in 
Fig.~\ref{fig4:Feynman} (c), 
which is given by
%
%== Eq ==%
%I_{l,l'}
\begin{align}
 I_{l,l'}(p_0,{\bf p},{\bf N})=&\int_{\rm BZ}\frac{d^2p'}{(2\pi)^2}
  \Biggl[
   \frac{\theta(\xi_l({\bf p}+{\bf p}'))\theta(-\xi_{l'}({\bf p}'))}
   {p_0+\xi_{l'}({\bf p}')-\xi_l({\bf p}+{\bf p}')+i\delta}
   -\frac{\theta(-\xi_{l}({\bf p}+{\bf p}'))\theta(\xi_{l'}({\bf p}'))}
   {p_0+\xi_{l'}({\bf p}')-\xi_l({\bf p}+{\bf p}')-i\delta}
  \Biggr]e^{-ip'_xN_y+ip'_yN_x}, 
\end{align}
in which the $p'_0$ integral has been performed. 
In Eq.~(\ref{eq4:response_stripe}), 
the first term and the second term  come from 
$\Delta S_3[a_\mu]$ and $\Delta S_2[a_\mu]$, respectively, 
and 
the second term is canceled with the $p=0$ part of the first term, as
expected from gauge invariance. 
Hence, $K^{i,i}(p=0,{\bf N})=0$ for $i=1,2$.

In the long wavelength limit, 
the largest contribution in the response function 
comes from the ${\bf N}=0$ part. 
In the case of $p_0=p_y=0$ and $p_x\to 0$, which is used in the next
section, 
the largest contribution comes from the lowest order term
in $K^{\mu\nu}_0(p_x)\equiv K^{\mu\nu}(0, p_x, 0)$ 
with respect to $p_x$.
Expanding $K^{\mu\nu}_0(p_x)$ up to the lowest order, 
we obtain the response functions in the long wavelength limit 
given by 
%
%== Eq ==%
%Response functions
\begin{align}
 K^{00}_0(p_x)=&
 -\frac{\sigma^{(\nu)}_{xy}}{\omega_c}p^2_x,\qquad 
 K^{0y}_0(p_x)=-i\sigma^{(\nu)}_{xy}p_x,\nonumber \\
 K^{y0}_0(p_x)=&i\sigma^{(\nu)}_{xy}p_x,\qquad 
 K^{yy}_0(p_x)=\alpha_K\omega_c p^2_x,
\end{align}
where $\sigma^{(\nu)}_{xy}=e^2\nu/2\pi$ 
and $\alpha_K=e^2\omega_c(l_0^2+2l_0\nu^\ast+\nu^\ast)/4\pi^2$. 
$\sigma_{xy}^{(\nu)}$ is identified as the Hall conductance since 
if we consider a static homogeneous electric field in the
$x$-direction generated by the gauge field $a_0^{\rm ex}(x)=xE_x$, then
the electric current in the $y$-direction $\langle j_y(x)\rangle $ is given in the long
wavelength limit by 
$\langle j_y(x) \rangle=\delta \Delta 
S_{\rm eff} / \delta a_y(x)=
K^{y0}_0(\partial_x)a_0(x)=-\sigma_{xy}^{(\nu)}E_x$, 
where the response
function transformed in the coordinate space is used.
The longitudinal resistivity becomes zero in the present calculation 
since the impurity potential 
is not included. 
If impurities are added, it is expected that 
the longitudinal resistivity becomes zero in
one direction and finite in the other
direction owing to %due to 
the anisotropic Fermi surface.

%%%%%%%%%%%%%%%%%%%%%%%%%%%%%%%%%%%%%%%%%%%%%%%%%%%%%%%%
% subsection 4-4-2 Response function of the ACDW state %
%%%%%%%%%%%%%%%%%%%%%%%%%%%%%%%%%%%%%%%%%%%%%%%%%%%%%%%%
\subsection{Response function of the anisotropic charge density wave state}
The action of the ACDW state at half-filling 
is given by
%
%== Eq ==%
%HF action for ACDW state 
\begin{align}
S_{\rm HF}[a_\mu,{\bf c},{\bf c}^\dag ]=\sum_{l,l'}\int_{\rm RBZ}\frac{d^3pd^3p'}{(2\pi)^6}
 {\bf c}_l^\dag(p)\Big[G^{-1}_l(p)
 \delta_{l,l'}(2\pi)^3\delta^3(p-p')
 -V^{(l,l')}_{a_1}(p;p')-V^{(l,l')}_{a_2}(p;p')\Big]{\bf c}_{l'}(p'), 
\end{align}
where $V_{a_j}^{(l, l')}(p;p')$ with $j=1,2$ is a $2\times 2$ matrix given by
%
%== Eq ==%
%Definition of V
\begin{align}
 %Va1a2
 V_{a_j}^{(l,l')}(p;p')=&U^{\dag}({\bf p}){\bf U}^{(l,l')}_{a_j}(p;
 p')U({\bf p}),
 \nonumber \\
 %Ua1a2
 {\bf U}_{a_j}^{(l,l')}(p;p')=&
 \left(
  \begin{array}{cc}
   U_{a_j}^{(l,l')}(p;p')& U_{a_j}^{(l,l')}(p;p'_0,p'_x,p'_y+\pi) \\
   U_{a_j}^{(l,l')}(p_0,p_x,p_y+\pi;p') 
    & U_{a_j}^{(l,l')}(p_0,p_x,p_y+\pi;p'_0,p'_x,p'_y+\pi)
  \end{array}
 \right), 
\end{align}
and $G_l(p)$ is a $2\times 2$ matrix Green function given by
%
%== Eq ==%
%Definition of G_l
\begin{equation}
 G_{l}(p)=\left\{
	   \begin{array}{ll}
	    \left(
	     \begin{array}{cc}
	      g^+_l(p)& 0\\
	      0 & g^-_l(p)
	     \end{array}
	    \right)
	    &{\rm for} \quad l=l_0\\
	    g^-_l(p){\bf 1}&{\rm for} \quad l<l_0\\
	    g^+_l(p){\bf 1}&{\rm for} \quad l>l_0
	   \end{array}
	   \right. .
\label{eq4:Green}
\end{equation}
In Eq.~(\ref{eq4:Green}), 
$g^{\pm}_l(p)=1/(p_0-\xi^\pm_l(p)\pm i\delta)$ is 
a one-particle Green function for the upper or lower band, 
${\bf 1}$ is a $2\times 2$ unit matrix,
and 
$\xi^{\pm}_l({\bf p})=E_{l_0}+\epsilon^{(l_0)}_{\pm}({\bf p})-\epsilon_{\rm F}$.

The partition function $Z[a_\mu]$ is calculated using path integrals by
%
%== Eq ==%
%Partition function
\begin{equation}
 Z[a_\mu]=\int\mathcal{D}{\bf c}^\dag\mathcal{D}{\bf c}\, 
 e^{iS_{\rm HF}[a_\mu,{\bf c},{\bf c}^\dag]} 
 =e^{{\rm Tr}\log[(-i)G^{-1}]}
 e^{{\rm Tr}\log[{\bf 1}-GV_{a_1}-GV_{a_2}]}. 
\end{equation}
The correction part of the effective action 
$\Delta S_{\rm eff}[a_\mu]$ is given by
%
%== Eq ==%
%Effective action
\begin{equation}
 \Delta S_{\rm eff}[a_\mu]=
  \Delta S_1[a_\mu] +\Delta S_2[a_\mu] +\Delta S_3[a_\mu] +\mathcal{O}[a_\mu^3],
  \label{eq4:Seff_ACDW}
\end{equation}
where
%
%== Eq ==%
%Delta_S's
\begin{align}
 %Delta S_1
 \Delta S_1[a_\mu]=&i{\rm Tr}[GV_{a_1}]
 =
 i\sum_l\int_{\rm RBZ}\frac{d^3p}{(2\pi)^3}
 {\rm Tr_{2\times 2}}[G_l(p)V^{(l,l)}_{a_1}(p;p)],
 \nonumber \\
 %Delta S_2
 \Delta S_2[a_\mu]=&i{\rm Tr}[GV_{a_2}]
 =
 i\sum_l\int_{\rm RBZ}\frac{d^3p}{(2\pi)^3}
 {\rm Tr_{2\times 2}}[G_l(p)V^{(l,l)}_{a_2}(p;p)],
 \nonumber \\
 %Delta S_3
 \Delta S_3[a_\mu]=&\frac{i}{2}{\rm Tr}[GV_{a_1}GV_{a_1}]
 =\frac{i}{2}\sum_{l,l'}
 \int_{\rm RBZ}\frac{d^3pd^3p'}{(2\pi)^6}
 {\rm Tr_{2\times 2}}
 [G_l(p)V^{(l,l')}_{a_1}(p;p')
 G_{l'}(p')V^{(l',l)}_{a_1}(p';p)]. 
 \label{eq4:Delta_ACDW}
\end{align}
Here, ${\rm Tr}$ denotes the trace with respect to the momentum
indices, the LL indices and
the $2\times 2$ matrix indices, 
and ${\rm Tr}_{2\times 2}$ denotes the
trace with respect to only the $2\times 2$ matrix indices. 
%%%
Substituting the expressions for $G_l(p)$, 
$V_{a_1}^{(l,l')}(p;p')$, 
$V_{a_2}^{(l,l')}(p;p')$ into
Eq.~(\ref{eq4:Delta_ACDW}),
we obtain the following expression for 
the ${\bf N}=0$ part of $\Delta S_{\rm eff}[a_\mu]$ 
as in the case of the striped Hall state, 
%
%== Eq ==%
%Effective action of the ACDW state
\begin{align}
 \Delta S_{\rm eff}[a_\mu]=&
 e(l_0+\frac{1}{2}) a_0(0,{\bf 0})
 -\frac{1}{2}\int\frac{d^3p}{(2\pi)^3}
 a_{\mu}(p_0,{\bf p})K_0^{\mu\nu}(p)a_\nu(-p_0,-{\bf p}), 
\end{align}
where $K^{\mu\nu}_0(p)$ is given by
%
%== Eq ==%
%
\begin{align}
 \label{eq4:response_ACDW}
 K^{\mu\nu}_0(p)=&
 -\sum_{l<l_0}\sum_{l'>l_0}
 \frac{e^2}{\omega_c(l'-l)}
 \left\{f^\mu_{l,l'}({\bf p})f^\nu_{l',l}(-{\bf p})
 +f^\mu_{l',l}({\bf p})f^\nu_{l,l'}(-{\bf p}))\right\}
 \nonumber \\
 &+\frac{1}{2}
 \left[-\sum_{l<l_0}+\sum_{l>l_0}\right]
 \frac{e^2}{\omega_c(l_0-l)}
 \left\{f^\mu_{l,l_0}({\bf p})f^\nu_{l_0,l}(-{\bf p})
 +f^\mu_{l_0,l}({\bf p})f^\nu_{l,l_0}(-{\bf p})
 \right\}
 \nonumber \\
 %%%
 &+\int_{\rm RBZ}\frac{d^2p'}{(2\pi)^2} 
 \frac{1}{p_0-(\epsilon({\bf p}')+\epsilon(\hat{\bf p}+{\bf p}'))}
 \nonumber \\
 %%%
 &\qquad \times 
 \left[1-
 \frac{A(\hat{\bf p}+{\bf p}')A({\bf p}')
 +{\rm Re}(B(\hat{\bf p}+{\bf p}')B^\ast({\bf p}')
 e^{-(i/2)r_s p_x})}
 {\epsilon(\hat{\bf p}+{\bf p}')\epsilon({\bf p}')}\right]
 e^2f^\mu_{l_0,l_0}({\bf p})f^\nu_{l_0,l_0}(-{\bf p})
 \nonumber \\
 &+(l_0+\frac{1}{2})\frac{e^2}{2\pi}\omega_c
 (\delta_{\mu,1}\delta_{\nu,1}+\delta_{\mu,2}\delta_{\nu,2}), 
\end{align}
up to $\mathcal{O}(\epsilon({\bf p})/\omega_C)$. 
In Eq.~(\ref{eq4:response_ACDW}), the last term is canceled with 
the $p=0$ term of the second term 
as expected from gauge invariance
again. Hence, $K_0^{i,i}(0)=0$ for $i=1,2$.

In the long wavelength limit $p_0=p_y=0$ and $p_x\to 0$, 
the largest contribution in the response function 
comes from the lowest order term with respect to $p_x$. 
The expressions of $K_0^{0y}(p_x)$ and $K_0^{y0}(p_x)$ become the same as in
the case of the striped Hall state. 
The expression of $K_0^{00}(p_x)$ becomes slightly different by the 
correction from the intra-LL effect 
at the uppermost partially-filled LL.
For the striped Hall state, 
the one-particle energy shown in Fig.~\ref{fig3:ene_stripeL2} has the 
inter-LL energy gap in the $p_x$-direction, and 
the inter-LL effect gives the response functions given 
in Eq.~(\ref{eq4:response_stripe}). 
For the ACDW state, 
the one-particle energy shown in Fig.~\ref{fig4:ene_ACDWL2} has the 
intra-LL energy gap in the $p_x$-direction as well as the inter-LL
energy gap. 
While the inter-LL effect gives the same expression of the response function
as that of the striped Hall state, 
the intra-LL effect causes some corrections to the response function. 
Including these corrections, we obtain $K_0^{00}(p_x)$ 
of the ACDW state in the long wavelength limit given by (see
Appendix \ref{appB:K00}) 
%
%== Eq ==%
%
\begin{equation}
 K^{00}_0(p_x)=-(1+\frac{\sqrt{B}}{\nu}\beta)
  \frac{\sigma_{xy}^{(\nu)}}{\omega_c}p_x^2, 
  \label{eq4:K_for_ACDW}
\end{equation}
where $a=\sqrt{2\pi/eB}$ is used explicitly in order to compare the
theoretical results with experimental data. 
The value of $\beta$ at each LL 
is shown in Table \ref{table4:beta}. 
Note that the unit of $\beta$ is $({\rm tesla})^{-1/2}$. 
%
%== Table ==%
%table
\begin{table}[tb]
\setlength{\extrarowheight}{1.05mm}
\newcolumntype{Y}{>{\centering\arraybackslash}X}
\begin{tabularx}{\linewidth}{YYY}\hline\hline
$l$ & parallel  & perpendicular \\\hline
0 & 0.497  & 0.497 \\
1 & 0.386  & 0.824 \\
2 & 0.472  & 1.044 \\
3 & 0.581  & 1.171 \\\hline\hline
\end{tabularx}
\caption{\label{table4:beta}The value of $\beta$ 
at each LL. {\it parallel} is the 
case in which the stripe direction faces the $y$-direction. 
{\it perpendicular} is the case in which the stripe direction faces the
 $x$-direction. The unit of $\beta$ is $({\rm tesla})^{-1/2}$.}
\end{table}

The Hall conductance is given by $\sigma_{xy}^{(\nu)}=e^2\nu/2\pi$, as in the case of
the striped Hall state. 
The longitudinal resistivity becomes zero in the present calculation 
since the impurity potential 
is not included. 
However it is expected that 
the longitudinal resistivity remains zero even in the system with
impurities because of the energy gaps.

%%%%%%%%%%%%%%%%%%%%%%%%%%%%%%%%%%%%%%%%%%%%%%%%%%%%%%%%%%%%%%%%%%
% section 4-5 Energy corrections due to finite electric currents % 
%%%%%%%%%%%%%%%%%%%%%%%%%%%%%%%%%%%%%%%%%%%%%%%%%%%%%%%%%%%%%%%%%%
\section{Energy corrections due to injected currents}
\label{chap4:energy_corrections}
In this section, we consider the quantum Hall system with an injected
electric current 
and 
investigate the current effect on the striped Hall state and the ACDW
state. 
Effects of impurities and metallic contacts are ignored in our
calculations.
For the striped Hall state, we only consider the current parallel to the
stripe direction since in this case, the current effect can be estimated
with no ambiguity even in the system with impurities. 
When the current flows in the stripe direction, 
charges accumulate around both edges of the sample in the direction
perpendicular to the stripe direction (perpendicular direction), 
as we will see later, 
and the electric field generates in the perpendicular direction. 
In this case, the impurity effect is negligible
since the Fermi surface has the inter-LL energy gap 
in the perpendicular direction. 
On the other hand, 
when the current flows in the perpendicular direction, 
the impurity effect becomes relevant since 
the electric field generates in the stripe direction 
while the Fermi surface is gapless in this direction. 
The current effect in this case is nontrivial and will be studied 
in future work.

In the system with an injected current, 
it is naively expected that the current flow causes the plus and minus 
charge accumulation at both edges of the sample with the opposite sign, 
as expected from the classical Hall effect. 
MacDonald et al.~have studied effects of an injected current on the 
integer quantum Hall state about two decades ago \cite{MacDonald}. 
They have calculated current and charge distributions 
and found that the charge accumulation occurs in the integer
quantum Hall state. 
The charge accumulation causes 
the energy correction via the Coulomb interaction between the accumulated
charges. 
It is expected that the same type of energy corrections exists even 
in the present highly correlated quantum Hall states. 
However, it has not been studied as far as the present authors know. 
In what follows, 
first, we derive current and charge distributions in the striped
Hall state and the ACDW state using the effective action, 
then, we estimate the current dependence of energy corrections of 
the two HF states. 
It is shown that the energy of the ACDW state increases faster than 
that of the striped Hall state as the injected current increases.

%%%%%%%%%%%%%%%%%%%%%%%%%%%%%%%%%%%%%%%%%%%%%%%%%%%%%
% subsection 4-5-1 Current and charge distributions %
%%%%%%%%%%%%%%%%%%%%%%%%%%%%%%%%%%%%%%%%%%%%%%%%%%%%%
\subsection{Current and charge distributions}
We study current and charge distributions of the striped Hall state and
the ACDW state. 
Let us begin with the system with no injected current. 
We denote the two-point function in the HF theory with no injected
current as $\langle \Psi^\dag({\bf x}, t)\Psi({\bf x}', t)\rangle_{I=0}$. 
If a finite electric current is injected, 
electromagnetic fields and the two-point function 
deviate from their original values. 
%These deviations are taken into account 
%in the calculation of the total energy. 
We define these deviations by 
%
%== Eq ==%
%Definition of deviations a, \delta\rho
\begin{align}
 \label{eq4:def_of_deviations}
 {\bm a}({\bf x}, t)=&{\bf A}({\bf x},t)-{\bf A}_{\rm ex}({\bf x}), \nonumber \\
 \delta\rho({\bf x},{\bf x}',t)=&
 \langle \Psi^\dag({\bf x},t)\Psi({\bf x}',t)\rangle - \langle
 \Psi^\dag({\bf x},t)\Psi({\bf x}',t)\rangle_{I=0}, 
\end{align}
where ${\bm a}({\bf x},t)$ and $\delta\rho({\bf x},{\bf x}',t)$ 
are unspecified for the moment and 
will be determined later. 
The total action in the Coulomb gauge 
$\nabla \cdot {\bf A}(x)=0$ is given as 
%
%== Eq ==%
%Total action
\begin{align}
 S_{\rm tot}[{\bf A}, \Psi^\dag, \Psi]=&\int dtd^3x
 \left( \frac{\epsilon}{2}{\dot {\bf A}}^2({\bf x},t)-\frac{1}{2\mu}
 (\nabla\times{\bf A}({\bf x},t))^2 \right)\nonumber \\
 &+\int dtd^3x \Psi^\dag({\bf x},t)
 \left(i\partial_t-\frac{({\bf p}+e{\bf
 A}({\bf x},t))^2}{2m^\ast}\right)\Psi({\bf x},t)\delta(z)\nonumber \\
 &-\frac{1}{2}\int dt d^3x d^3x'
 \Psi^\dag({\bf x},t)\Psi^\dag({\bf x}',t)V({\bf x}-{\bf x}')
 \Psi({\bf x}',t)\Psi({\bf x},t)\delta(z)\delta(z').
\label{eq4:total_action}
\end{align}
where $\mu$ is the magnetic constant, $m^\ast$ is the effective electron
mass, and the dot means the
time derivative. 
This total action consists of 
the three-dimensional electromagnetic field term and the two-dimensional 
electron field term. 
In the Coulomb gauge, the 
interaction between electric fields is expressed by the Coulomb
interaction. 
By applying the HF approximation to the Coulomb interaction part 
and substituting Eq.~(\ref{eq4:def_of_deviations}) into 
Eq.~(\ref{eq4:total_action}), 
we rewrite the total action as %is rewritten as 
%
%== Eq ==%
%Total action
\begin{equation}
 S_{\rm tot}[\delta\rho,{\bm a},\Psi^\dag,\Psi]=S_{\rm EM}[{\bm a}]
  +S_{\rm HF}[\delta\rho,{\bm a},\Psi^\dag,\Psi],
\end{equation}
where
%
%
%== Eq ==%
%
\begin{align}
 S_{\rm EM}[{\bm a}]=&\int dtd^3x
 \left(\frac{\epsilon}{2}\dot{{\bm a}}^2({\bf x},t)
 -\frac{1}{2\mu}(\nabla\times{\bm a}({\bf x},t))^2
 \right), \nonumber \\
 S_{\rm HF}[\delta\rho,{\bm a},\Psi^\dag,\Psi]=
 &\int dtd^3x\Psi^\dag({\bf x},t)
 \left(i\partial_t-\frac{({\bf p}+e{\bf A}_{\rm ex}({\bf x})
 +e{\bm a}({\bf x},t))^2}{2m^\ast}\right)\Psi({\bf x},t)\delta(z)\nonumber \\
 &-\int dtd^3xd^3x'\Big[
 (\langle \Psi^\dag({\bf x},t) \Psi({\bf x},t)\rangle_{I=0}
 +\delta\rho({\bf x},{\bf x},t))
 V({\bf x}-{\bf x'})\Psi^\dag({\bf x}',t)\Psi({\bf x}',t)\nonumber \\
 &-(\langle\Psi^\dag({\bf x},t)\Psi({\bf x}',t)\rangle_{I=0}
 +\delta\rho({\bf x},{\bf x}',t))
 V({\bf x}-{\bf x'})\Psi^\dag({\bf x}',t)\Psi({\bf x},t)\Big]\delta(z)\delta(z').
\label{eq4:total_action2}
\end{align}
In the expression of $S_{\rm EM}[{\bm a}]$, 
the term of the uniform external magnetic field is dropped
since it gives only the same energy constant to the two HF states. 
In Eq.~(\ref{eq4:total_action2}), the term including 
$\delta\rho({\bf x,}{\bf x},t)$ 
and the term including $\delta\rho({\bf x},{\bf x}',t)$ are 
the Hartree term and the Fock term, respectively. 
As seen in Appendix \ref{appB:Fock_term}, 
the Fock term becomes negligible compared to the Hartree term 
in the long wavelength limit 
since in the momentum space, 
the Hartree term is proportional to the Coulomb potential $V({\bf k})$, 
which is $\mathcal{O}(1/k)$, and gives a larger contribution than the Fock
term for 
the small momentum $k$.
In the following calculation, the deviation of
the Fock term is dropped. 
If we introduce the potential generated by the electron density deviation as
%
%== Eq ==%
%Definition of a_0
\begin{equation}
 a_0({\bf x},t)\equiv \int d^3x'
  \frac{(-e) \delta\rho({\bf x}',{\bf x}',t)}
  {4\pi\epsilon|{\bf x}-{\bf x}'|}
  \delta(z').
  \label{eq4:scalar_potential}
\end{equation}
$S_{\rm HF}$ is rewritten as
%
%== Eq ==%
%
\begin{align}
 S_{\rm HF}[a_0,{\bm a},\Psi^\dag,\Psi]=&\int dtd^3x\Psi^\dag({\bf x},t)
 \left(i\partial_t+e a_0({\bf x},t)-
 \frac{({\bf p}+e {\bf A}_{\rm ex}({\bf x})
 +e {\bm a}({\bf x},t))^2}{2m^\ast}\right)\Psi({\bf x},t)\delta(z)\nonumber \\
 &-\int dtd^3xd^3x'\Big[
 \langle \Psi^\dag({\bf x},t)\Psi({\bf x},t)\rangle_{I=0}
 V({\bf x}-{\bf x'})\Psi^\dag({\bf x}',t)\Psi({\bf x}',t)\nonumber \\
 &\qquad -\langle\Psi^\dag({\bf x},t)\Psi({\bf x}',t)\rangle_{I=0} 
 V({\bf x}-{\bf x'})
 \Psi^\dag({\bf x}',t)\Psi({\bf x},t)\Big]\delta(z)\delta(z'). 
\end{align}
This HF action $S_{\rm HF}$ has the same form as the 
%
%The same form of the action has been already obtained from the 
Hamiltonian in the system with the infinitesimal external 
gauge field shown in Eq.~(\ref{eq4:Hamiltonian_with_a}) 
if we apply the
HF approximation to the Coulomb interaction part. 
The important difference is that 
$a_\mu({\bf x},t)$ in the present case represents not the infinitesimal 
external gauge field but 
the finite gauge field induced by the current flow. 
Although the meaning of $a_\mu({\bf x},t)$ is different, 
the effective action obtained in the previous section is 
applicable as long as $a_\mu({\bf x},t)$ is small.

The partition function is given by 
%
%== Eq ==%
%Partition function
\begin{equation}
 Z=\int {\mathcal D}{\bm a}\int {\mathcal D} 
  \Psi^\dag {\mathcal D}\Psi\, 
  e^{iS_{\rm EM}[{\bm a}]+iS_{\rm HF}[a_0,{\bm a},\Psi^\dag,\Psi]}. 
\end{equation}
Integrating out electron fields and expanding the results 
up to second order of $a_0({\bf x},t)$ and ${\bm a}({\bf x},t)$, 
we obtain the effective action $S_{\rm eff}[a_0,{\bm a}]$ as
%
%== Eq ==%
%
\begin{equation}
 Z=\int {\mathcal D}{\bm a}\, e^{iS_{\rm EM}[{\bm a}]+
  iS_0+i\Delta S_{\rm eff}[a_0,{\bm a}]}.
\end{equation}
The functional derivative of 
$(S_{\rm EM}[{\bm a}]+S_0+\Delta S_{\rm eff}[a_0,{\bm a}])$ with respect to
${\bm a}({\bf x}, t)$ gives the Maxwell's equation for ${\bm a}({\bf
x},t)$, 
%
%== Eq ==%
%Maxwell's equation for a and j
\begin{equation}
 (\epsilon \partial^2_t-\frac{1}{\mu}\nabla^2){\bm a}({\bf x},t)
  =\langle {\bf j}({\bf x},t) \rangle_a\delta(z),
  \label{eq4:vector_potential}
\end{equation}
where ${\bf j}({\bf x},t)$ is a current operator and 
$\langle \hat{O}(x) \rangle_a$ means an expectation value of an operator
$\hat{O}(x)$ for the system with finite $a_\mu({\bf x},t)$ (or
equivalently finite currents). 
The solution of this equation gives the stationary point of the action 
with respect to $a_\mu({\bf x},t)$. 
We use the action 
into which the solution of Eq.~(\ref{eq4:vector_potential}) is
substituted as the effective action. 
$\langle {\bf j}({\bf x},t) \rangle_a$ and 
$\delta\rho({\bf x},t) \equiv \delta\rho({\bf x},{\bf x},t)$ are
calculated from the effective action by
%
%== Eq ==%
%
\begin{align}
 \frac{\delta \Delta S_{\rm eff}
 [a_0, {\bm a}]}{\delta{\bm a}({\bf x},t)}=&\langle {\bf j}({\bf x},t)
 \rangle_a\delta(z), \nonumber \\
 -\frac{\delta \Delta S_{\rm eff}[a_0, {\bm a}]}{\delta a_0({\bf x},t)}=&
 (-e)[\rho_0({\bf x})+\delta\rho({\bf x},t)]\delta(z),
 \label{eq4:eff_to_j}
\end{align}
where the $\rho_0({\bf x})$ is the expectation value of the density operator in
the system with no injected current. Equations (\ref{eq4:scalar_potential}), 
(\ref{eq4:vector_potential}) and (\ref{eq4:eff_to_j}) determine 
$a_0({\bf x},t)$ and ${\bm a}({\bf x},t)$, 
or equivalently 
$\delta\rho({\bf x},t)$ and $\langle {\bf j}({\bf x},t) \rangle_a$, 
self-consistently.

\begin{figure}[tb]
\begin{center}
\includegraphics[width=5cm]{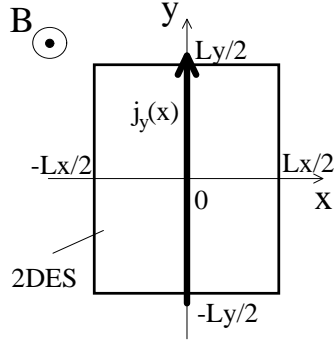}
\end{center}
\caption{\label{fig4:Current}
Schematic view of the 2D electron system in a magnetic field 
with the injected current. 
The current flows in the $y$-direction and has only 
the $x$-coordinate dependence.
}
\end{figure}

We concentrate on the finite system with  
the static injected current flowing in the
$y$-direction and depending only on $x$. 
The lengths of the 2D electron system in the $x$-direction and the $y$-direction are 
$L_x$ and $L_y$, respectively (Fig.~\ref{fig4:Current}). 
%
%== Fig ==%
%figure
In this case, the electron density also depends only on $x$ and 
Eqs.~(\ref{eq4:scalar_potential}) and (\ref{eq4:vector_potential})
give the following solutions at $z=0$: 
%
%== Eq ==%
%
\begin{align}
 a_0(x)=&-\frac{1}{2\pi\epsilon}\int^{L_x/2}_{-L_x/2} dx'\ln|x-x'|
 (-e)\delta\rho(x')\nonumber ,\\
 a_y(x)=&\frac{\mu}{2\pi}\int^{L_x/2}_{-L_x/2} dx'\ln|x-x'|
 \langle j_y(x') \rangle _a.
 \label{eq4:self_eq1}
\end{align} 
As shown in the previous section, 
the effective action can be divided into 
the non-perturbed part and the correction part due to currents,
and in the long wavelength limit, 
the correction part $\Delta S_{\rm eff}$ is given by 
%
%== Eq ==%
%
\begin{align}
 \Delta S_{\rm eff}[a_0,{\bm a}]=&
  -TL_y\int^{L_x/2}_{-L_x/2} dx (-e)\bar{\rho}_0 a_0(x)-
  \frac{TL_y}{2}\int^{L_x/2}_{-L_x/2} dx\, 
  a_\mu(x)K^{\mu\nu}_0(\partial_x) a_\nu(x),
  \label{eq4:Delta_S_eff}
\end{align}
where $\bar{\rho}_0$ is a uniform part of the density, and 
$T$ is the total time. 
$K_0^{\mu\nu}(\partial_x)$ is the Fourier transformed form 
of the response function obtained in the previous section. 
Substitution of Eq.~(\ref{eq4:Delta_S_eff}) into
Eq.~({\ref{eq4:eff_to_j}}) gives 
$\delta\rho(x)$ and $\langle j_y(x) \rangle_a$ as 
%given as
%
%== Eq ==%
%
\begin{align}
 (-e)\delta\rho(x)=&K^{00}_0(\partial_x)a_0(x)
 -K^{0y}_0({\partial_x})a_y(x)\nonumber, \\
 \langle j_y(x)\rangle_a=&
 K^{y0}_0(\partial_x)a_0(x)-K^{yy}_0(\partial_x)a_y(x).
 \label{eq4:self_eq2}
\end{align}
Equations (\ref{eq4:self_eq1}) and ({\ref{eq4:self_eq2}}) determine
the current and charge density distributions up to an overall constant. 
The overall constant is determined by requirement of %requiring 
the following constraints: 
%
%== Eq ==%
%
\begin{equation}
 \int^{L_x/2}_{-L_x/2}dxj_y(x)=I,\qquad \int^{L_x/2}_{-L_x/2}dx\delta\rho(x)=0, 
\end{equation}
where $I$ is a total current. 
Using the explicit form of the response functions derived 
in Sec.~\ref{chap4:response_functions}, 
we obtain the integral equations to determine 
the current and charge distribution.

\begin{figure}[tb]
\begin{center}
\includegraphics[width=7cm]{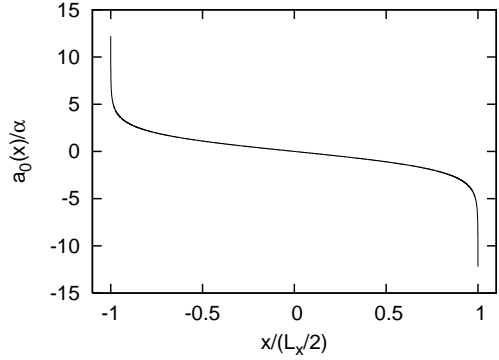}
\end{center}
\caption{\label{fig4:potential}Potential distribution $a_0(x)$. 
The first derivative of $a_0(x)$ gives 
the current distribution and the second derivative gives 
the charge distribution.}
\end{figure}

The same type of the integral equations has already been solved for the 
integer quantum Hall state \cite{MacDonald,Thouless,Beenakker1,Beenakker2}. 
Their results are summarized as follows: 
\begin{enumerate}
 \item In Eq.~(\ref{eq4:self_eq2}), the terms including the vector potential
       $a_y(x)$ give a very small effect in the integral equations 
       compared to the terms including the scalar potential $a_0(x)$ 
       and the vector potential terms are negligible in a good approximation.   
 \item The analytical solution of the integral equation without the vector
       potential term 
       is obtained by means of the Wiener-Hopf technique. 
 \item $a_0(x)={\rm const}\times \ln|(x-L_x/2)/(x+L_x/2)|$ is the good
       approximate form of the analytical solution except near the edge 
       and the constant coefficient is determined 
       from the constraint for the total current. 
\end{enumerate}
The same results hold in the present case. 
The integral equation for the potential is given by 
%
%== Eq ==%
%
\begin{equation}
 a_0(x)=-\gamma
  \int^{L_x/2}_{-L_x/2}dx'\ln|x-x'|\partial^2_{x'}a_0(x'),
  \label{eq4:integral_eq} 
\end{equation}
where $\gamma=(1+\beta \sqrt{B}/\nu)\sigma^{(\nu)}_{xy}/2\pi\epsilon\omega_c$ 
($\beta=0$ for the striped Hall state). 
$\gamma$ has the dimension of length and is very small for the magnetic
fields of the order of 
several tesla in the quantum Hall regime. 
For example, 
if $\epsilon=13\epsilon_0$, $m^\ast=0.067m_e$ (these are parameters in GaAs), 
and $\beta=0$, then $\gamma$ is of the order of $10^{-8}$ m. 
The current and charge distributions are obtained from $a_0(x)$ by 
%
%== Eq ==%
%
\begin{equation}
 \label{eq4:rho}
 (-e)\delta\rho(x)=2\pi\epsilon\gamma
 \partial^2_x a_0(x),\quad 
 \langle j_y(x)\rangle_a=-\sigma^{(\nu)}_{xy}\partial_x a_0(x).
\end{equation}
The approximate solution of Eq.~(\ref{eq4:integral_eq}) is given by 
(Fig. \ref{fig4:potential})
%
%== Fig ==%
%figure
%
%
%== Eq ==%
%
\begin{equation}
 a_0(x)=\alpha\ln\left|\frac{x-L_x/2}{x+L_x/2}\right|\qquad
  {\rm for}\quad |x|\le \frac{L_x}{2}-\gamma,
  \label{eq4:a_0}
\end{equation}
with a linear extrapolation of $a_0$ to $\pm IR_H/2$ 
in the interval within $\gamma$ from the edge, 
where $\alpha=IR_H/2(1+\ln(L_x/\gamma))$ and $R_H=1/\sigma^{(\nu)}_{xy}$
is the Hall resistivity. 
One may verify that Eq.~(\ref{eq4:a_0})
is indeed the approximate solution of the integral 
Eq.~(\ref{eq4:integral_eq}), by substituting 
Eq.~(\ref{eq4:a_0}) into Eq.~(\ref{eq4:integral_eq}) and performing one
partial integration.

%%%%%%%%%%%%%%%%%%%%%%%%%%%%%%%%%%%%%%%
% subsection 4-5-2 Energy corrections %
%%%%%%%%%%%%%%%%%%%%%%%%%%%%%%%%%%%%%%%
\subsection{Energy corrections}
The energy correction due to the injected current per 
unit space-time volume is calculated from the effective action by 
$(S_{\rm EM}[{\bm a}]+\Delta S_{\rm eff}[a_0,{\bm a}])/TL_xL_y$. 
Since in the present case of $\nu^\ast=1/2$, 
the area occupied by one particle at 
the uppermost partially-filled LL is $2a^2$ 
(here, the vNL constant $a$ is written explicitly), 
the energy correction per particle $\delta E$ is given by 
$[(S_{\rm EM}[{\bm a}]+\Delta S_{\rm eff}[a_0,{\bm a}])/TL_xL_y]\times 2a^2$. 
Substituting Eqs.~(\ref{eq4:self_eq1}) and ({\ref{eq4:self_eq2}}) into
this expression, 
we obtain the energy correction per particle given by 
%
%== Eq ==%
%
\begin{equation}
 \delta E[I]=-\frac{e^2}{2\pi\epsilon L_x}
  \int^{L_x/2}_{-L_x/2}dxdx'\delta\rho(x)\ln|x-x'|\delta\rho(x'). 
  \label{eq4:energy_correction}
\end{equation}
Substituting Eq.~(\ref{eq4:rho}) into Eq.~(\ref{eq4:energy_correction}) and
using Eq.~(\ref{eq4:integral_eq}), 
we rewrite the energy correction as 
%
%== Eq ==%
%
\begin{equation}
 \label{eq4:energy_correction2}
 \delta E[I]=\frac{(\sigma_{xy}^{(\nu)})^2}
  {2\pi\epsilon\gamma L_x\omega_c^2}
  \int^{L_x/2}_{-L_x/2}dx\, a_0(x)\partial_x^2a_0(x). 
\end{equation}
Substituting Eq.~(\ref{eq4:a_0}) into 
Eq.~(\ref{eq4:energy_correction2}) %this expression 
and performing the $x$ integral, 
we obtain the final result; 
%
%== Eq ==%
%
\begin{equation}
 \delta E[I]=\frac{\pi\epsilon}{L_x(\sigma_{xy}^{(\nu)})^2}\times
  \frac{\ln(2/b)-1}{(\ln(2/b)+1)^2}\times I^2. 
\end{equation}
where $b$ is a dimensionless constant given by $b=\gamma/(L_x/2)(\ll 1)$. 
This expression depends on the filling factor, 
the magnetic field strength and experimental parameters. 
Since the actual filling factor includes the spin degree of freedom, 
we use $\nu_{\rm ex}= 2l_0+\nu^\ast$ for lower spin
bands and $\nu_{\rm ex}= (2l_0+1)+\nu^\ast$ for upper spin bands 
instead of $\nu$. 
The magnetic field strength is related to the filling factor by 
$B=h n_e/e \nu_{\rm ex}$ ($n_e$ is an electron density). 
For example, if $n_e=2.67\times 10^{15} {\rm m}^{-2}$, then the magnetic
field strengths are 4.42 T ($\nu_{\rm ex}=5/2$), 3.15 T ($\nu_{\rm ex}=7/2$), 
2.45 T ($\nu_{\rm ex}=9/2$), 2.01 T ($\nu_{\rm ex}=11/2$), 1.70 T
($\nu_{\rm ex}=13/2$) 
and so on. 
We use $\epsilon=13\epsilon_0$, $m^\ast=0.067m_e$, 
$n_e=2.67\times 10^{15}$ m, and $L_x=5\times 10^{-3}$ m in order to estimate the values of
energy corrections, which are the parameters used in the experiment by
Lilly et al.~\cite{Lilly}. 
Then the energy correction is given by 
$\delta E[I]=C\times I^2 (q^2/l_B)$ with the coefficient $C$ shown
in Table \ref{table4:C}. 
%
%== Table ==%
%======================= Table ==========================%
\begin{table}[tb]
\setlength{\extrarowheight}{1.05mm}
\newcolumntype{Y}{>{\centering\arraybackslash}X}
\begin{tabularx}{\linewidth}{YYYYY}\hline\hline
$\nu_{\rm ex}$ & stripe & parallel  & perpendicular & $I_c$\\ \hline
5/2  & 325.0  & 330.6   & 335.6   & 0.041 \\
7/2  & 204.4  & 206.6   & 209.0   & 0.065 \\
9/2  & 144.7  & 146.1   & 147.6   & 0.040 \\
11/2 & 109.9  & 110.7   & 111.6   & 0.053 \\
13/2 & 87.44  & 88.08   & 88.68   & 0.047 \\ \hline\hline
\end{tabularx}
\caption{Values of the coefficient $C$ in units of ${\rm nA}^{-2}$ and the
 critical current $I_C$ in units of nA. 
{\it parallel} is the case in which the stripe direction 
is parallel to the current. 
{\it perpendicular} is the case in which the stripe direction is 
perpendicular to the current. 
}
\label{table4:C}
\end{table}
%========================= Table ==========================%

\begin{figure}[tb]
\begin{center}
\includegraphics[width=7cm]{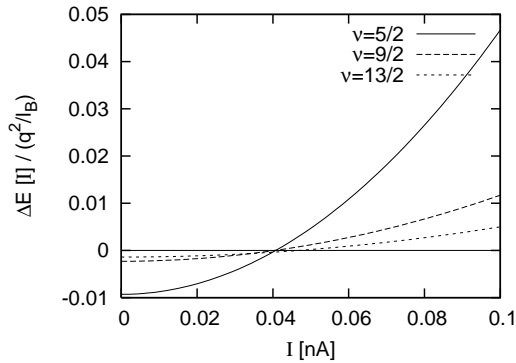}
\end{center}
\caption{\label{fig4:ene_diff}Energy differences $\Delta E[I]$
between the striped Hall state 
and the ACDW state. The results at 
$\nu=5/2, 9/2, 13/2$ are shown. 
When $\Delta\epsilon[I]$ is positive, 
the striped Hall state has lower energy.}
\end{figure}

As shown in Sec.~\ref{chap4:Hartree_Fock}, in the system with no injected current 
the energy of the ACDW state is slightly lower than 
that of the striped Hall state. The differences of energy per particle
$\Delta E_0$ 
are 
$9.3\times 10^{-3}$ ($l_0=1$), $2.3\times 10^{-3}$ ($l_0=2$), 
$1.4\times 10^{-3}$ ($l_0=3$) and so on in units of $q^2/l_B$. 
When the finite current is injected, 
charges are accumulated in both edges with the opposite sign. 
The accumulated charges give 
the energy corrections $\delta E[I]$ which depend on the value of
current $I$. Including these corrections, 
the energy difference between the striped Hall state and the ACDW state 
$\Delta E[I]=-\Delta E_0 + 
(\delta E_{\rm ACDW}[I] - \delta E_{\rm stripe}[I])$ 
varies depending on $I$. 
The current dependence of $\Delta E[I]$ is shown in
Fig.~\ref{fig4:ene_diff}. 
In Fig.~\ref{fig4:ene_diff}, 
only the parallel case is plotted for the ACDW states since 
it has a weaker current dependence than the perpendicular case does. 
%
%== Fig ==%
%figure
The signs of the energy differences change 
at the critical values of current $I_c$. 
The critical values are shown in Table \ref{table4:C}. 
The critical values are about $0.04$ - $0.05$ nA for $l=2$ and $l=3$ LLs. 
The current used in the experiments \cite{Lilly,Du} 
is above $1$ nA and is much larger than the critical value. 
At $1$ nA, the energy differences $\Delta \epsilon[I]$ become 
$1.4$ $(\nu=9/2)$ and $0.63$ $(\nu=13/2)$ in units of
$q^2/l_B$ 
which are much larger than the original energy differences at a zero
injected current. 
Hence, the striped Hall state becomes the lower energy state and 
should be realized in the experiments. 
We note that for $I=0.1$ nA, the density deviation around edges is of the order of 
0.001\% of the mean value of the density, which ensures that 
the electromagnetic fields induced by currents are kept small enough 
for the current of the order of $0.1$ nA.

%%%%%%%%%%%%%%%%%%%%%%%
% section 4-6 Summary %
%%%%%%%%%%%%%%%%%%%%%%%
\section{Summary}
\label{chap4:summary}
In this chapter, 
we have investigated the effect of the finite electric current on the 
striped Hall state and the ACDW state 
in the system with no impurities and no metallic contacts 
within the self-consistent HF approximation. 
First, at zero injected current, the energies of the two HF states have
been calculated and we have confirmed that the ACDW state has slightly 
lower energy than the striped Hall state as pointed out in 
Ref.~\cite{Cote_Fertig}. 
Next, within the same HF approximation, energy corrections due to finite
injected currents flowing in the stripe direction have been calculated 
and the energies of the two HF states including the energy corrections 
have been compared. 
For this purpose, 
we calculated electromagnetic response functions, and 
using them, 
current and charge distributions were 
determined for both states in the system with the injected current.
It has been found that the charge accumulation occurs 
around both edges with the opposite sign, 
just as in the case of the integer quantum Hall state studied by 
MacDonald et.al and other authors 
\cite{MacDonald,Thouless,Beenakker1,Beenakker2}. 
We hope that current and charge distributions 
will be observed in experiments for anisotropic states. 
The charge accumulation results in the energy enhancement 
via the Coulomb interaction 
between the accumulated charges. 
The energy enhancement was estimated 
from the current and charge distributions. 
It is found that the energy of the ACDW state increases 
faster than that of the striped Hall state does 
as the injected current increases. 
%
%In the system with no injected current, the energy of 
%the ACDW state is lower 
%than that of the striped Hall state. 
Hence, the striped Hall state becomes the lower energy state 
when the current exceeds the critical value.
The critical value is estimated at about $0.04$ - $0.05$ nA. 
The current used in the experiments for 
the anisotropic states \cite{Lilly,Du} is 
above $1$ nA. 
This result suggests that the striped Hall state is realized in the
experiments. 
In addition, 
the striped Hall state has the anisotropic Fermi surface, which 
naively explains 
the experimental features of the anisotropic states, i.e., 
the anisotropic longitudinal resistivities and the un-quantized Hall
resistivities. 
Hence, we conclude that 
the striped Hall state is realized in
the experiment rather than the ACDW state 
and 
predict that the ACDW state is realized if an experiment 
is performed with a current smaller than the critical value.

%=== Appendix ===%
\appendix

\chapter{von Neumann lattice formalism}
%%%%%%%%%%%%%%%%%%%%%%%%%%%%%%%%%%%%%%%%%%%%
% Appendix A von Neumann lattice formalism %
%%%%%%%%%%%%%%%%%%%%%%%%%%%%%%%%%%%%%%%%%%%%

%%%%%%%%%%%%%%%%%%%%%%%%%%%%%%%%%%%%%%%%%%%%%%%%%%%%%%%%%%%%%%
% section A-1 Wave function of the von Neumann lattice basis %
%%%%%%%%%%%%%%%%%%%%%%%%%%%%%%%%%%%%%%%%%%%%%%%%%%%%%%%%%%%%%%
\section{Wave function of the von Neumann lattice basis}
\label{appA:wave_function}
%EIGENSTATE OF THE MAGNETIC ANGULAR MOMENTUM OPERATOR
%
By using the von Neumann lattice basis introduced in Chapter \ref{chapter2}, 
the Hilbert space of a one-particle state of a free electron in a
magnetic field ${\bf B}=(0,0,B)$ is spanned by 
$|l,{\bf N}\rangle=|f_l\rangle \otimes |\alpha_{m,n}\rangle$.
The wave function in the coordinate space 
$\psi_{l,{\bf N}}({\bf x})=\langle {\bf x}|l,{\bf N}\rangle$ 
is given by 
%
%== Eq ==%
%
\begin{equation}
 \psi_{l,{\bf N}}({\bf x})=e^{i\pi(m+n+mn)}e^{-\pi|z_{m,n}|^2/2a^2}
  \sum_{M=0}^{\infty}\frac{1}{\sqrt{M!}}
  \left(\frac{\sqrt{\pi}\, z_{m,n}}{a}\right)^M\phi_{l,M}({\bf x}), 
\end{equation}
where $\phi_{l,M}({\bf x})=\langle {\bf x}|l,M\rangle$ with 
%
%== Eq ==%
%
\begin{equation}
 |l,M\rangle=\frac{(L_A^\dag)^l}{\sqrt{l!}}
  \frac{(L_B^\dag)^M}{\sqrt{M!}}|0,0\rangle, \quad 
  L_A|0,0\rangle = L_B |0,0\rangle =0.
\end{equation}
The explicit form of $\phi_{l,M}({\bf x})$ is obtained by using the
coordinate expression of $L_A$ and $L_B$, 
%
%== Eq ==%
%
\begin{equation}
 L_A=\frac{1}{\sqrt{2}\, l_B}
  \left[ -\frac{i}{2}(x-iy)-il_B^2\left(\frac{\partial}{\partial
   x}-i\frac{\partial}{\partial y}\right)\right], \quad 
  L_B=\frac{1}{\sqrt{2}\, l_B}
  \left[\frac{1}{2}(x+iy)+l_B^2\left(\frac{\partial}{\partial x}+
   i\frac{\partial}{\partial y}\right)\right].
\end{equation}
These expressions are rewritten by using the complex coordinates
$z=(x-iy)/l_B$ and $z^\ast=(x+iy)/l_B$ as 
%
%== Eq ==%
%
\begin{gather}
 L_A=-i\sqrt{2}\,
 e^{-|z|^2/4}\frac{\partial}{\partial{z^\ast}}e^{|z|^2/4}, \quad 
 L_A^\dag=\frac{i}{\sqrt{2}}
 e^{-|z|^2/4}(z^\ast-2\frac{\partial}{\partial z})e^{|z|^2/4}, \notag \\
 L_B=\sqrt{2}\, e^{-|z|^2/4}\frac{\partial}{\partial z}e^{|z|^2/4},\quad  
 L_B^\dag=\frac{1}{\sqrt{2}}
 e^{-|z|^2/4}(z-2\frac{\partial}{\partial z^\ast})e^{|z|^2/4}, 
\end{gather}
where the derivations act on not only the exponential function but also
the wave function on which these operators act. 
Using these expressions and performing some calculations, 
we obtain the explicit form of $\phi_{l,M}({\bf x})$;  
%
%== Eq ==%
%
\begin{equation}
 \phi_{l,M}({\bf x})=C_{l,M}e^{i(l-M)\theta-r^2/4l_B^2}
  \left(\frac{r}{l_B}\right)^{|l-M|}L^{|l-M|}_{(l+M-|l-M|)/2}
  \left(\frac{r^2}{2l_B^2}\right), 
\label{eqA:phi_lM}
\end{equation}
where $C_{l,M}$ is a normalization factor, and 
$L^n_m(x)=(e^x x^{-n}/m!)(d/dx)^m(e^{-x}x^{m+n})$ is a Laguerre
polynomial. 
In Eq.~(\ref{eqA:phi_lM}), 
the polar coordinate representation of $z$, 
$z=r(\cos\theta-i\sin\theta)/l_B$, 
is used.
If $l=0$, $\phi_{0,M}({\bf x})$ is given by 
%
%== Eq ==%
%
\begin{equation}
 \phi_{0,M}({\bf x})=\frac{1}{\sqrt{2\pi 2^M M!}\, l_B}
  \left(\frac{x-iy}{l_B}\right)^M e^{-r^2/4l_B^2}.
\label{appA:phi_0M}
\end{equation}
Equation (\ref{appA:phi_0M}) gives 
%
%== Eq ==%
%
\begin{equation}
 \phi_{0,{\bf N}}({\bf x})=\frac{1}{a}e^{i\pi(m+n+mn)}e^{-i(\pi/a)(mr_s y
  -nx/r_s)}
  e^{-(\pi/2a^2)\{(x-amr_s)^2+(y-an/r_s)^2\}}, 
\end{equation}
%
%
%== Eq ==%
%
\begin{equation}
 |\phi_{0,{\bf N}}({\bf x})|^2=\frac{1}{a^2}e^{-(\pi/a^2)
\{(x-amr_s)^2+(y-an/r_s)^2\}}.
\end{equation}
Hence, for $l=0$, the probability density is localized
at the lattice points, $a(mr_s, n/r_s)$. 
A similar argument can be applied for the case of $l\neq 0$.

%%%%%%%%%%%%%%%%%%%%%%%%%%%%%%%%%%%%%%%%%%%%%%%%%%%%%%%%%%%%%%
% section A-2 Explicit form of the                           %
%             Inner product <\alpha_{\bf p}|\alpha_{\bf p}'> %
%%%%%%%%%%%%%%%%%%%%%%%%%%%%%%%%%%%%%%%%%%%%%%%%%%%%%%%%%%%%%%
\section{Explicit form of the inner product 
$\langle \alpha_{\bf p}|\alpha_{{\bf p}'}\rangle$}
\label{appA:inner_product}
By using the momentum representation of $|\alpha_{m,n}\rangle$, 
$|\alpha_{\bf p}\rangle=\sum_{m,n}e^{imp_x+inp_y}|\alpha_{m,n}\rangle$, 
the inner product is given by 
%
%== Eq ==%
\begin{equation}
 \langle \alpha_{\bf p}|\alpha_{{\bf p}'}\rangle
  =\sum_{m,n,m',n'}e^{-i{\bf p}\cdot {\bf N}+i{\bf P}'\cdot {\bf N}'}
  \langle \alpha_{m,n}|\alpha_{m',n'}\rangle.
\label{eqA:inner_product} 
\end{equation}
Since the $\langle \alpha_{m,n}|\alpha_{m',n'}\rangle$ is a function of 
$(m-m')$ and $(n-n')$, Eq.~(\ref{eqA:inner_product}) can be rewritten as 
%
%== Eq ==%
%
\begin{gather}
 \langle \alpha_{\bf p}|\alpha_{{\bf p}'}\rangle=\alpha({\bf p})
 (2\pi)\sum_{\bf N}\delta^2({\bf p}-{\bf p}'-2\pi{\bf N}),\notag \\
 \alpha({\bf p})\equiv 
 \sum_{M,N}e^{-ip_xM-ip_yN} \langle \alpha_{M,N}|\alpha_{0,0}\rangle ,
\label{eqA:inner_product2}
\end{gather}
where the relation, 
$\sum_{n=-\infty}^{\infty}e^{ipn}=(2\pi)
\sum_{n=-\infty}^{\infty}\delta(p-2\pi n)$,  
is used.
In what follows, we derive the explicit form of $\alpha({\bf p})$.

From Eq.~(\ref{eq2:inner_product}), 
$\langle \alpha_{m,n}|\alpha_{0,0}\rangle$ is given by 
%
%== Eq ==%
%
\begin{equation}
 \langle
  \alpha_{m,n}|\alpha_{0,0}\rangle=\exp(-\frac{\pi}{2}x^tAx+i\pi(m+n)), 
\end{equation}
where 
%
%== Eq ==%
%
\begin{equation}
 x=\left(
  \begin{array}{c}
   m\\
   n
  \end{array}
  \right),\quad 
 A=\left(
    \begin{array}{cc}
     r_s^2 & -i\\
     -i & 1/r_s^2
    \end{array}
   \right).
\end{equation}
By using the identity, 
%
%== Eq ==%
%
\begin{equation}
 1=\frac{1}{2}(\det A)^{-1/2}\int^{\infty}_{-\infty}d^2s\, 
  \exp(-\frac{\pi}{2}(s-iAx)^tA^{-1}(s-iAx)), 
\label{eqA:identity}
\end{equation}
with 
%
%== Eq ==%
%
\begin{equation}
 s=\left(
    \begin{array}{c}
     s_x\\
     s_y
    \end{array}
   \right),\quad 
 \det A=2,\quad 
 A^{-1}=\frac{1}{2}\left(
		    \begin{array}{cc}
		     1/r_s^2 & i \\
		     i & r_s^2
		    \end{array}
		   \right),
\end{equation}
the index of the exponential function in Eq.~(\ref{eqA:identity}) 
can be transformed 
from the quadratic function to the linear function in $m$ and $n$ as
follows, 
%
%== Eq ==%
%
\begin{align}
 \langle \alpha_{m,n}|\alpha_{0,0}\rangle&=
 \frac{1}{2\sqrt{2}}\int^{\infty}_{-\infty}d^2s\, 
 \exp(-\frac{\pi}{2}(s-iAx)^tA^{-1}(s-iAx))\exp(-\frac{\pi}{2}x^tAx+i\pi(m+n))
 \nonumber \\
 &=\frac{1}{2\sqrt{2}}\int^{\infty}_{-\infty}d^2s\, 
 \exp(-\frac{\pi}{2}s^tA^{-1}s+i\pi (ms_x+ns_y)+i\pi (m+n)). 
 \label{eqA:inner_product3}
\end{align}
Substitution of Eq.~(\ref{eqA:inner_product3}) into 
Eq.~(\ref{eqA:inner_product2}) yields 
%
%== Eq ==%
%
\begin{equation}
 \alpha({\bf
  p})=\frac{1}{2\sqrt{2}}\int^{\infty}_{-\infty}e^{-(\pi/2)s^tA^{-1}s}\sum_{m,n}e^{-ip_xm-ip_yn+i\pi(s_x+1)m+i\pi(s_y+1)n}.  
\end{equation}
The summation of $m$ and $n$ gives the delta
function. The integration of $s_x$ and $s_y$ rewrites 
$\alpha({\bf p})$ as 
%
%== Eq ==%
%
\begin{align}
 \alpha({\bf p})=&\sqrt{2}\, 
 \sum_{m,n}\exp\left[-\frac{\pi}{2}(\frac{p_x}{\pi}+2m-1, 
 \frac{p_y}{\pi}+2n-1)A^{-1}
 \left(
 \begin{array}{c}
  p_x/\pi + 2m-1\\
  p_y/\pi + 2n-1
 \end{array}
 \right)\right]\nonumber \\
 =&\sqrt{2}\,
 \exp\left[
 -\frac{\pi}{4}
 \left\{
 \frac{1}{r_s^2}\left(\frac{p_x}{\pi}-1\right)^2+
 r_s^2\left(\frac{p_y}{\pi}-1\right)^2
 \right\}
 -
 i\frac{\pi}{2}\left(\frac{p_x}{\pi}-1\right)\left(\frac{p_y}{\pi}-1\right)
 \right]
 \nonumber \\
 &\times  
 \sum_{m}\exp\left[
 i\pi\left\{-\frac{p_y}{\pi}+1+\frac{i}{r_s^2}
 \left(\frac{p_x}{\pi}-1\right)\right\}m+
 i\pi\left(\frac{i}{r_s^2}\right)m^2\right]\nonumber \\
 &\times
 \sum_{n}\exp\left[
 i\pi\left\{-\frac{p_x}{\pi}+1+ir_s^2
 \left(\frac{p_y}{\pi}-1\right)\right\}n+
 i\pi(ir_s^2)n^2 \right]. 
\label{eqA:alpha}
\end{align} 
By using the relation between the summation of an integer $n$ and
the definition of a Jacobi's theta function of the first kind $\theta_1(v|\tau)$, 
%
%== Eq ==%
%
\begin{equation}
 \sum_{n=-\infty}^{\infty}e^{i\pi(1-\tau+2v)n+i\pi\tau n^2}=
  -ie^{-i\pi(\tau/4-v)}\theta_1(v|\tau), 
\end{equation}
the summation of $m$ in Eq.~(\ref{eqA:alpha}) is rewritten as  
%
%== Eq ==%
%
\begin{equation}
 -ie^{-i\pi(\tau_1/4-v_1)}\theta_1(v_1|\tau_1),\quad 
 \hbox{with}\quad v_1=\frac{i}{2\pi r_s^2}(p_x+ir_s^2p_y), 
 \quad \tau_1=\frac{i}{r_s^2}, 
\end{equation}
and the summation of $n$ in Eq.~(\ref{eqA:alpha}) is rewritten as 
%
%== Eq ==%
%
\begin{equation}
 -ie^{-i\pi(\tau_2/4-v_2)}\theta_1(v_2|\tau_2),\quad 
 \hbox{with}\quad v_2=-\frac{1}{2\pi}(p_x-ir_s^2p_y), 
 \quad \tau_2=\frac{i}{r_s^2}. 
\end{equation}
In addition, by using the relations between theta functions, 
%
%== Eq ==%
%theta
\begin{gather}
 \theta_1(v|\tau)=-\theta_1^\ast(-v^\ast|\tau), \\
 \theta_1(v|\tau)=e^{i3\pi/4}\tau^{-1/2}
 e^{i\pi v^2/\tau}\theta_1(v/\tau |-1/\tau),
\end{gather}
here we provide that $\tau$ is a pure imaginary, 
we obtain the explicit form of $\alpha({\bf p}))$ given by 
%
%== Eq ==%
%alpha
\begin{equation}
 \alpha({\bf p})=\beta({\bf p})\beta^\ast({\bf p}), 
\end{equation}
where 
%
%== Eq ==%
%definition of beta 
\begin{equation}
 \beta({\bf p})\equiv (\sqrt{2}\, r_s)^{1/2}e^{-(r_sp_y)^2/4\pi}
  \theta_1\left(\frac{p_x+ir_s^2p_y}{2\pi}\Biggm|ir_s^2\right). 
\end{equation}
%

%%%%%%%%%%%%%%%%%%%%%%%%%%%%%%%%%%%%%%%%%%%%%%%%%%%%%%%%%%%%%%%%% 
% section A-3 Density operator on the von Neumann lattice basis %
%%%%%%%%%%%%%%%%%%%%%%%%%%%%%%%%%%%%%%%%%%%%%%%%%%%%%%%%%%%%%%%%%
\section{Density operator on the von Neumann lattice basis}
\label{appA:density_operator}
On the von Neumann lattice basis, an electron density operator 
$\rho({\bf k})=
\int d^2x\, \Psi^\dag({\bf x})\Psi({\bf x})e^{i{\bf k}\cdot {\bf x}}$ 
is expressed as 
%
%== Eq ==%
%density operator
\begin{equation}
 \rho({\bf k})=\sum_{l,l'}\int_{\rm BZ}\frac{d^2p\, d^2p'}{(2\pi)^4}\, 
  b^\dag_l({\bf p})b_{l'}({\bf p}')\int d^2x\, 
  \langle l,{\bf p}|{\bf x}\rangle \langle {\bf x}|l',{\bf p}'\rangle
  e^{i{\bf k}\cdot{\bf x}}.  
\end{equation}
The integral over ${\bf x}$ is performed as 
%
%== Eq ==%
%inner product
\begin{equation}
 \int d^2x\langle l,{\bf p}|{\bf x}\rangle \langle {\bf x}|l',{\bf p}\rangle 
  e^{i{\bf k}\cdot {\bf x}}=\langle l,{\bf p}|
  e^{i{\bf k}\cdot({\bf X}+{\bm \xi})}
  \left[\int d^2x |{\bf x}\rangle \langle {\bf x}|\right]
  |l',{\bf p}'\rangle
  =\langle f_l|e^{i{\bf k}\cdot {\bm \xi}}|f_{l'}\rangle 
  \langle \beta_{\bf p}|e^{i{\bf k}\cdot{\bf X}}|\beta_{{\bf
  p}'}\rangle,  
\label{eqA:integral}
\end{equation}
where the operation form of ${\bf x}$, 
${\bf x}={\bf X}+{\bm \xi}$, 
and the identity ${\bf 1}=\int d^2x|{\bf x}\rangle \langle {\bf x}|$ 
are used. 
In Eq.~(\ref{eqA:integral}), 
$\langle f_l|e^{{\bf k}\cdot {\bm \xi}}| f_{l'}\rangle
\equiv f^0_{l,l'}({\bf k})$ 
is given in Appendix \ref{appA:LL_matrix}. 
Whereas, 
$\langle \beta_{\bf p}|e^{i{\bf k}\cdot {\bf X}}|\beta_{{\bf p}'}\rangle
\equiv
F({\bf p},{\bf p}',{\bf k})$ is 
calculated in what follows.

By using the relation $|\beta_{\bf p}\rangle =|\alpha_{\bf
p}\rangle/\beta({\bf p})$, 
$F({\bf p},{\bf p}',{\bf k})$ is expressed as 
%
%== Eq ==%
%F
\begin{equation}
 F({\bf p},{\bf p}',{\bf k})=
  \frac{1}{\beta^\ast({\bf p})\beta({\bf p}')}
  \langle \alpha_{\bf p}|e^{i{\bf k}\cdot {\bf X}}
  |\alpha_{{\bf p}'}\rangle.  
\end{equation}
Using the relations $X=(a/2\sqrt{\pi})(L_B+L^\dag_B)$ and
$Y=-i(a/2\sqrt{\pi})(L_B-L^\dag_B)$, 
we rewrite $e^{i{\bf k}\cdot{\bf X}}$ as 
%
%== Eq ==%
%exponential
\begin{equation}
 e^{i{\bf k}\cdot {\bf X}}=
  e^{(a/2\sqrt{\pi})(ik_x+k_y)L_B+(a/2\sqrt{\pi})(ik_x-k_y)L_B^\dag}
  =e^{-(a^2/8\pi)k^2}e^{(a/2\sqrt{\pi})(ik_x-k_y)L_B^\dag}
  e^{(a/2\sqrt{\pi})(ik_x+k_y)L_B}, 
\label{eqA:exp}
\end{equation}
here, the Cambell-Hausdorff formula Eq.~(\ref{eq2:CH_formula}) is used. 
By using Eq.~(\ref{eqA:exp}) and the relations, 
%
%== Eq ==%
%L_B
\begin{equation}
 L_B|\alpha_{m,n}\rangle
  =\frac{\sqrt{\pi}}{a}z_{m,n}|\alpha_{m,n}\rangle,\quad 
  \langle \alpha_{m,n}|L^\dag_B=\langle
  \alpha_{m,n}|\frac{\sqrt{\pi}}{a}z^\ast_{m,n}, 
\end{equation}
the matrix element $\langle \alpha_{\bf p}|e^{i{\bf k}\cdot {\bf X}}
|\alpha_{{\bf p}'}\rangle $ is calculated as  
%
%== Eq ==%
%inner product of alpha 0
\begin{align}
 \langle \alpha_{\bf p}|e^{i{\bf k}\cdot {\bf X}}
|\alpha_{{\bf p}'}\rangle =&
 \sum_{{\bf N},{\bf N}'}
 e^{-(1/8\pi)k^2}e^{(1/2)(ik_x-k_y)z^\ast_{m,n}+(1/2)(ik_x+k_y)z_{m',n'}}
 e^{-i{\bf p}\cdot {\bf N}+i{\bf p}'\cdot {\bf N}'}\nonumber \\
 =&e^{-(1/8\pi)k^2}\sum_{\bf N}
 e^{-i\{(1/2)(p_x-(r_s/2)(k_x+ik_y))\}m-i\{p_y+(1/2r_s)(ik_x-k_y)\}n}
 \langle \alpha_{m,n}|\nonumber \\
 &\times \sum_{{\bf N}'}
 e^{i\{p_x'+(r_s/2)(k_x-ik_y)\}m'+i\{p_y'+(1/2r_s)(ik_x+k_y)\}n'}
 |\alpha_{m',n'}\rangle \nonumber \\
 =&e^{-(1/8\pi)k^2}
 \langle \alpha_{p_x-(r_s/2)(k_x+ik_y),\, p_y+(1/2r_s)(ik_x-k_y)}|
 \alpha_{p_x'+(r_s/2)(k_x-ik_y),\, p_y'+(1/2r_s)(ik_x+k_y)}\rangle 
 \nonumber \\
 =&e^{-k^2/8\pi}f(p_x-\frac{r_s}{2}(k_x+ik_y),\,  
 p_y+\frac{1}{2r_s}(ik_x-k_y))
 (2\pi)^2\sum_{\bf N}\delta^2({\bf p}-{\bf p}'-\hat{\bf k}-2\pi{\bf N}),
\label{eqA:matrix_element} 
\end{align}
where we set $a=1$ and define $f({\bf p})\equiv |\beta({\bf p})|^2$.
Since the explicit form of $f({\bf p})$ is given by 
%
%== Eq ==%
%f(p) 1
\begin{equation}
 f({\bf p})=(\sqrt{2}\, r_s)e^{-(r_s p_y)^2/2\pi}
  \left[-\theta_1
   \left(-\frac{p_x-ir_s^2p_y}{2\pi}\Biggm|
    ir_s^2\right)\right]
  \theta_1\left(\frac{p_x+ir_s^2p_y}{2\pi}\Biggm|ir_s^2\right), 
\end{equation}
the function $f(p_x-(r_s/2)(k_x+ik_y),\,
p_y+(1/2r_s)(ik_x-k_y))$ 
in Eq.~(\ref{eqA:matrix_element}) is
given by 
%
%== Eq ==%
%f(p) 2
\begin{align}
 &f(p_x-\frac{r_s}{2}(k_x+ik_y),\,  
 p_y+\frac{1}{2r_s}(ik_x-k_y))\nonumber \\
 &\qquad =
 e^{k^2/8\pi}e^{-(r_s p_y)^2/4\pi}e^{-r_s^2(p_y-k_y/r_s)^2/4\pi}
 e^{-(ir_s/4\pi)k_x(2p_y-k_y/r_s)}\nonumber \\
 &\qquad \times 
 \left[-\theta_1\left(-\frac{p_x-ir_s^2p_y}{2\pi}\Biggm|ir_s^2\right)\right]
 \theta_1\left(\frac{(p_x-r_s k_x)+
 ir_s^2(p_y-k_y/r_s)}{2\pi}\Biggm|ir_s^2\right).
\end{align}
Hence, 
%
%== Eq ==%
%inner product of alpha
\begin{equation}
 \langle \alpha_{\bf p}|e^{i{\bf k}\cdot {\bf X}}|\alpha_{{\bf
  p}'}\rangle =\beta^\ast({\bf p})\beta({\bf p}-\hat{\bf k})
  e^{-(ir_s/4\pi)k_x(2p_y-k_y/r_s)}(2\pi)^2\sum_{\bf N}
  \delta^2({\bf p}-{\bf p}'-\hat{\bf k}-2\pi{\bf N}), 
\end{equation}
and 
%
%== Eq ==%
%inner product of beta
\begin{equation}
 \langle \beta_{\bf p}|e^{i{\bf k}\cdot{\bf X}}|\beta_{{\bf p}'}\rangle 
  =e^{-(ir_s/4\pi)k_x(2p_y-k_y/r_s)}(2\pi)^2
  \sum_{\bf N}\delta^2({\bf p}-{\bf p}'-2\pi{\bf N})e^{i\phi({\bf
  p}',{\bf N})}, 
\end{equation}
where the boundary condition for $\beta({\bf p})$ is used. 
As a result, the density operator $\rho({\bf k})$ is expressed on the
vNL basis as 
%
%== Eq ==%
%Density on vNL
\begin{equation}
 \rho({\bf k})=\sum_{l,l'}
  \int_{\rm BZ}\frac{d^2p}{(2\pi)^2}b^\dag_l({\bf p})
  b_{l'}({\bf p}'-\hat{\bf k})f^0_{l,l'}({\bf k})
  e^{-(ir_s/4\pi)k_x(2p_y-k_y/r_s)}. 
\end{equation}
%

%%%%%%%%%%%%%%%%%%%%%%%%%%%%%%%%%%%%%%%%%%%%
% section A-4 Landau level matrix elements %
%%%%%%%%%%%%%%%%%%%%%%%%%%%%%%%%%%%%%%%%%%%% 
\section{Landau level matrix elements}
\label{appA:LL_matrix}
The matrix elements $\langle l_1 |e^{i{\bf q}\cdot {\bm \xi}}|l_2
\rangle$ are given as follows: 
%
%== Eq ==%
%LL_matrix
\begin{equation}
 \langle l_1 |e^{i {\bf q}\cdot{\bm \xi}}| l_2
  \rangle\\
 =
  \left\{
   \begin{array}{ll}
    \sqrt{\frac{l_1!}{l_2!}}
     \left(
      \frac{q_x+iq_y}{\sqrt{4\pi}}
     \right)^{l_2-l_1}e^{-q^2/8\pi}L_{l_1}^{l_2-l_1}
     \left(
      \frac{q^2}{4\pi}
     \right)
     & \hbox{for $l_2>l_1$}
     \\
    \sqrt{\frac{l_2!}{l_1!}}
     \left(
      \frac{q_x-iq_y}{\sqrt{4\pi}}
     \right)^{l_1-l_2}e^{-q^2/8\pi}L_{l_2}^{l_1-l_2}
     \left(
      \frac{q^2}{4\pi}
      \right)
     & \hbox{for $l_1>l_2$}\\
    e^{-q^2/8\pi}L_{l_1}
     \left(
      \frac{q^2}{4\pi}
       \right)
     & \hbox{for $l_2=l_1$},
   \end{array}
       \right.,  
  \label{eq:def_of_LL}
\end{equation}
where $L_l^{l'}(x)$ is a Laguerre polynomial. 
For $l'=0$, the Laguerre polynomial is given by 
%
%== Eq ==%
%Laguerre polynomial
\begin{align}
 &L^0_l(x)=\sum^{l}_{n=0}(-1)^n\frac{l!}{(l-n)!n!n!}x^n,\nonumber \\
 &L^0_0(x)=1, \nonumber \\
 &L^0_1(x)=1-x, \nonumber \\
 &L^0_2(x)=1-2x+\frac{1}{2}x^2, \nonumber \\
 &L^0_3(x)=1-3x+\frac{3}{2}x^2-\frac{1}{6}x^3.
\end{align}
From Eq.~(\ref{eq:def_of_LL}), 
the matrix elements $f^\mu_{l_1,l_2}({\bf q})$ defined by 
Eq.~(\ref{eq4:def_of_f}) are given by 
$f^0_{l_1, l_2}({\bf q})=
\langle l_1|e^{i{\bf q}\cdot {\bm \xi}}|l_2\rangle$, 
$f^x_{l_1,l_2}({\bf q})=i\omega_c\partial_{q_y}
\langle l_1|e^{i{\bf q}\cdot {\bm \xi}}|l_2\rangle$ 
and 
$f^y_{l_1,l_2}({\bf q})=-i\omega_c\partial_{q_x}
\langle l_1|e^{i{\bf q}\cdot {\bm \xi}}|l_2\rangle$. 
Note that $\{f^\mu_{l_1,l_2}(-{\bf k})\}^\ast=f^\mu_{l_2,l_1}({\bf k})$ holds 
following from its definition.

The values of $f^\mu_{l_1,l_2}(0)$ and its derivatives are given by 
%
%== Eq ==%
%Derivation of f^{\mu\nu}
\begin{align}
 &f_{l_1,l_2}^{0}(0)=\delta_{l_1,l_2},\nonumber \\
 &f_{l_1,l_2}^{x}(0)=\left.
 i\omega_c \frac{\partial f_{l_1,l_2}^{0}(q)}{\partial q_y}\right|_{q=0}
 =
 \left\{
 \begin{array}{ll}
  -\omega_c \sqrt{\frac{l_1+1}{4\pi}}
   \delta_{l_2,l_1+1}& \hbox{for $l_2>l_1$}\\
  \omega_c\sqrt{\frac{l_1}{4\pi}}
   \delta_{l_1,l_2+1}& \hbox{for $l_1>l_2$}\\
  0 & \hbox{for $l_1=l_2$}
 \end{array}
 \right. ,\\
 &f_{l_1,l_2}^{y}(0)=
 \left.
 -i\omega_c \frac{\partial f_{l_1,l_2}^{0}(q)}{\partial q_x}\right|_{q=0}
 =
 \left\{
 \begin{array}{ll}
  -i\omega_c\sqrt{\frac{l_1+1}{4\pi}}
   \delta_{l_2,l_1+1}& \hbox{for $l_2>l_1$}\\
  -i\omega_c\sqrt{\frac{l_1}{4\pi}}
   \delta_{l_1,l_2+1}& \hbox{for $l_1>l_2$}\\
  0 & \hbox{for $l_1=l_2$}
 \end{array}
 \right., 
\end{align}
%
%
%== Eq ==%
%Derivation of f^{\mu\nu} 2
\begin{align}
 &\left.
 \frac{\partial f_{{l_1,l_2}}^{x}(q)}{\partial q_y}
 \right|_{q=0}
 =
 \left\{
 \begin{array}{ll}
  -i\frac{\omega_c}{4\pi}
   \sqrt{(l_1+1)(l_1+2)}\delta_{l_2,l_1+2}
   &  \hbox{for $l_2>l_1$}\\
  -i\frac{\omega_c}{4\pi}
   \sqrt{l_1(l_1-1)}\delta_{l_1,l_2+2}
   &  \hbox{for $l_1>l_2$}\\
  -i\frac{\omega_c}{4\pi}(l_1+\frac{1}{2}) & \hbox{for $l_1=l_2$}\\
 \end{array}
 \right. ,\\
 %%%
 &\left. 
 \frac{\partial f_{{l_1,l_2}}^{y}(q)}{\partial q_x}\right|_{q=0}
 =
 \left\{
 \begin{array}{ll}
  -i\frac{\omega_c}{4\pi}
   \sqrt{(l_1+1)(l_1+2)}\delta_{l_2,l_1+2}
   &  \hbox{for $l_2>l_1$}\\
  -i\frac{\omega_c}{4\pi}
   \sqrt{l_1(l_1-1)}\delta_{l_1,l_2+2}
   &  \hbox{for $l_1>l_2$}\\
  i\frac{\omega_c}{4\pi}(l_1+\frac{1}{2})
   & \hbox{for $l_1=l_2$}
 \end{array}
 \right..
\end{align}
%

%%%%%%%%%%%%%%%%%%%%%%%%%%%%%%
% section A-5 HF Hamiltonian %
%%%%%%%%%%%%%%%%%%%%%%%%%%%%%%
\section{Hartree-Fock Hamiltonian}
\label{appA:HF_Hamiltonian}
In the HF theory, the Coulomb potential term given by
Eq.~(\ref{eq2:Coulomb_int_term}) is approximated by
Eq.~(\ref{eq2:Coulomb_int_term_HF}). 
In the summation in Eq.~(\ref{eq2:Coulomb_int_term_HF}), 
both the Hartree term and the Fock term are 
proportional to the density operator $\bar{\rho}_{l_3,l_4}(\tilde{\bf
k})$. 
This is a characteristic of the quantum Hall system \cite{HF0,HF}. 
In what follows, we derive Eq.~(\ref{eq2:Coulomb_int_term_HF}) 
from Eq.~(\ref{eq2:Coulomb_int_term}). 

\subsection{Hartree term}
The Hartree term is given by 
%
%== Eq ==%
% 
\begin{equation}
 \mathcal{V}_{{\rm Hartree}}=\sum_{l_1,l_2,l_3,l_4}
  \int\frac{d^2 k}{(2\pi)^2}\int_{{\rm BZ}}
  V({\bf k})f^0_{l_1,l_2}({\bf k})f^0_{l_3,l_4}(-{\bf k})
  \langle \bar{\rho}_{l_1,l_2}({\bf k})\rangle 
  \bar{\rho}_{l_3,l_4}(-{\bf k}). 
\end{equation}
Transforming the variable ${\bf k}$ to $-\tilde{\bf k}$ and using the
relation $V(-{\bf k})=V({\bf k})$, 
we obtain 
%
%== Eq ==%
\begin{equation}
 \mathcal{V}_{{\rm Hartree}}=\sum_{l_1,l_2,l_3,l_4}
  \int\frac{d^2 k}{(2\pi)^2}\int_{{\rm BZ}}
  V(\tilde{\bf k})f^0_{l_1,l_2}(-\tilde{\bf k})f^0_{l_3,l_4}(\tilde{\bf k})
  \langle \bar{\rho}_{l_1,l_2}(-\tilde{\bf k})\rangle 
  \bar{\rho}_{l_3,l_4}(\tilde{\bf k}). 
\end{equation}

\subsection{Fock term}
Fock term is given by  
%
%== Eq ==%
%
\begin{equation}
 \mathcal{V}_{{\rm Fock}}=
 -\sum_{l_1,l_2,l_3,l_4}\int\frac{d^2k}{(2\pi)^2}\int_{{\rm BZ}}
 \frac{d^2pd^2p'}{(2\pi)^4}
 \langle b^\dag_{l_1}({\bf p})b_{l_2}({\bf p}')\rangle
 b^\dag_{l_3}({\bf p}'-{\bf k})b_{l_4}({\bf p}-{\bf k})v^{\rm
 Fock}_{l_1,l_2,l_3,l_4}(\tilde{\bf k}),  
\end{equation}
with 
%
%== Eq ==%
\begin{equation}
v^{\rm Fock}_{l_1,l_2,l_3,l_4}({\bf k})=V({\bf k})f^0_{l_1,l_4}(-{\bf k})
 f^0_{l_3,l_2}({\bf k})
 e^{-(i/2\pi) k_x(p_y-p'_y)}.
\end{equation}
For simplicity, we define $\mathcal{V}_{l}$ by 
$\mathcal{V}_{\rm
Fock}=-\sum_{l_1,l_2,l_3,l_4}\mathcal{V}_{l}$ and 
calculate $\mathcal{V}_{l}$.

If we insert the identity 
%
%== Eq ==%
\begin{equation}
 1=\int\frac{d^2 q}{(2\pi)^2}(2\pi)^2\delta({\bf q}-{\bf k})
  e^{(i/2\pi)k_y(p'_x-p_x)-(i/2\pi)k_y(p'_x-p_x)}
\end{equation}
into $\mathcal{V}_{l}$, we obtain 
%
%== Eq ==%
\begin{align}
 \mathcal{V}_{l}=&\int\frac{d^2q}{(2\pi)^2}\int\frac{d^2k}{(2\pi)^2}
 \int_{\rm BZ}\frac{d^2p d^2p'}{(2\pi)^4}
 \langle b^\dag_{l_1}({\bf p})b_{l_2}({\bf p}')\rangle 
 b^\dag({\bf p}'-{\bf k})b({\bf p}-{\bf k})v^{\rm
 Fock}_{l_1,l_2,l_3,l_4}(\tilde{\bf k}) \nonumber \\
 &\times e^{-(i/2\pi)q_x(p_y-p'_y)-(i/2\pi)q_y(p'_x-p_x)}(2\pi)^2
 \delta^2({\bf q}-{\bf k})e^{(i/2\pi)k_y(p'_x-p_x)}. 
\end{align}
Dividing the integral range of the ${\bf k}$ integral by 
%
%== Eq ==%
\begin{equation}
 \int\frac{d^2 k}{(2\pi)^2} F({\bf k})=\sum_{\bf N}\int_{\rm BZ}
\frac{d^2 k}{(2\pi)^2}F({\bf k}-2\pi{\bf N}), 
\end{equation}
and using the relation derived from the boundary condition 
Eq.~(\ref{eq2:boundary_condition_for_b}) 
%
%== Eq ==%
\begin{equation}
 b^\dag({\bf p}'-{\bf k}+2\pi{\bf N})b({\bf p}-{\bf k}+2\pi{\bf N})=
  b^\dag({\bf p}'-{\bf k})b({\bf p}-{\bf k})e^{i(p'_x-p_x)N_y}, 
\end{equation}
we rewrite $\mathcal{V}_l$ as 
%
%== Eq ==%
\begin{align}
 \mathcal{V}_l=&\int_{\rm BZ}
 \frac{d^2 q}{(2\pi)^2}\frac{d^2k d^2p d^2p'}{(2\pi)^6}
 \langle b_{l_1}^\dag({\bf p})b_{l_2}({\bf p}')\rangle 
 b^\dag_{l_3}({\bf p}'-{\bf k})b_{l_4}({\bf p}-{\bf k})
 v^{\rm Fock}_{l_1,l_2,l_3,l_4}(\tilde{\bf q}) \nonumber \\
 &\times e^{-(i/2\pi)q_x(p_y-p'_y)-(i/2\pi)q_y(p'_x-p_x)}
 \sum_{\bf N}(2\pi)^2\delta^2({\bf q}-{\bf k}+2\pi{\bf N})
 e^{(i/2\pi)k_y(p'_x-p_x)}. 
\end{align}
Using the relation 
%
%== Eq ==%
\begin{equation}
 \sum_{\bf N}(2\pi)^2\delta^2({\bf q}-{\bf k}+2\pi{\bf N})=\sum_{\bf N}
  e^{-i(q_x-k_x)N_y+i(q_y-k_y)N_x}, 
\end{equation}
and rewriting the variable ${\bf k}$ as ${\bf p}'$ and ${\bf p}'$ as
${\bf k}$, we obtain 
%
%== Eq ==%
\begin{align}
 \mathcal{V}_l=&
 \sum_{\bf N}\int\frac{d^2 q}{(2\pi)^2}
 \int_{\rm BZ}\frac{d^2k d^2p d^2p'}{(2\pi)^6}\langle 
 b_{l_1}^\dag({\bf p})b_{l_2}({\bf k})\rangle 
 b^\dag_{l_3}({\bf k}-{\bf p}')b({\bf p}-{\bf p}')
 v^{\rm Fock}_{l_1,l_2,l_3,l_4}(\tilde{\bf q})\nonumber \\
 &\times 
 e^{-(i/2\pi)q_x(p_y-k_y)-(i/2\pi)q_y(k_x-p_x)+(i/2\pi)p'_y(k_x-p_x)}
 e^{-i(q_x-p_x')N_y+i(q_y-p_y')N_x}.
\end{align}
Transforming ${\bf k}$ to $-{\bf k}+{\bf p}$ and ${\bf p}'$ to $-{\bf
p}'$, 
we obtain 
%
%== Eq ==%
\begin{align}
 \mathcal{V}_l=&
 \sum_{\bf N}\int\frac{d^2 q}{(2\pi)^2}
 \int_{\rm BZ}\frac{d^2k d^2p d^2p'}{(2\pi)^6}
 \langle b_{l_1}^\dag({\bf p})b_{l_2}({\bf p}-{\bf k})\rangle 
 b^\dag_{l_3}(-{\bf k}+{\bf p}+{\bf p}')b_{l_4}({\bf p}+{\bf p}')
 v^{\rm Fock}_{l_1,l_2,l_3,l_4}(\tilde{\bf q})\nonumber \\
 &\times e^{-(i/2\pi)q_xk_y+(i/2\pi)q_yk_x}e^{(i/2\pi)p'_yk_x}
 e^{-i(q_x+p'_x)N_y+i(q_y+p'_y)N_x}.
\label{eqA:V_Fock1}
\end{align}

Using the relation 
%
%== Eq ==%
\begin{equation}
 b_{l_2}({\bf p}-{\bf k})b^\dag_{l_3}(-{\bf k}+{\bf p}+{\bf p}')=
  b_{l_2}({\bf p}-{\bf k}-2\pi{\bf N})
  b^\dag_{l_3}(-{\bf k}-2\pi{\bf N}+{\bf p}+{\bf p}')e^{ip'_xN_y}, 
\end{equation}
we rewrite Eq.~(\ref{eqA:V_Fock1}) as 
%
%== Eq ==%
\begin{align}
 \mathcal{V}_{l}=&\sum_{\rm N}\int_{\rm BZ}\frac{d^2 k}{(2\pi)^2}
 \int\frac{d^2q}{(2\pi)^2}\frac{d^2p d^2p'}{(2\pi)^4}
 \langle b_{l_1}^\dag({\bf p})b_{l_2}({\bf p}-{\bf k}-2\pi{\bf
 N})\rangle 
 b^\dag_{l_3}(-{\bf k}-2\pi{\bf N}+{\bf p}+{\bf p}')b_{l_4}({\bf p}+{\bf p}')
 \nonumber \\ 
 &\times v^{\rm Fock}_{l_1,l_2,l_3,l_4}(\tilde{\bf q})
 e^{-(i/2\pi)q_x(k_y+2\pi N_y)+(i/2\pi)q_y(k_x+2\pi N_x)}
 e^{(i/2\pi)p'_y(k_x+2\pi N_x)}
\end{align}
Using the relation 
%
%== Eq ==%
\begin{equation}
 \sum_{\bf N}\int_{\rm BZ}\frac{d^2 k}{(2\pi)^2}F({\bf k}+2\pi{\bf N})=
  \int \frac{d^2 k}{(2\pi)^2}F({\bf k}), 
\end{equation}
and transforming ${\bf p}'$ to ${\bf p}'-{\bf p}$, 
we finally obtain 
%
%== Eq ==%
\begin{align}
 \mathcal{V}_l=&\int\frac{d^2 k}{(2\pi)^2}\frac{d^2 p d^2p'}{(2\pi)^4}
 \langle b^\dag_{l_1}({\bf p})b_{l_2}({\bf p}-{\bf k})\rangle 
 b^\dag_{l_3}({\bf p}')-{\bf k})b_{l_4}({\bf p}')e^{(i/2\pi)(p'_y-p_y)k_x}
 \nonumber \\
 &\times \int\frac{d^2q}{(2\pi)^2}v^{\rm
 Fock}_{l_1,l_2,l_3,l_4}(\tilde{\bf q})
 e^{-(i/2\pi)(q_xk_y-q_yk_x)}. 
\end{align}
Thus, the combination of the Hartee term and the Fock term gives 
\begin{equation}
 \mathcal{V}_{\rm HF}=\sum_{l_1,l_2,l_3,l_4}
  \int\frac{d^2k}{(2\pi)^2}v^{\rm HF}_{l_1,l_2,l_3,l_4}(\tilde{\bf k})
  \langle\bar{\rho}_{l_1,l_2}(-\tilde{\bf k})\rangle
  \bar{\rho}_{l_3,l_4}(\tilde{\bf k}), 
\end{equation}
with the HF potential 
%
%== Eq ==%
%HF potential
\begin{equation}
 v^{\rm HF}_{l_1,l_2,l_3,l_4}({\bf k})=
 V({\bf k})
 f_{l_1,l_2}^0(-{\bf k})f_{l_3,l_4}^0({\bf k})-
 \int\frac{d^2k'}{(2\pi)^2}V({\bf k}')f_{l_1,l_4}^0(-{\bf k}')
 f_{l_3,l_2}^0({\bf k}')e^{-(i/2\pi)(k'_xk_y-k'_yk_x)}.  
\end{equation}

\chapter{Highly anisotropic states}
%%%%%%%%%%%%%%%%%%%%%%%%%%%%%%%%%%%%%%%%%%%%%%%%%%%%
%%%%%%%%%%%%%%% Appendix %%%%%%%%%%%%%%%%%%%%%%%%%%%
%%%%%%%%%%%%%%%%%%%%%%%%%%%%%%%%%%%%%%%%%%%%%%%%%%%%
\section{Deviation of the Hartree-Fock Hamiltonian due to currents}
\label{appB:Fock_term}
In Sec.~\ref{chap4:energy_corrections}, 
the deviations of the magnetic field and the two-point function 
are taken into account in the long
wavelength limit in order to consider the current effect on the HF
states. The deviation of the two-point function is caused by 
the deviation of $\langle \bar{\rho}_{l_1,l_2}(-\tilde{\bf k}) \rangle$. 
We only consider the deviation at the partially filled LL $l_0$ 
since it would give the largest contribution in our calculation 
and denote it as $\delta\bar\rho_{l_0}(-\tilde{\bf k})$.  
Then, the deviation of $\mathcal{H}_{\rm HF}$ is given by 
%
%== Eq ==%
%
\begin{equation}
 \delta\mathcal{H}_{\rm HF}=\sum_{l_3,l_4}
 \int\frac{d^2k}{(2\pi)^2}v_{l_0,l_0,l_3,l_4}^{\rm HF}(\tilde{\bf k})
 \delta\bar{\rho}_{l_0}(-\tilde{\bf k})\bar{\rho}_{l_3,l_4}(\tilde{\bf k}). 
\end{equation} 
In the long wavelength limit, 
$\delta\bar\rho_{l_0}(-\tilde{\bf k})$ is relevant only for the small
momentum. 
When we expand $v_{l_1,l_2,l_3,l_4}^{\rm HF}(\tilde{\bf k})$ with respect to
${\bf k}$, 
the largest contribution comes from the lowest order term  
with respect to ${\bf k}$. 
For each set of the LLs $(l_3,l_4)$, 
the Hartree term of the HF potential gives the lower order term with respect
to ${\bf k}$ since the Hartree term has $V({\bf k})$ which is
$\mathcal{O}(1/k)$. Hence, the Hartree term gives the main contribution 
and the Fock term is negligible in the long wavelength limit.

\section{Calculation of $K^{00}$ for the ACDW state}
\label{appB:K00}
When $p_0=p_y=0$ and $p_x\to 0$, the response function $K^{00}_0(p_x)$
is Taylor expanded with respect to $p_x$ as 
%
%== Eq ==%
%
\begin{equation}
 K^{00}_0(p_x)=K^{00}_0(0)+p_x\partial_{p_x}K^{00}_0(0)+
  \frac{p^2_x}{2}\partial^2_{p_x}K^{00}_{0}(0)+\dots
\end{equation} 
The first and second terms become zero. 
The third term includes the corrections from the inter-LL term and the
intra-LL term. The inter-LL term gives the same expression for
$K^{00}_0$ as that in the striped Hall state. 
The intra-LL term gives the extra correction $\Delta K^{00}_0(p_x)$
given by 
%
%== Eq ==%
%
\begin{equation}
\Delta K^{00}_0(p_x)=\kappa \times p^2_x, 
\end{equation}
where $\kappa$ is given by 
%
%== Eq ==%
%
\begin{equation}
 \kappa=\frac{e^2}{2}
  \int_{\rm RBZ}\frac{d^2p'}{(2\pi)^2}\frac{1}{2\epsilon({\bf p'})}
  \partial^2_{p_x}
  \left. \left[
   \frac{A(\hat{\bf p}+{\bf p}')A({\bf p}')+
   {\rm Re}(B(\hat{\bf p}+{\bf p}')B^\ast({\bf p}')e^{-(i/2)\hat{p}_x})}
   {\epsilon(\hat{\bf p}+{\bf p}')\epsilon({\bf p}')}
  \right]
  \right|_{p_x=0}.
\end{equation}
$\beta$ in Eq.~(\ref{eq4:K_for_ACDW}) is defined by 
$\beta\equiv - \kappa \times \nu \omega_c/\sigma_{xy}^{(\nu)} \sqrt{B}$. 
The finite $\kappa$ is the result of the band formation at 
the partial filled LL, 
while the band structure is generated by 
the density modulation of the ACDW state in both directions. 
Hence, it may be considered that 
the finite $\kappa$ reflects 
the remaining density modulation effect of the ACDW state 
in the long wavelength limit.

%=== Bibliography ===%
%%%%%%%%%%%%%%%%%%%%%%%%%%%%%%%%%%%%%%%%%%
%%%%%%%%%%% References %%%%%%%%%%%%%%%%%%%
%%%%%%%%%%%%%%%%%%%%%%%%%%%%%%%%%%%%%%%%%%

\end{document}